\documentclass[acmsmall, noacm]{acmart}
\usepackage[linesnumbered,ruled,vlined]{algorithm2e}
\usepackage{multirow}
\usepackage{wrapfig,lipsum,booktabs}
\usepackage{colortbl}
\usepackage{adjustbox}
\usepackage{subcaption} 
\usepackage{amsmath}
\usepackage[normalem]{ulem}
\usepackage{tabularx}
\usepackage{ragged2e}
\AtBeginDocument{%
	}

\usepackage{tcolorbox}

\setcopyright{none}
\acmDOI{}
\acmISBN{}
\renewcommand\footnotetextcopyrightpermission[1]{}
\newtcolorbox{takeawaybox}{
	colback=gray!5,
	colframe=black!30,
	boxrule=0.5pt,
	coltitle=black,
	arc=2pt,
	left=6pt,
	right=6pt,
	top=4pt,
	bottom=4pt,
	title=\textbf{Key takeaway},
	fonttitle=\small,
}

\newcommand{\ciao}[1]{{\setlength\fboxrule{0pt}\fbox{\tcbox[colframe=black,colback=white,shrink tight,boxrule=0.2pt,extrude by=0.5mm]{\small #1}}}}

\newcommand{\ie}{{\textit{i.e.,} }}
\newcommand{\eg}{{\textit{e.g.,} }}

\newcommand{\myverb}[1]{%
	\unskip%
	\hspace{0.2em}%
	\mbox{%
		\fontsize{9}{10}%
		\usefont{OT1}{lmtt}{b}{n}%
		\spaceskip=0.2em\relax%
		#1%
	}%
	\ignorespaces%
}

\newcommand{\myverbS}[1]{%
	\unskip%
	\hspace{0.2em}%
	\mbox{%
		\fontsize{8}{9}%
		\usefont{OT1}{lmtt}{b}{n}%
		\spaceskip=0.2em\relax%
		#1%
	}%
	\ignorespaces%
}

\begin{document}

\title{Assessing Resilience in Authoritative DNS Infrastructure Supporting Government Services}

\author{Agung Septiadi}
\affiliation{%
  \institution{University of New South Wales}
  \city{Sydney}
  \country{Australia}
}

\author{Minzhao Lyu}
\affiliation{%
  \institution{University of New South Wales}
  \city{Sydney}
  \country{Australia}
}\thanks{This is the preprint version of our ACM SIGMETRICS 2026 paper \cite{ASeptiadiSIGMETRICS26}. \\Corresponding author: Minzhao Lyu (minzhao.lyu@unsw.edu.au).}

\author{Hassan Habibi~Gharakheili}
\affiliation{%
  \institution{University of New South Wales}
  \city{Sydney}
  \country{Australia}
}

\author{Vijay Sivaraman}
\affiliation{%
  \institution{University of New South Wales}
  \city{Sydney}
  \country{Australia}
}

\begin{abstract}
Online government services, such as taxation, civil services, and immigration, are increasingly regarded as critical national infrastructure. Because these services directly influence public trust, any disruption can have significant societal and political consequences. Yet their supporting infrastructures remain vulnerable to outages from natural disasters, geopolitical tensions, and targeted attacks. Central to their operation is the authoritative Domain Name System (DNS) infrastructure, the single source of truth that maps government domain names to service endpoints. While indispensable, this infrastructure also represents a potential and critical point of system failure. 
In this paper, we introduce a \textbf{comprehensive assessment framework} with purpose-designed mechanisms to systematically evaluate the operational resilience of authoritative DNS infrastructure supporting government services. Complementing prior studies on website hosting, recursive resolution, and DNS record integrity, our work provides a holistic view of authoritative DNS operation.
Our first contribution develops a multi-sourced data schema that characterizes a (government) domain's authoritative DNS infrastructure across four hierarchical levels: physical hosting infrastructure, server functionality, name servers, and individual hosting instances. Using data collected from six representative countries, our second contribution identifies resilience attributes at their finest applicable hierarchy across three operational phases: infrastructure placement, service configuration, and DNS record dispatch. Our method assigns numerical scores to each attribute and aggregates them algorithmically to enable consistent and cross-domain comparisons. Lastly, we apply our method to government domains in the six countries, highlighting their strengths and weaknesses in authoritative DNS resilience and pinpointing operational practices that require improvement.
\end{abstract}

\maketitle

\section{Introduction}

Public services delivered by government agencies support essential societal functions, including taxation, healthcare, social welfare, civil service, immigration, and non-profit initiatives. As these services increasingly transition from in-person delivery to online platforms worldwide, an evolution reflected globally in the United Nation's E-government Development Index \cite{united_e-gov_2024}, the underlying digital systems have become critical infrastructure and the focus of substantial public investment. For example, in 2025, the U.S. government allocated approximately US\$75.1 billion for civilian federal information technology (IT) systems \cite{congress_information_2024}, and the European Union earmarked roughly €208.1 billion to advance the digital transformation across public service infrastructure \cite{european_digital_2025}.

Given this centrality, disruptions to  government digital infrastructure constitute high-priority incidents for national and local authorities. Unfortunately, such disruptions are not uncommon. Between January and May 2025 alone, 21 documented attacks targeted government digital infrastructure \cite{csis_significant_2025}, many propagating across interconnected  systems and causing prolonged service outages \cite{pallarito_cloudflare_2022}. Australia, for example, reported an 11\% annual rise in cyber incidents affecting government digital infrastructure in 2025 \cite{acsc_annual_2025}.
These disruptions arise from various causes, ranging from natural disasters that physically damage network assets to cyberattacks (by criminal or state-linked actors) that degrade access or compromise data integrity.

Among the digital systems required for government service availability, the \textbf{authoritative Domain Name System} (DNS) infrastructure is especially critical. It is the exclusive mechanism for mapping  a service's domain names to the IP address of its hosting server instance. If this infrastructure becomes unreachable, such as during the large-scale Amazon AWS DNS outage in 2025 \cite{robinson_single_2025}, users cannot access the service, 
even if the website itself remains operational. If authoritative DNS records are tampered with, attackers can redirect users to fraudulent sites and harvest sensitive information.

Given these risks, industry groups have long promoted operational best practices to strengthen service resilience \cite{mitchell_what_2025}, and many governments, including those of the U.S., the U.K., and Australia, fund programs aimed at improving the robustness of their digital infrastructure \cite{congress_information_2024, agor_quarterly_2025, uksecretary_state_2025}. Despite these efforts, evaluating authoritative DNS infrastructures remains a challenge. 
First, as we will discuss in \S\ref{sec:background}, operating authoritative DNS for a service domain is more complex than often assumed.
Second, public-service DNS infrastructures within a country are operated in a highly distributed
manner by government-owned entities, trusted third-party local operators, or foreign (cloud) service providers.
As a result, assessing DNS resilience at scale requires integrating diverse information from multiple distributed data sources.

Motivated by these challenges, prior research has examined various dimensions of network resilience across infrastructure supporting (government) digital services. For example, Habib \textit{et al.} \cite{habib_formalizing_2025} developed a statistical framework that quantifies Internet dependence using measures of centralization and regionalization across website hosting, DNS resolution, and certificate authority ecosystems, demonstrating their influence on infrastructure resilience. From a placement perspective, Kumar \textit{et al.} \cite{kumar_choices_2024} analyzed the organizational types, such as third-party providers or state-owned enterprises, that host government websites across countries. Geopolitical factors have also been studied, including DNS, PKI, and hosting infrastructures supporting Russian government domains \cite{jonker_where_2022}. 
Work focused specifically on DNS resilience has investigated redundancy in recursive resolution \cite{sommese_assessing_2022} and the availability of authoritative DNS records \cite{houser_comprehensive_2022}. Despite 
previous studies on DNS resilience, a methodology that comprehensively evaluates the operational resilience of authoritative DNS infrastructure is lacking, as is the application to government services.
To address the gap, we assess the network resilience of authoritative DNS infrastructure through three operational phases that have not been considered previously: infrastructure placement, service configuration, and DNS record dispatch.

In this paper, we develop a \textbf{comprehensive assessment framework} that empirically evaluates the
operational resilience of authoritative DNS infrastructure for a target (\eg government) domain.
Our framework models resilience across the three phases of the network operation process: hosting infrastructure placement, authoritative service configuration, and DNS record dispatch. This phased analysis provides infrastructure operators, domain owners, and administrative agencies with clear, quantitative, and actionable evidence to diagnose improper or suboptimal authoritative DNS settings that are often obscured by operational complexity. Coupled with an automated, multisource data-processing pipeline, our framework has been applied to the authoritative DNS infrastructure supporting government domain names across six representative countries, revealing valuable and previously unseen insights into their operational resilience.
Specifically, the contributions of this paper are threefold.

Driven by our qualitative model (\S\ref{sec:background}) of the three-phase network operational process for authoritative DNS infrastructure and informed by industry best practices for operational resilience, our \textbf{first} contribution (\S\ref{sec:sectiondataSchema}) captures information relevant to each phase for each element in the four hierarchical levels of authoritative DNS supporting a domain name, from infrastructure and server functionality to name servers and hosting instances. The schema is populated with data from authoritative name servers, Internet resource registries, and trusted IP intelligence sources. 
We automate this data collection and compile a dataset of 7,307 government services across six countries spanning diverse geographic and economic contexts. Using Australia as a representative country, we present operational practices that impact the resilience of government services.

To enable an interpretable and comparable evaluation of authoritative DNS resilience, our \textbf{second} contribution (\S\ref{sec:assessment}) develops a systematic methodology to assess resilience per domain. We define 18 operational resilience attributes that span the three operational phases of placement, configuration, and dispatch, each mapped to its relevant hierarchy within the authoritative DNS infrastructure. Each attribute is assigned a five-point numerical score, ranging from worst to best, based on its impact on resilience and its alignment with industry-recommended practices. These scores are hierarchically aggregated using an impact-driven algorithm, producing both fine-grained visibility into operational settings and a consolidated resilience summary suitable for executive interpretation.

Our \textbf{third} contribution (\S\ref{sec:insights}) applies our methodology to 7,307 government domain names in six countries (Australia, Brazil, France, Indonesia, the U.K., and the U.S.), allowing a cross-national assessment of authoritative DNS operational resilience.
Comparative analysis reveals unique national patterns in the three phases (placement, configuration, and dispatch) shaped by differing infrastructure strategies and policy environments.
For example, Australia and the U.K. rely heavily on foreign cloud providers even for primary authoritative functionality; France and Indonesia demonstrate strong placement resilience due to local hosting policies, but exhibit lower resilience in service configuration and DNS record dispatch; and the U.S. government domains generally score well overall, yet exhibit some weaknesses such as exposure of primary name servers to the public Internet, insufficient subnet redundancy, and not recommended DNSSEC algorithms.
These findings demonstrate that our assessment framework provides actionable insights for both DNS operators and government agencies, enabling them to systematically benchmark their current practices and strengthen operational resilience to better serve their citizens.

\section{Preliminary on Network Operation of Authoritative DNS Infrastructure}\label{sec:background}
In this section, we introduce the three-phase network operation process that governs the authoritative DNS infrastructure of a given domain name (\S\ref{subsec:authoritativeDNS}) and summarize industry-recommended practices that strengthen operational resilience across these phases (\S\ref{subsec:resilience}).

\subsection{Authoritative DNS Infrastructure of (Government) Domain Names}\label{subsec:authoritativeDNS}

\begin{figure}[t!]
	\centering
	\includegraphics[width=0.95\linewidth]{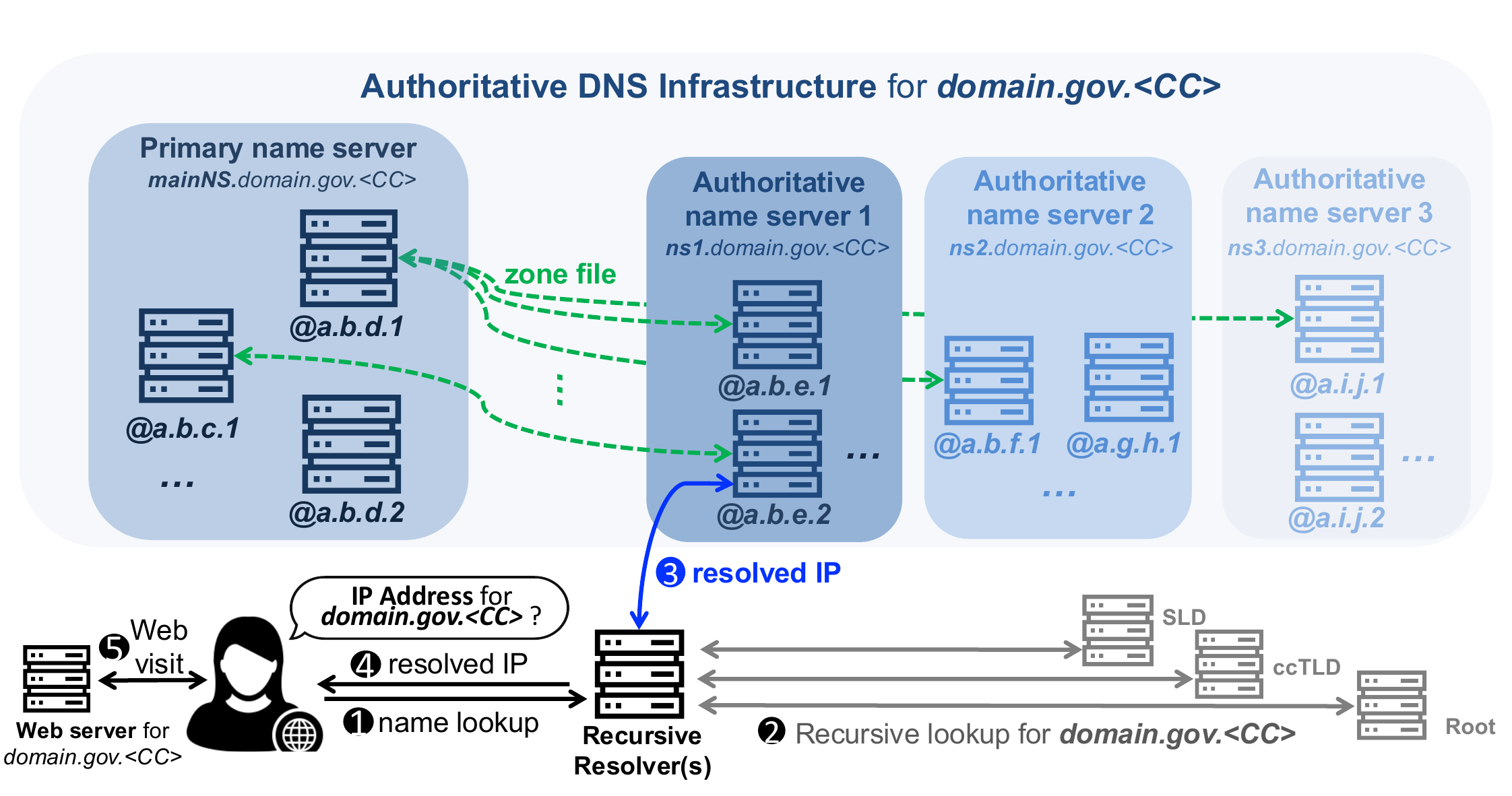}
	\vspace{-3mm}
	\caption{A simplified view of the authoritative DNS infrastructure for a domain and its role in resolving and serving user web requests.}
	\label{fig:thedns}
	\vspace{-2mm}
\end{figure}

The authoritative DNS infrastructure is the sole authority responsible for mapping a domain name to the IP address(es) of the service that hosts it. 
As illustrated at the bottom of Fig.~\ref{fig:thedns}, a user who attempts to access a website on {\myverb{domain.gov.<CC>}} first resolves its IP address through a recursive resolver. This resolution proceeds through a sequence of lookups, from the root, to the country code top-level domain (ccTLD), and then to the second-level domain (SLD), as shown in steps \ciao{1} and \ciao{2}. The process continues until the resolver reaches an authoritative name server instance on the IP address {\myverb{a.b.e.2}} corresponding to {\myverb{ns1.domain.gov.<CC>}} in step \ciao{3}. This server instance hosts a copy of DNS records (\ie zone file) for the requested domain. 
Once the recursive resolver obtains the IP address of the web server hosting {\myverb{domain.gov.<CC>}} (step \ciao{4}), the user's browser connects to the service in \ciao{5}.

\textbf{\textit{The authoritative DNS infrastructure:}} Supporting step \ciao{3} requires a nontrivial authoritative DNS infrastructure operated for the domain name.   As shown in the upper part of Fig.~\ref{fig:thedns}, this infrastructure consists of two categories of servers: the ``\textbf{primary}'' name server and multiple  ``\textbf{authoritative}'' name servers. The primary name server (\eg {{\myverb{mainNS.domain.gov.<CC>}}) hosts the source version of the DNS zone file for the domain name and may run on multiple instances, each with its own IP address (\eg {\myverb{a.b.d.1}}). The authoritative name servers (\eg {\myverb{ns1.domain.gov.<CC>}}) maintain synchronized copies of this zone file retrieved from the primary name server and are responsible for answering DNS lookup requests from users. 
	
	\textbf{\textit{The network operational process:}} From the perspective of the government department responsible for the service, operating the authoritative DNS infrastructure involves three phases: infrastructure placement, service configuration, and DNS record dispatch.
	In the \textbf{infrastructure placement} phase, administrators determine where the primary and authoritative name servers will be hosted, each with a unique server name (\eg {\myverb{ns1.domain.gov.<CC>}}) and a corresponding physical or virtual instance. As studied in prior work \cite{habib_formalizing_2025}, this decision includes selecting hosting organizations, choosing physical geolocations, and assigning logical IP addresses to each server instance.
	In the \textbf{service configuration} phase, the deployed server instances are configured to fulfill their designated roles as primary and/or authoritative name servers. From a resilience perspective \cite{steurer_measuring_2025}, redundancy in both server types is essential. Hosting organizations must also decide whether to co-locate or separate primary and authoritative functionalities, balancing operational cost against resilience.
	In the DNS \textbf{record dispatch} phase, servers begin transmitting DNS data to their intended clients:  the primary server synchronizes zone files with authoritative servers, and authoritative servers respond to user queries. Operators may adopt standardized security mechanisms to strengthen resilience against malicious interference, including Asynchronous Full Zone Transfer (AXFR) \cite{rfc5936} for controlled zone-file synchronization and DNS Security Extensions (DNSSEC) to ensure that DNS responses are cryptographically verifiable and resistant to hijacking or manipulation.

	\subsection{Industry Best Practices for Resilient Operation of Authoritative DNS Infrastructure}\label{subsec:resilience}
	Having outlined the three phases of the network operational process for authoritative DNS infrastructure, we now summarize industry-recommended practices that enhance resilience against physical disruptions, such as natural disasters and power outages \cite{moss_magnitude_2025}, and network-based threats, such as IP- or DNS-layer DDoS attacks \cite{hiesgen_age_2024} and integrity-compromising exploits \cite{heftrig_harder_2024}. Adopting these practices typically increases infrastructure cost and operational complexity, but significantly strengthens service robustness.
	
	From the perspective of \textbf{infrastructure placement}, the Internet community recommends \cite{habib_formalizing_2025} that organizations hosting authoritative DNS servers and their instances maintain a close administrative relationship with the responsible government department, such as through State-Owned Enterprises (SOE). Hosting on reputable international cloud providers (\eg Amazon AWS) or trusted domestic providers is also considered acceptable. In contrast, hosting DNS infrastructure on small, foreign-operated enterprises or on entities lacking formal registration poses substantial resilience risks. Physical placement also matters: for government services primarily accessed by local residents, both primary and authoritative DNS instances should ideally be hosted within the country with locally registered/operated IP addresses to ensure administrative proximity and reduce exposure to geopolitical disruptions.
	
	During the \textbf{service configuration} phase, best practices articulated in standards such as RFC 2182 \cite{rfc2182} emphasize shielding the primary DNS function from public exposure and ensuring sufficient redundancy across both primary and authoritative roles. The primary name server should only be reachable by trusted authoritative servers and should not be co-hosted with publicly exposed authoritative functionality. Redundancy can be achieved by enabling IP Anycast \cite{rfc4786} for the primary service so that authoritative servers connect to the nearest available instance. If Anycast is not used, the primary service should operate on multiple instances, preferably more than two, each with a unique IP address and located in different autonomous systems (ASes) or subnets. Authoritative DNS functions should likewise employ multiple server names (\eg {\myverb{{\{ns1, ns2, ns3\}.domain.gov.<CC>}}} in Fig.~\ref{fig:thedns}), each implemented using Anycast or backed by multiple IP addresses.
	
	Finally, in the \textbf{record dispatch} phase, which includes zone-file transfers from the primary server to authoritative servers and DNS responses served to clients, standardized security mechanisms should be enforced to defend against integrity and hijacking attacks. Specifically, Asynchronous Full Zone Transfer (AXFR) should be used to distribute zone files only to trusted authoritative servers securely, and DNS Security Extensions (DNSSEC) should be enabled to ensure that DNS responses are cryptographically verifiable, as discussed earlier in \S\ref{subsec:authoritativeDNS}.

	\section{Analyzing Authoritative DNS Resilience with Multisourced Data Schema}\label{sec:sectiondataSchema}
	Building on the three-phase operational model of authoritative DNS infrastructure and the associated industry best practices, this section introduces our multisourced data schema that characterizes the network operational resilience of a domain name (\S\ref{subsec:dataSchema}). We then describe the collection of a large-scale dataset covering government-listed public service names in six countries (\S\ref{subsec:dataCollection}), followed by an illustrative analysis using one representative medium-sized country as a case study (\S\ref{subsec:analysis}).

	\subsection{Structured Data Schema for a (Government) Domain Name}\label{subsec:dataSchema}
	
	To fully capture the network resilience of the authoritative DNS infrastructure for a domain name, we design a data schema that aggregates operational information that reflects the practices across the three phases of infrastructure placement, service configuration, and DNS record dispatch. 
	As illustrated on the left side of Fig.~\ref{fig:thedataschema}, the schema is organized to mirror the hierarchical structure of an authoritative DNS infrastructure. It is populated through automated scripts that query three distinct categories of data sources.
	
	\begin{figure}[t!]
		\centering
		\includegraphics[width=\linewidth]{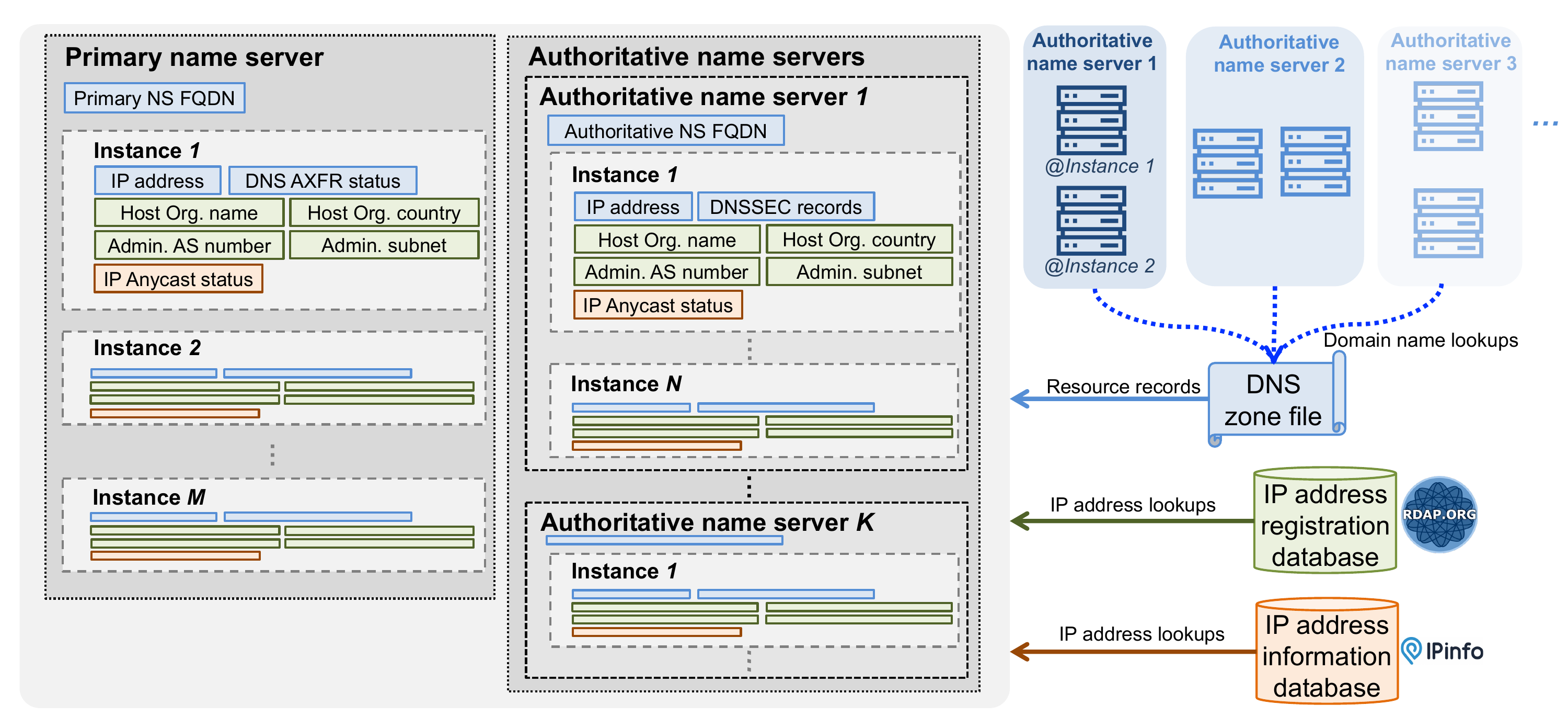}
		\vspace{-5mm}
		\caption{Our structured \textbf{data schema} (the left region in gray) integrating DNS resource records, IP registration data, and IP operational information to represent the resilience-relevant attributes of authoritative DNS infrastructure for a government domain.}
		\label{fig:thedataschema}
	\end{figure}
	
	To represent resilience factors with each operational phase (\S\ref{sec:background}), the schema models the two core components of the authoritative DNS infrastructure:  the primary name server and the set of authoritative name servers. Each name server is uniquely identified by its fully qualified domain name (\ie {\myverb{Primary NS FQDN}} and {\myverb{Authoritative NS FQDN}}) and may consist of one or more hosting instances. For every instance, the schema records its IP address, hosting organization, network configuration, and support for relevant data-security protocols. These elements appear as data components (rectangular boxes in Fig.~\ref{fig:thedataschema}) that are indexed according to their position within the hierarchy.
	
	The values of these components are obtained from three complementary data sources, which together provide a comprehensive view of each server instance. 
	As shown on the right side of Fig.~\ref{fig:thedataschema}, color-coded data fields correspond to the source from which each item is obtained. The process begins by querying DNS resource records from all authoritative name-server instances, retrieving information such as server roles, advertised IP addresses, and DNSSEC settings. AXFR accessibility is measured directly by issuing zone-transfer requests to all primary name-server instances from our measurement nodes. 
	Next, for each server instance, we retrieve administrative registration records by querying its  IP address through the Registration Data Access Protocol (RDAP) \cite{rdap_registration_2025}, which provides information on the hosting organization, subnet assignment, and autonomous system. Since some operational attributes, like Anycast deployment, are not exposed by the registration authorities, we supplement these records with data from the commercial IP intelligence services  IPInfo \cite{ipinfo_ipinfo_2025}, which actively tracks operational configurations and is widely used by the Internet measurement studies \cite{kumar_choices_2024, osali_sibling_2025, griffioen_have_2024, zhou_regional_2023, xing_yesterday_2024}.
	
	The complete schema is represented in JSON format, with each data component indexed within the hierarchical structure. For each domain name, the data collection module generates a single JSON object by integrating the three data sources, which is then processed by the resilience assessment module. Details of the measurement setup and processing pipeline are provided in Appendix \S\ref{appendix:dataCollection}.
	
	As underlying data sources, we assume that IP registration information obtained via the Registration Data Access Protocol (RDAP), standardized by the Internet Engineering Task Force (IETF), is sufficiently accurate for identifying infrastructure ownership. We also use the IPInfo database as the sole source to determine IP Anycast status. While alternative IP intelligence databases could be used for validation or substitution, exploring such extensions is beyond the scope of this study, which focuses on the resilience modeling schema and assessment methodology.
	
	We further note that our proposed schema can be extended. In its current form, it captures the categorical characteristics of hosting infrastructure, such as hosting organization type and binary geolocation (\ie internal or external to the respective country). Additional dimensions, such as concentration across cloud providers or finer-grained geolocation, could support more detailed modeling.

	\subsection{Dataset Collection for Government-Listed Public Service Domains at National Scale}\label{subsec:dataCollection}
	We collected our dataset in November 2025 to assess all public-service domain names listed by federal or central governments in six representative countries.
	The six countries represent diverse geographic regions (North America, Europe, and Asia-Pacific) and economic contexts (developed and developing economies). In addition, their government agencies maintain public services domain lists in English, ensuring accessibility and consistency of records for measurement.
	
	Table~\ref{tab:dataset_summary} summarizes these countries, the government authorities responsible for publishing the official service lists, the total number of listed public-service domains, and the number of unique parent domain names, each supported by a distinct authoritative DNS infrastructure.
	By comparing the number of fully qualified domain names (\eg {\myverb{www.ready.gov}}) with the number of unique parent domains (\eg  {\myverb{ready.gov}}), we observe three distinct operational strategies for authoritative DNS used at the government level in the six countries.

	\begin{wraptable}{l}{8cm}
		\vspace{-3mm}
		\scriptsize
		\caption{Summary of government-listed public-service domains and the corresponding unique parent domain names for each of the six studied countries.}
		\vspace{-4mm}
		\label{tab:dataset_summary}     
		\renewcommand{\arraystretch}{1.1}
		\begin{adjustbox}{width=0.58\textwidth}
			\begin{tabular}{llrr}
				\toprule
				\textbf{Country} & \textbf{Authority} & \textbf{\# services} & \textbf{\# parent domains} \\
				\midrule
				AU & {\myverbS{finance.gov.au}} \cite{agor_quarterly_2025} & 790 & 273 \\ \hline
				BR & {\myverbS{gov.br}} \cite{brazil_orgaos_2025} & 41 & 2 \\ \hline
				FR & {\myverbS{data.gouv.fr}} \cite{datagouv_list_2022} & 659 & 192 \\ \hline
				ID & {\myverbS{setkab.go.id}} \cite{setkab_kabinet_2924} & 60 & 54 \\ \hline
				UK & {\myverbS{gov.uk}} \cite{ukgov_list_2025} & 5,548 & 5,475 \\ \hline
				US & {\myverbS{get.gov}} \cite{cisa_getgov_2025} & 1,349 & 1,308 \\
				\bottomrule
			\end{tabular}
		\end{adjustbox}      
		\vspace{-2mm}
	\end{wraptable}
	
	The \textbf{first} strategy, used by the U.S., the U.K., and Indonesia, assigns a dedicated domain name to nearly every online public-service endpoint, with only a small number of services sharing the same domain. This decentralized model offers individual government departments substantial flexibility in managing the authoritative DNS infrastructures that support their respective services.
	The \textbf{second} strategy, adopted by Australia and France, groups multiple services under a small set of shared parent domains, typically two or three services per domain, often reflecting departmental structure. For example, both {\myverb{consult.treasury.gov.au}} and {\myverb{ttaasag.treasury.gov.au}} fall under the parent domain {\myverb{treasury.gov.au}} for the Department of Treasury.
	Brazil represents the \textbf{third} strategy, in which authoritative DNS operations are highly centralized. All 41 listed public-service domains map to only two parent domains {\myverb{bcb.gov.br}} and {\myverb{www.gov.br}}.
	
	For every domain name identified across the six countries, we collected the full set of data components defined by our schema in \S\ref{subsec:dataSchema}.
	These data support the manual analysis presented in \S\ref{subsec:analysis} and form the basis of our systematic resilience assessment in \S\ref{sec:assessment}. As we show later, the resulting assessments reveal distinct resilience profiles across countries that correlate strongly with their administrative structures and the operational strategies adopted for authoritative DNS infrastructure.

	\subsection{Analytical Insights for a Medium-Sized Country}\label{subsec:analysis}
	
	Analysis of our dataset reveals a wide variety of network operational practices within authoritative DNS infrastructures supporting government services, with implications for resilience in terms of diversity and complexity. To illustrate these characteristics, we present selected observations for Australia.
	Australia represents a medium-sized authoritative DNS infrastructure for government services, with moderate complexity in placement, configuration and dispatch practices. This makes it a suitable illustrative case to motivate the resilience modeling and assessment methods introduced in \S\ref{sec:assessment}.
	
	\subsubsection{Infrastructure placement}\label{subsubsec:placement}
	We begin by examining the two fields in our data schema that reflect infrastructure for each domain name: the ``host Org. name'' and the ``hosting country'', which together indicate the administrative ownership and physical location of each primary or authoritative name-server instance.
	Host organizations fall broadly into four categories, State-Owned enterprises (SOE), local private enterprises, foreign large enterprises, and foreign small-and-medium (SME) enterprises, each associated with varying levels of operational resilience \cite{kumar_choices_2024}. In addition, a small number of instances appear with incomplete registration records; these are classified as ``unregistered'', representing the least trustworthy category.
	
	\textbf{\textit{Primary name servers by enterprise category:}} For the primary name server, which serves as the sole authoritative source for a domain's zone file, instances are deployed on one or three physical or virtual servers, each assigned a unique IP address. 
	Fig.~\ref{fig:primaryInfrastructure} shows the number of domain names whose primary name servers are hosted in each enterprise category, with stacked bars indicating the number of hosting instances and segmented by unique hosting organization.
	Among the 273 government-listed service domains, 37 (13.55\%) have their primary name servers directly operated by 8 SOEs, typically considered the most resilient arrangement. However, all SOE-hosted primary servers run on a single instance, reducing resilience from a service-configuration standpoint. These observations motivate the definition of resilience attributes that capture enterprise types and the number of hosting instances in \S\ref{sec:assessment}.
	
	\begin{figure}[!t]
		\centering
		\hspace*{-3.4cm}
		\begin{subfigure}{0.485\linewidth}
			\centering
			\includegraphics[height=4.1cm]{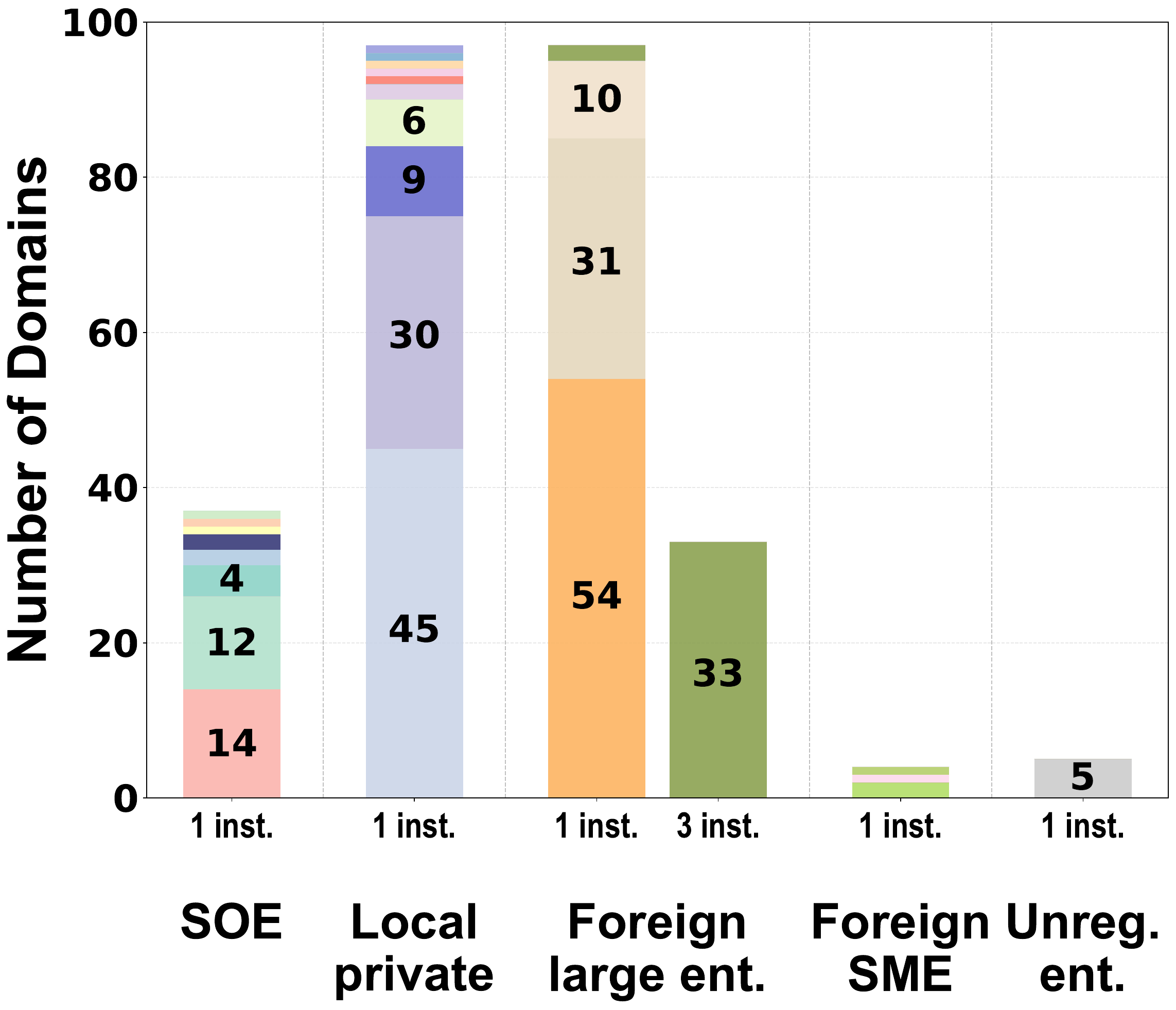}
			\caption{Primary name servers.}
			\label{fig:primaryInfrastructure}
		\end{subfigure}
		\hspace{-0.9cm}
		\begin{subfigure}{0.465\linewidth}
			\centering
			\includegraphics[height=4.1cm]{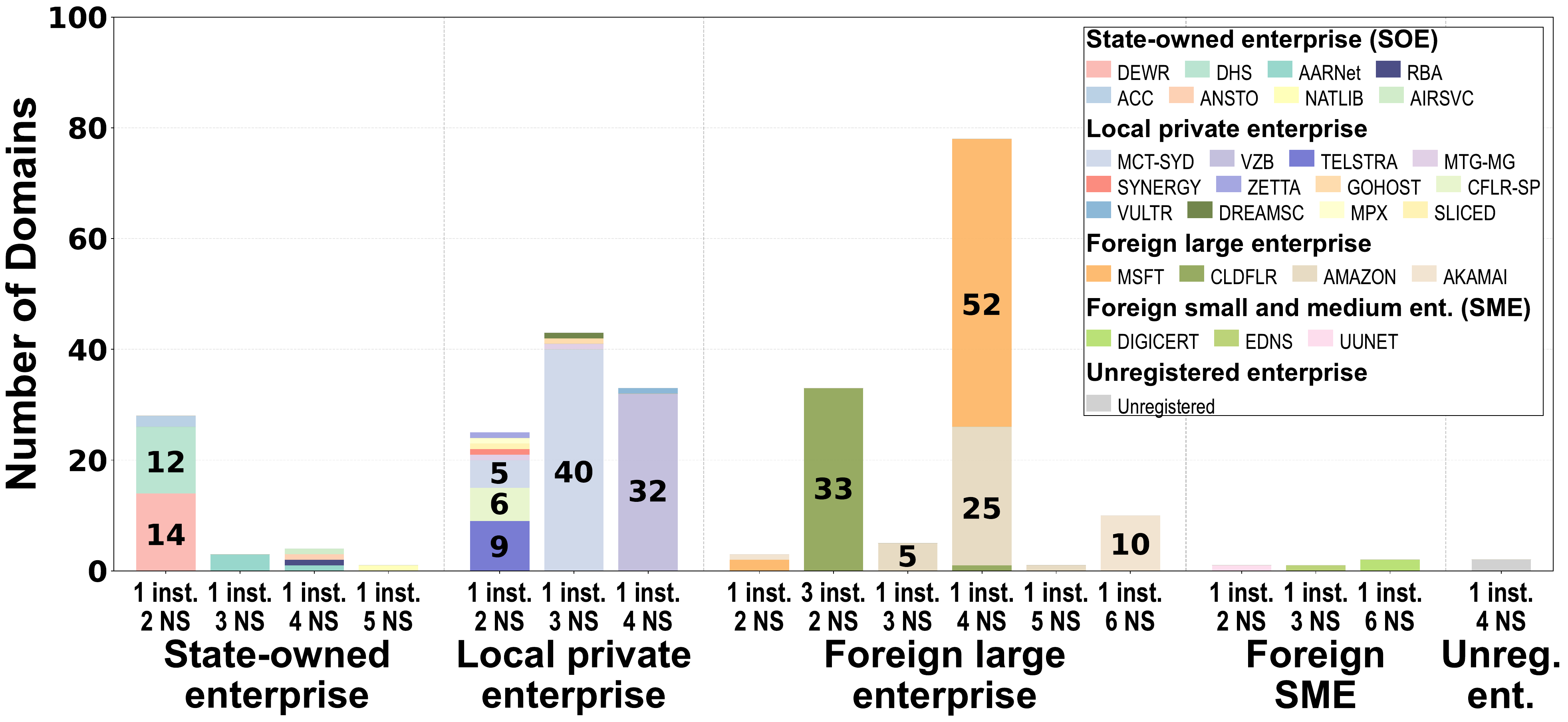}
			\caption{Authoritative name servers.}
			\label{fig:authoratitiveInfrastructure}
		\end{subfigure}
		\vspace{-3mm}
		\captionsetup{skip=4mm, belowskip=-1mm}
		\caption{Hosting-organization categories for (a) primary and (b) authoritative name servers supporting government-listed domain names in Australia. Each server may run on one or more hosting instances. Hosting enterprises are grouped into five categories: state-owned enterprises (SOE), local private enterprises, foreign large enterprises, foreign SMEs, and unregistered entities, with different implications for resilience.}
		\label{fig:orgAndCountry}
		\vspace{-2mm}
	\end{figure}
	
	A further 97 domains (35.53\%) operate their primary name server on a single instance hosted by local private enterprises, offering reasonable resilience against foreign influence \cite{jain_ukrainian_2022}. Another 97 domains (35.53\%) rely on large international cloud providers, including Microsoft, Cloudflare, Amazon, Google, or Akamai, for their primary DNS hosting, while 33 domains (12.09\%) use Cloudflare across three instances. As will be discussed later in \S\ref{subsubsec:service_configuration} and \S\ref{subsubsec:data_dispatch}, hosting on major cloud platforms generally improves resilience in the configuration and dispatch phases. However, these providers are foreign entities, raising concerns about digital sovereignty. 
	A small number of domains (4, 1.47\%) use foreign SMEs to host their primary name servers, a configuration that introduces substantial operational risks for such a critical function. The worst cases involve five domains (1.83\%) whose primary servers are hosted on instances without complete registration information; notably, these domains also co-host their authoritative name servers on the same instance, thereby exposing primary functionality, intended to remain shielded from public access, to unnecessary cyber risk.
	
	\textbf{\textit{Authoritative name servers by enterprise category:}} Unlike the sole primary name server required to maintain a consistent zone-file origin, each domain typically advertises multiple authoritative name servers (\eg two to six in this country, as shown by the ``auth'' x-ticks in Fig.~\ref{fig:authoratitiveInfrastructure})  to improve regional (and global) accessibility. Each authoritative server is itself hosted on one or three instances, though different patterns appear in other countries. For example, the authoritative server {\myverb{aliza.ns.cloudflare.com}} for  {\myverb{aifs.gov.au}} operates across three distinct IP addresses.

	The combination of multiple authoritative servers, each with multiple hosting instances, introduces significantly more complexity than the configuration of a single primary server on multiple instances. In this country, 270 domains (98.90\%) have all authoritative instances hosted by a single enterprise category, while 3 domains (1.10\%) distribute their authoritative hosting across two categories. We show a simplified count of domain names in Fig.~\ref{fig:authoratitiveInfrastructure} summarizing the dominant enterprise type for authoritative hosting per domain. It can be seen that most domains rely primarily on foreign large enterprises (47.62\%), followed by local private enterprises (37.00\%) and SOEs (13.19\%). A small number of domains depend on foreign SMEs for authoritative hosting, representing the least resilient placement.

	\subsubsection{Service configuration}\label{subsubsec:service_configuration}
	
	In the service-configuration phase, each hosting instance (whether it supports the primary or authoritative name server) is configured to control how it is reached by Internet users. To understand how practices in this phase affect network resilience, we analyze the \myverb{IP address}, \myverb{Admin. subnet}, \myverb{Admin. AS number}, and {\myverb{IP Anycast status}}  fields in our data schema, which together describe the network address (IP) configuration of each server instance. 
	Redundancy in both name servers and hosting instances, already discussed in \S\ref{subsubsec:placement}, also plays a central role in this phase. 
	
	\textbf{\textit{Primary name servers by the hosting redundancy:}} Overall, redundancy remains limited for most government domain names in this country. As shown in Fig.~\ref{fig:primaryInfrastructure} and Fig.~\ref{fig:authoratitiveInfrastructure}, 240 domains (87.91\%) host their primary name servers on single instances, and 57 domains (20.88\%) advertise only two authoritative name servers, each deployed on a single instance.
	Such configurations offer little protection against localized failures or targeted network attacks.
	
	\begin{figure*}[t]
		\vspace{-2mm}
		\centering
		\begin{subfigure}[t]{0.45\textwidth}
			\centering
			\includegraphics[width=\linewidth]{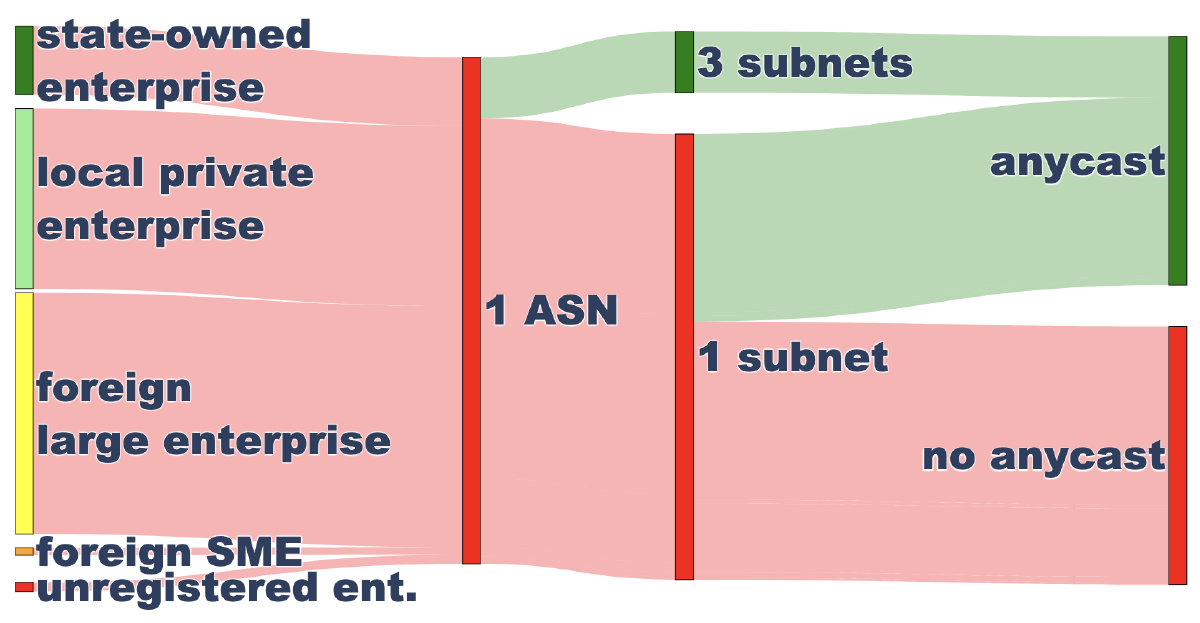}
			\caption{Primary name servers.}
			\Description{Relationship between the hosting organization types of primary name servers and the routing configurations deployed within the DNS infrastructure.}
			\label{fig:sankey_prim}
		\end{subfigure}
		\hfill
		\begin{subfigure}[t]{0.45\textwidth}
			\centering
			\includegraphics[width=\linewidth]{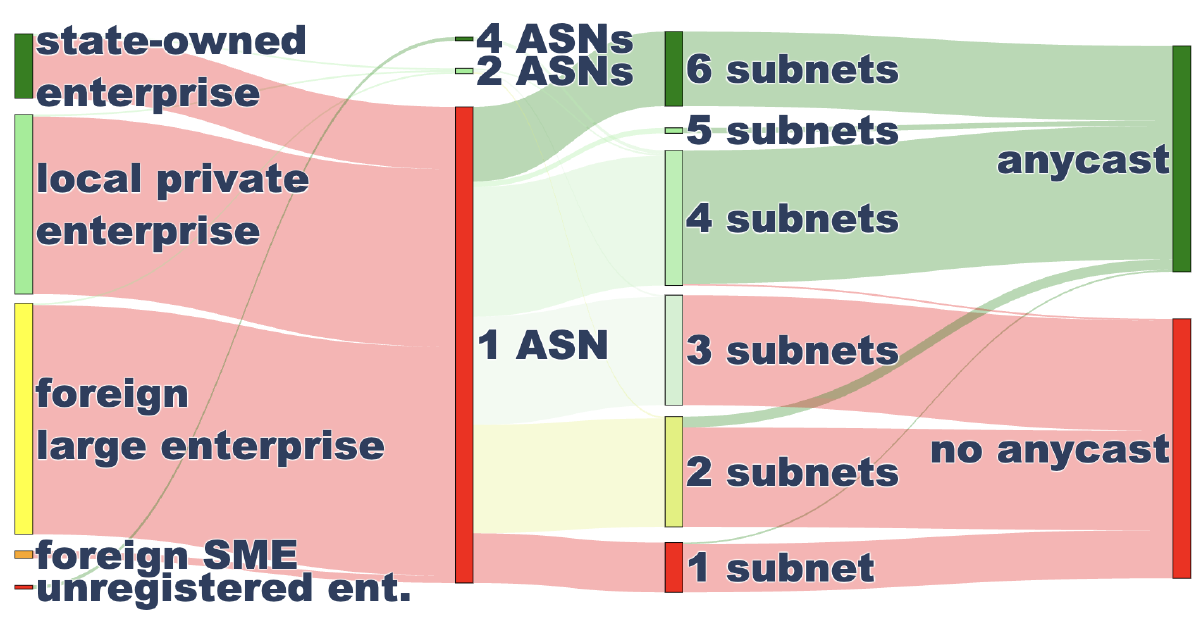}
			\caption{Authoritative name servers.}
			\Description{Relationship between the hosting organization types of authoritative name servers and the routing configurations deployed within the DNS infrastructure.}
			\label{fig:sankey_auth}
		\end{subfigure}
		\vspace{-3mm}
		\caption{Overview of operational practices in (a) primary and (b) authoritative name servers of Australian government public-service domains. Hosting-enterprise types, administrative ASN and subnet diversity, and Anycast configurations vary substantially across domains, leading to differing resilience profiles.}
		\label{fig:sankey_combined}
		\vspace{-4mm}
	\end{figure*}
	
	By contrast, the 33 domains whose primary name servers span three instances, all rely on cloud providers (specifically, Cloudflare) that configure these instances using Anycast routing \cite{rfc4786}. Anycast enables user requests to reach the nearest Cloudflare node, providing strong resilience against network failures and geographically distributed attacks. 
	However, this contrasts with the 37 domains (13.55\%) whose primary name servers are operated by State-Owned Enterprises. While SOE operations offer resilience against geopolitical and jurisdictional risks, all of these instances reside within a single administrative subnet, and none use Anycast, significantly limiting flexibility and redundancy in access paths.
	The relationship among enterprise type, administrative subnet/ASN diversity, and Anycast use for primary servers is shown in Fig.~\ref{fig:sankey_prim}.
	Similar patterns hold for domains whose primary servers are hosted by local private enterprises: 96 out of 97 domains are placed within a single subnet without Anycast, with one exception {\myverb{aafcans.gov.au}} using Anycast across a single subnet. In \S\ref{sec:assessment}, we will define resilience attributes that capture hosting redundancy and access flexibility to characterize these service configuration practices.
	
	\textbf{\textit{Authoritative name servers by the hosting redundancy:}} The configuration of authoritative name servers is even more complex, as illustrated in Fig.~\ref{fig:sankey_auth}. The interplay among hosting-organization type, addressing configuration, subnet/ASN diversity, and Anycast deployment varies widely across domains, highlighting the need for a systematic, multi-attribute resilience assessment, far beyond what single-aspect analysis in prior work typically capture \cite{steurer_measuring_2025}.

	\subsubsection{DNS Record Dispatch}\label{subsubsec:data_dispatch}
	
	In the record-dispatch phase, we examine the two fields in our data schema (\ie {\myverb{AXFR status}} and {\myverb{DNSSEC records}}) that indicate whether secure protocols are enforced to protect DNS record integrity. AXFR governs the transfer of zone files from the primary name server to authoritative servers, while DNSSEC protects responses served by authoritative servers to Internet users.
	
	\begin{wrapfigure}{l}{0.58\textwidth}
		\vspace{-5mm}
		\includegraphics[width=\linewidth]{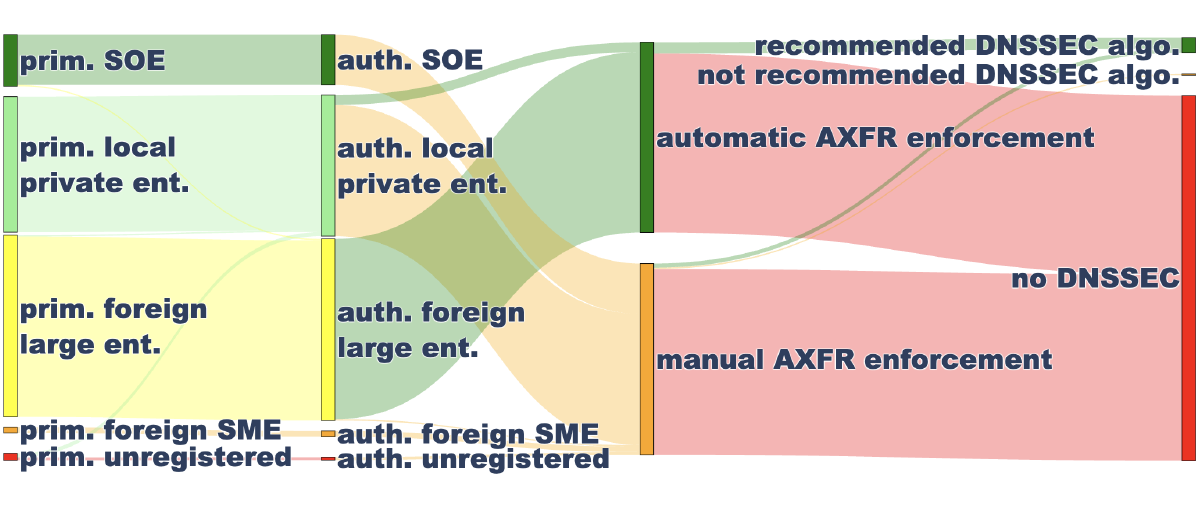}
		\vspace{-10mm}
		\caption{Operational practices in the DNS record-dispatch phase for Australian government public-service domains, showing the relationship between primary-server hosting enterprise type, AXFR enforcement method, and the worst-case DNSSEC configuration among authoritative server instances.}
		\label{fig:sankeygrey}
		\vspace{-4mm}
	\end{wrapfigure}
	
	\textbf{\textit{Dispatching from the primary to authoritative name servers:}} For the primary name servers of government domains in this country, \textbf{restricted distribution} of zone files via AXFR is consistently enforced: only authorized authoritative server instances are permitted to fetch zone data. 
	As shown in Fig.~\ref{fig:sankeygrey}, this enforcement is automatic for 136 domains (49.82\%), all of which rely on specialized managed DNS services provided by Google, Akamai, and Cloudflare \cite{google_cloud_2025, akamai_edge_2025, cloudflare_cloudflare_2025}. 
	The remaining 137 domains (50.18\%) use manual (\ie not managed DNS services) AXFR configuration by IT operators,  a practice that carries a higher risk of misconfiguration \cite{joshi_biggest_2025}. 
	In particular, in all six countries studied, every public-service domain enforces AXFR on its primary servers, which is captured as a resilience attribute in the systematic assessment in \S\ref{sec:assessment}.
	
	\textbf{\textit{Dispatching from authoritative servers to clients:}} However, DNSSEC deployment shows substantial variation between hosting-enterprise categories.
	As also shown in Fig.~\ref{fig:sankeygrey}, none of the 37 domains (13.55\%) operated by State-Owned Enterprises enforce DNSSEC on their authoritative servers, and all rely on manually configured AXFR. Among the 97 domains (35.53\%) managed by local private enterprises, 90 (32.97\%) also use manual AXFR and do not deploy DNSSEC. A small subset of six defense-service domains uses automatically enforced AXFR but likewise omits DNSSEC. One domain uses standard zone file delivery and deploys DNSSEC, but with cryptographic algorithms that do not meet IETF recommendations \cite{rfc8624}. In contrast, 130 domains (47.62\%) managed by foreign large enterprises rely on automatic AXFR enforcement, reflecting mature operational practices, yet only 8 domains (2.93\%) in this group deploy DNSSEC, all of which use recommended algorithms. This highlights an operational gap: while zone-transfer protection is widely implemented, DNSSEC adoption remains limited even among domains hosted by technically sophisticated providers.

	\section{Systematic Assessment of Authoritative DNS Resilience for a Domain Name}\label{sec:assessment}
	Building on the comprehensiveness of our data schema and the complexity revealed in the preceding analysis, we now develop a systematic assessment methodology for authoritative DNS resilience. Our approach defines a set of descriptive attributes ({\S\ref{subsec:attribute}}) and converts them into comparable indicative scores for each domain's authoritative DNS infrastructure (\S\ref{subsec:scores}). These scores are aggregated at four hierarchical levels of the infrastructure  ({\S\ref{subsec:scoreAggregation}}), allowing domain owners and administrative agencies to obtain a clear and comprehensive view of the resilience posture of the authoritative DNS system of a government domain.
	
	\subsection{Attributes for Authoritative DNS Resilience at their Lowest Hierarchical Levels}\label{subsec:attribute}
	
	As discussed in \S\ref{subsec:authoritativeDNS}, the authoritative DNS infrastructure for a domain name can be viewed at four hierarchical levels: the overall infrastructure, the two name-server functionalities (primary or authoritative), the name servers themselves, and their hosting instances. As shown in \S\ref{subsec:resilience} and empirically illustrated in \S\ref{subsec:analysis},  placement and configuration decisions at any of these levels can materially affect the operational resilience of the entire infrastructure. To capture these effects, we define resilience attributes at the lowest applicable hierarchical level, aligned with the operational phases, and depicted at the leftmost color-coded blocks in Fig.~\ref{fig:attribute}.
	
	\begin{figure}[t!]
		\centering
		\includegraphics[width=\linewidth]{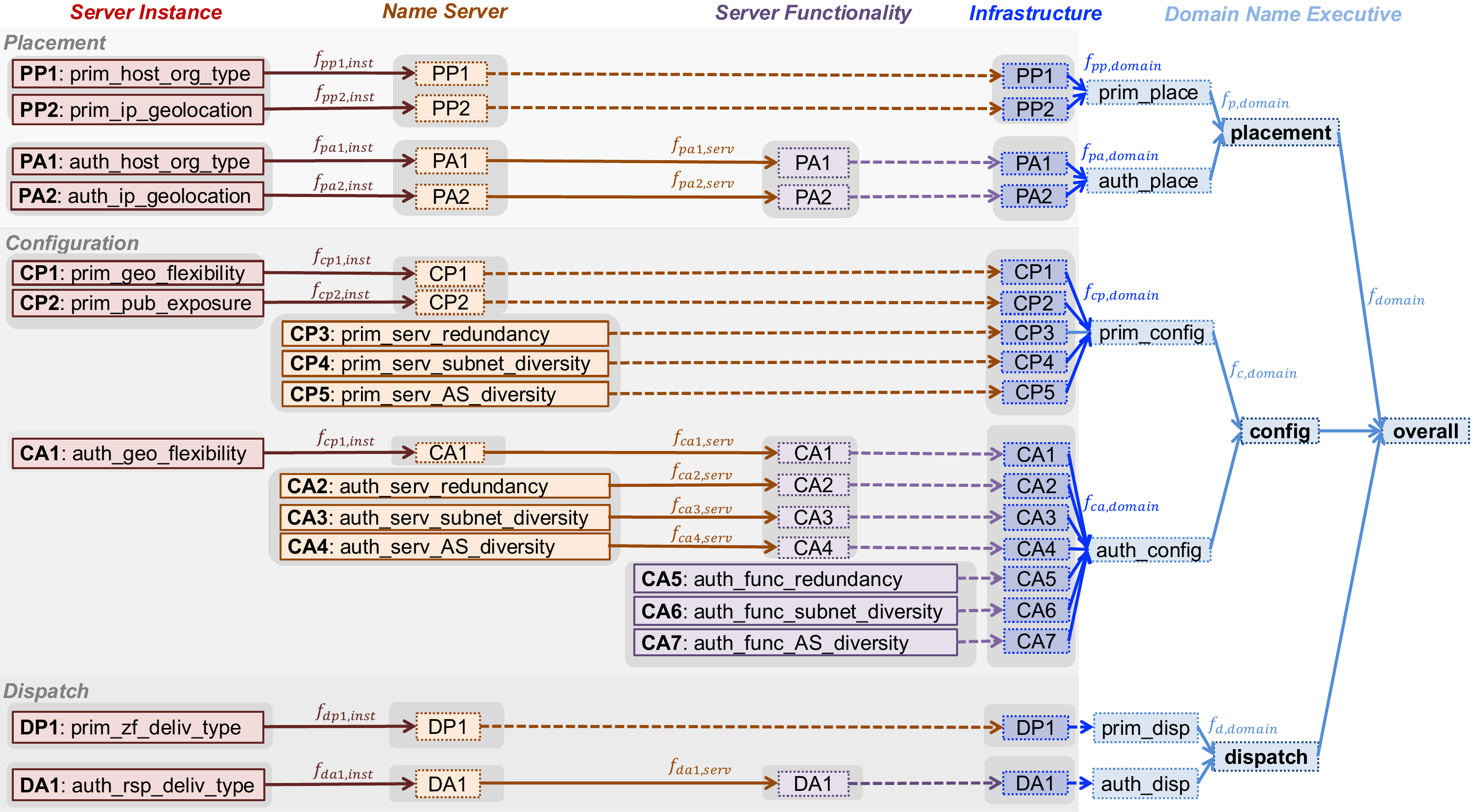}
		\vspace{-5mm}
		\caption{Attributes used to characterize the network operational resilience of an authoritative DNS infrastructure, organized at four hierarchical levels (server instance, name server, server functionality, and overall infrastructure), as outlined in the gray region.}
		\label{fig:attribute}
		\vspace{-5mm}
	\end{figure}
	
	\subsubsection{Placement of Authoritative DNS Infrastructure}\label{subsubsec:phase1}
	For the infrastructure placement phase, we define four attributes, namely ``prim\_host\_org\_type'', ``prim\_ip\_geolocation'', ``auth\_host\_org\_type'', and ``auth\_ip\_geolocation'', abbreviated as {\myverb{PP1}}, {\myverb{PP2}}, {\myverb{PA1}}, and {\myverb{PA2}}, that describe the hosting enterprise type and the IP geolocation for primary and authoritative name-server instances.
	
	Based on our observations in \S\ref{subsubsec:placement}, the hosting enterprise attributes {\myverb{PP1}} and {\myverb{PA1}} take one of five values: ``state-owned enterprise'', ``local private enterprise'', ``foreign large enterprise'', ``foreign small-and-medium enterprise'', and ``unregistered enterprise''. These categories imply different levels of resilience with respect to geopolitical stability and administrative control.
	
	For geolocation attributes {\myverb{PP2}} and {\myverb{PA2}}, we use two-letter country codes derived from the IPInfo geolocation database \cite{ipinfo_ipinfo_2025}. Noting that Anycast-enabled IP addresses operated by international cloud providers may appear with foreign registration information (\ie the ``Host Org. country'' field) but serve clients locally; in such cases,  the local country code is assigned to reflect their operational geography. 
	In our implementation (Appendix~\S\ref{appendix:dataCollection}), {\myverb{PP2}} and {\myverb{PA2}} are derived by mapping the ``Host Org. name'' field of each server instance to an enterprise-classification catalog maintained for each studied country.
	
	\subsubsection{Configuration of Name Server Functionalities}
	In the middle block of Fig.~\ref{fig:attribute}, we define twelve attributes describing configuration practices that impact resilience during the service configuration phase.
	The first five attributes {\myverb{CP1-5}} apply to the primary name server, and the remaining seven {\myverb{CA1-7}} describe the authoritative name servers.
	
	For primary-server instances, {\myverb{CP1}} (``prim\_accessibility'') indicates whether Anycast is enabled, providing geographic flexibility and improved fault tolerance.
	{\myverb{CP2}} (``prim\_pub\_exposure'') indicates whether the primary instance is co-hosted with authoritative functionality, a practice that increases exposure to public-facing threats.
	Attributes {\myverb{CP3-5}}, defined at the name-server level, capture redundancy and diversity: the number of IP addresses configured with primary functionality, the diversity of hosting subnets, and the diversity of autonomous systems (ASes).
	
	Analogous attributes are defined for authoritative name servers: {\myverb{CA1}} reflects regional accessibility through Anycast usage at each instance, {\myverb{CA2}} captures redundancy at the authoritative name-server level, and {\myverb{CA3}} and {\myverb{CA4}} measure subnet and AS redundancy per authoritative server, and {\myverb{CA5}} captures redundancy in authoritative functionality (\ie the number of distinct authoritative server names such as {\myverb{ns1.gov.com.<CC>}}).
	Finally, {\myverb{CA6}} and {\myverb{CA7}} describe the redundancy of authoritative servers on different subnets and ASes, respectively.
	
	\subsubsection{Dispatch of Authoritative DNS Records}
	For the record-dispatch phase, we define two attributes:  {\myverb{DP1}} and {\myverb{DA1}} to capture the enforcement of secure protocols governing zone-file delivery and DNS-response integrity. 
	
	{\myverb{DP1}} (``prim\_zf\_deliv\_type'') describes how AXFR is enforced for distributing DNS zone files from the primary server instance to authorized authoritative servers. As discussed in \S\ref{subsec:analysis}, the attribute takes one of three values: ``automatic cloud-based enforcement'', ``manual enforcement'', or ``no enforcement''. This classification is determined by comparing the hosting enterprise with our maintained list of cloud DNS providers known to automatically enforce AXFR, along with the AXFR-status field in the data schema.
	
	{\myverb{DA1}} (``auth\_rsp\_deliv\_type'') describes DNSSEC deployment on authoritative name server instances and takes one of three values: ``recommended algorithms'', ``not recommended algorithms'', and ``not configured''. 
	These values indicate whether DNSSEC is enabled and, if so, whether cryptographic algorithms comply with the IETF recommendations \cite{rfc8624}.
	In our processing pipeline (Appendix~\S\ref{appendix:dataCollection}), {\myverb{DA1}} is assigned by comparing the algorithms listed in each DNSSEC record with the recommended-algorithm set.
	
	\subsection{Assessment Scores for Each Attribute}\label{subsec:scores}
	Having defined a comprehensive set of attributes and possible values that describe authoritative DNS operational practices for government domains, we now introduce a quantitative scoring scheme to evaluate each attribute. We adopt a five-point Likert scale \cite{mcleod_likert_2025}, which is widely used in scoring frameworks across disciplines \cite{englbrecht_towards_2020, lee_chatfive_2024}. Each attribute in our model has at most five meaningful value options (or can be mapped to five), making a larger scale unnecessary.
	
	Our primary motivation for assigning a Likert score to each attribute is to enable concise and quantitative comparison of operational practices at the executive level. These scores serve as as a diagnostic heuristic, allowing managerial and policy stakeholders to interpret resilience landscape through simplified numerical indicators. Each score corresponds to a specific operational practice and retains an interpretable meaning for detailed analysis. These scores are not intended for direct comparison across attributes, as each attribute reflects a distinct operational context and may vary across assessors.
	
	Appendix Table~\ref{tab:resilience_framework} presents the assessment criteria for the attributes at their lowest hierarchical levels (instance, name server, or name server functionality). Each attribute describes one aspect of the operational resilience for primary or authoritative DNS functionality. As discussed intuitively in \S\ref{subsec:analysis}, these attribute values impact resilience in different ways. We formalize this by assigning each attribute option a score from 1 to 5, where 1 represents the least resilient practice and 5 represents the most resilient one. This enables both comparative analysis across domains and meaningful aggregation to higher hierarchical levels.
	
	\textbf{Attributes with five finite value options:}  Attributes {\myverb{PP1}} and {\myverb{PA1}}, which describe the hosting-enterprise type for primary and/or authoritative server instances, each have five finite options. These options are scored from five (most resilient, \eg state-owned enterprise) to one (least resilient, \eg an unregistered entity in the IP registry).
	
	\begin{wrapfigure}{l}{0.5\textwidth}
		\centering
		\vspace{-1mm}
		\includegraphics[width=\linewidth]{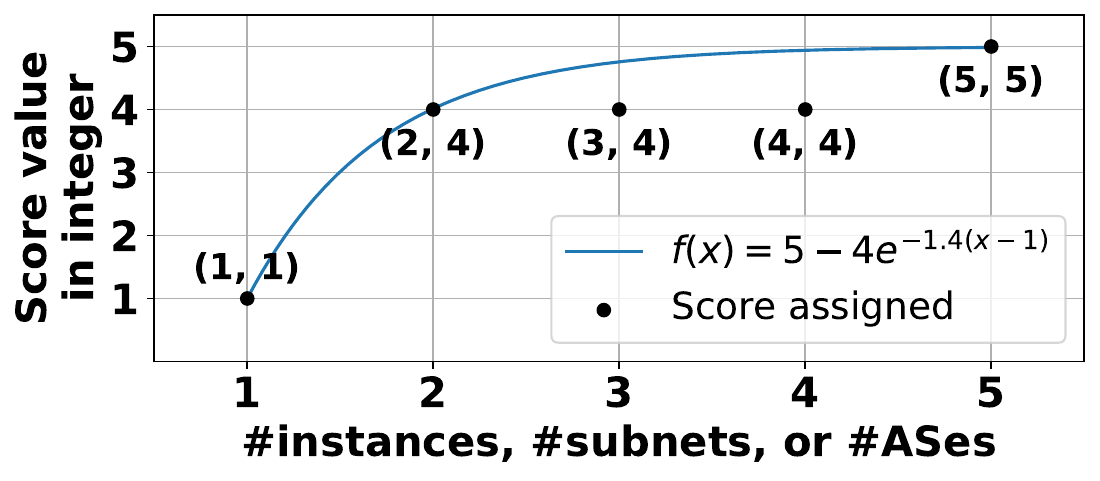}
		\vspace{-8.5mm}
		\caption{Exponential saturation function used to assign scores for redundancy/diversity attributes, capturing diminishing resilience returns as the number of instances, subnets, or ASes increases. The function is constrained by the minimum score (1,\textbf{1}) and the recommended operational practice (2,\textbf{4}).}
		\vspace{-4mm}
		\label{fig:curve}
	\end{wrapfigure}
	
	\textbf{Attributes with three finite value options:} For DNS record dispatch-phase attributes {\myverb{DP1}} and {\myverb{DA1}}, each takes one of three options:  strong and recommended enforcement of data-integrity mechanisms, insufficient enforcement (\eg manual AXFR configuration for {\myverb{DP1}} and weak DNSSEC algorithms for {\myverb{DA1}}), and no enforcement. These are mapped to scores 5, 3, and 1, respectively.
	
	\textbf{Attributes with two finite value options:} Five attributes, {\myverb{PP2}}, {\myverb{PA2}}, {\myverb{CP1}}, {\myverb{CP2}}, and {\myverb{CA1}}, have binary values that clearly reflect good or poor resilience practices. Therefore, these are assigned a value of 1 or 5. For example, {\myverb{CP2}} indicates whether a primary server instance shares its IP address with an authoritative name server instance. Co-location reduces operational costs but exposes the primary instance to the same risks as the publicly reachable authoritative server. Thus, co-location is scored 1 (worst practice), whereas using a dedicated IP address is scored 5 (best practice).
	
	\textbf{Attributes with infinite value options:} Nine attributes (\ie {\myverb{CP3-5}} and {\myverb{CA2-7}}) describe the counts of server instances, name servers, subnets, or ASes, and therefore have theoretically unbounded value ranges. For example, the attribute {\myverb{CP3}} represents the number of hosting instances for the primary name server.
	Although an extremely large number of instances would maximize redundancy, operational experience shows diminishing returns, with most DNS operators recommending \textbf{two} instances/subnets/ASes as the practical and cost-effective standard for resilience \cite{rfc2182, akamai_designing_2025, cloudflare_improving_2020}.
	Therefore, a value of 1 (no redundancy/diversity) is assigned 1 out of 5, and a value of 2 is assigned 4 out of 5, representing recommended best practice.

	To determine a realistic threshold for full scoring, we apply a bounded exponential growth function \cite{arghavani_suss_2024}, commonly used to model diminishing returns, where outputs are constrained by the minimum and maximum values in our scoring scheme. The fitted curve is shown in Fig.~\ref{fig:curve}. It can be seen that the function approaches the maximum score of 5 as soon as the attribute value reaches five. Consequently, in our scoring scheme, these seven attributes receive: a full 5/5 score when the count of instances/subnets/ASes is five or more, or a score of 4/5 for values from 2 to 4.

	\subsection{Aggregating Scores for Comparative Assessment Across Domain Names}\label{subsec:scoreAggregation}
	While attribute scores defined at their finest applicable hierarchy provide detailed visibility into operational practices, comparing resilience across domain names remains challenging because authoritative DNS infrastructures vary in their hierarchical complexity. To address this, we develop impact-driven aggregation strategies that combine scores of \textbf{each attribute} from lower levels (\ie server instance, name server, or NS functionality) into a single attribute-level score at the infrastructure level (the blue boxes in Fig.~\ref{fig:attribute}). 
	Such aggregations beyond attribute-level scores provide a higher-level summary of operational resilience for a given domain name. This reduces the effort and technical expertise required to assess overall resilience, supporting initial scoping and subsequent drill-down analysis by domain operators and policy makers.
	
	Aggregation follows the hierarchical structure indicated by the solid arrows in Fig.~\ref{fig:attribute}, representing how poor or strong operational practices propagate upward through the infrastructure. Drawing on industry operational practices, we observe two dominant forms of propagational impact: one in which aggregation is governed by the best available option (\S\ref{subsubsec:aggregation-best}) and one in which it is governed by the worst option (\S\ref{subsubsec:aggregation-worst}). These cases collectively form the basis of our aggregation strategies, formalized in Algorithm~\ref{algo:1}.
	
	Some attributes do not require aggregation at specific levels because they inherently produce a single score, \eg {\myverb{PP1}} at the name server level. These are directly mapped to the higher level, as shown by the dashed arrows in Fig.~\ref{fig:attribute}.
	
	After attribute-level scores are obtained at the authoritative DNS infrastructure level, they can be further summarized into phase-level scores. For example, $f_{pp,domain}$ and $f_{pa,domain}$ for the primary and authoritative placement attributes, and $f_{cp,config}$ and $f_{ca,config}$ for the configuration attributes. 
	These phase-level scores are then aggregated into three operational-phase summaries: 
	$f_{p,domain}$ for infrastructure placement, $f_{c,domain}$ for service configuration, and $f_{d,domain}$ for DNS record dispatch. 
	The weighting of attributes and phases is inherently subjective and depends on assessor priorities. In our evaluation (\S\ref{sec:insights}), we adopt equal weighting across all attributes at the infrastructure-level and throughout all phases, using simple averaging to maintain interpretability and comparability. 
	These weights represent the relative importance assigned to each attribute-level score in assessing the overall resilience of an authoritative DNS infrastructure (the ``Domain Name Executive'' region in Fig.~\ref{fig:attribute}). Therefore, they are inherently subjective and may vary across assessors, potentially resulting in different overall scores.
	
	\RestyleAlgo{ruled}
	\SetNlSty{textbf}{(}{)}
	\begin{algorithm}[t!]
		\fontsize{9}{10}\selectfont
		\textbf{Input:} $\textbf{s}_{a,h,n}$ $\leftarrow$ the resilient scores of the attribute $a$ for \textbf{all} items at the hierarchical level $h$, which belongs to the $n$th item at the upper $h+1$ level; 
		$\Delta D$ $\leftarrow$ value step; 
		\textit{BestStrategy} $\leftarrow$ boolean for adopting a strategy dominated by the best or the worst score;
		
		\textbf{Output:} $s_{a,h+1,n}$ $\leftarrow$ the resilient score for the attribute $a$ of the $n$th item at the hierarchical level $h+1$;
		
		\tcp{\color{brown}best-score dominant strategy by decreasing the highest score with small factors.}
		\nl \If{BestStrategy is $true$}{
			$s_{a,h+1,n,base}   \leftarrow   max(\textbf{s}_{a,h,n}); s_{a,h+1,n,tmp}   \leftarrow   max(\textbf{s}_{a,h,n})$;
			
			\tcp{counting the numbers of each score value in the decreasing order.} 
			$\textbf{scoreVal,scoreCt}  \leftarrow counting_{\downarrow,1}(\textbf{s}_{a,h,n})$; 
			
			\For{$val,ct$ in \textbf{scoreVal,scoreCt}}{
				\If{val is \textbf{not} $s_{a,h+1,n,base}$}{
					\tcp{decrease from the highest score.}
					$s_{a,h+1,n,tmp} \leftarrow  s_{a,h+1,n,tmp} - (ct*\Delta D/\sum(\textbf{scoreCt}))^{val}$; 
				}
			}
		}
		
		\tcp{\color{cyan}worst-score dominant strategy by increasing the lowest score with small factors.}
		\nl \If{BestStrategy is $false$}{
			$s_{a,h+1,n,base} \leftarrow min(\textbf{s}_{a,h,n}); s_{a,h+1,n,tmp}   \leftarrow   min(\textbf{s}_{a,h,n})$; 
			
			\tcp{counting the numbers of each score value in increasing order.}
			$\textbf{scoreVal,scoreCt}  \leftarrow counting_{\uparrow,1}$
			$(\textbf{s}_{a,h,n})$; 
			
			\For{$val,ct$ in \textbf{scoreVal,scoreCt}}{
				\If{val is \textbf{not} $s_{a,h+1,n,base}$}{
					\tcp{increase from the lowest score.}
					$s_{a,h+1,n,tmp} \leftarrow  s_{a,h+1,n,tmp} + (ct*\Delta D/\sum(\textbf{scoreCt}))^{(6-val)}$;
				}
			}
		} 
		\textbf{return} $s_{a,h+1,n} \leftarrow s_{a,h+1,n,tmp}$;
		\caption{Aggregation of resilience scores for an attribute to the next hierarchical level using \textbf{\textit{impact-driven strategies}} dominated by either the best or the worst score.}
		\label{algo:1}
	\end{algorithm}
	
	\subsubsection{Aggregating scores dominated by the best case}\label{subsubsec:aggregation-best}
	Eight attributes, including {\myverb{PA1-2}}, {\myverb{CP1}}, and {\myverb{CA1-4}}, exhibit propagational behavior in which overall resilience is governed primarily by the best operational practices. 
	Consider {\myverb{PA1}}, which describes the type of hosting company for an authoritative name-server instance. Authoritative servers do not edit or originate DNS records; they only serve users with a copy of the records.
	Thus, having at least one instance operated by a highly trusted enterprise (\eg an SOE) can preserve service availability even if other instances are hosted by less trustworthy entities. In such cases, the presence of a highly resilient instance dominates the overall resilience of the authoritative name server.
	
	Consequently, when aggregating instance-level scores into a single score for the authoritative name server, the aggregated value should remain the best score and should never fall below the next-lower resilience level.
	For example, if all instances are operated by SOEs (each scoring 5), then the aggregated score is unequivocally 5. If one instance scores 5 and the remaining instances score 1, the aggregate score will be $(4,5]$, convergent to 4 as the number of poorly configured instances increases.
	The score cannot fall below 4, because a configuration that includes one instance operated by a state-owned enterprise (SOE) is inherently more resilient than a configuration where all instances are hosted by local private enterprises (which score 4). This reflects the realistic operational difficulty of remediating a large number of poorly configured instances while still benefiting from the uplift provided by a highly resilient one.

	To algorithmically capture this behavior, we develop a best-score-dominant aggregation strategy, in the first block in Algorithm~\ref{algo:1} (lines \#4-12). As shown in line \#9, for each unique score value $val$ that is lower than the best value $s_{a,h+1,base}$ in the set, the aggregated score is reduced by a fractional exponent $\left( ct *\Delta D \big/ \sum (\mathbf{scoreCt}) \right)^{val}$, proportional to the count $ct$ of that score among the instances. 
	
	This formulation reflects our \textbf{design principles}: all instances with the same score have equivalent resilience impact and are interchangeable; instances with higher scores dominate the resilience contribution of the group; the penalty introduced by a less resilient instance decreases exponentially with increasing $val$, ensuring that multiple moderately good instances never outweigh the impact of one high-quality instance; and with the diminishing factor $\Delta D$ set to 1, consistent with the granularity of the five-point scale, the aggregated score is guaranteed to never drop below the next lower meaningful score. 
	
	This best-score-dominant aggregation strategy applies to all attributes whose resilience behavior follows this pattern, including $f_{pa1,inst}$, $f_{pa1,serv}$, $f_{pa2,inst}$, $f_{pa2,serv}$, $f_{cp1,inst}$, $f_{ca1,inst}$, $f_{ca1,serv}$, $f_{ca2,serv}$, $f_{ca3,serv}$ and $f_{ca4,serv}$, as indicated in Fig.~\ref{fig:attribute}.

	\subsubsection{Aggregating scores dominated by the worst case}\label{subsubsec:aggregation-worst}
	Five attributes, including {\myverb{PP1-2}}, {\myverb{CP2}}, {\myverb{DP1}}, and {\myverb{DA1}}, exhibit propagational behavior in which overall resilience at the upper level is predominantly determined by the worst operational practice. These attributes require an aggregation strategy opposite to the best-case-dominant logic described in \S\ref{subsubsec:aggregation-best}. The corresponding algorithmic procedure is implemented in the second block of Algorithm~\ref{algo:1} (lines \#14–22).

	Consider \myverb{PP1}, which characterizes the hosting-enterprise type of a primary name-server instance. Because every primary instance has authority over the source zone file of a domain, an instance operated by an untrustworthy organization can propagate compromised records to all authoritative servers, potentially without detection by the domain owner. Therefore, the resilience of the primary name server is fundamentally limited by its least secure instance. 
	
	A similar pattern arises for {\myverb{CP2}}, which indicates whether a primary instance is co-hosted with authoritative functionality. If any instance is co-located and therefore exposed to the public Internet, the entire primary server (\eg \myverb{ns1.domain.gov}) becomes an attractive target for availability and integrity attacks \cite{examlabs_understanding_2025, evans_why_2022}. In such cases, the weakest configuration determines the resilience of the functionality.
	
	Motivated by these considerations, we define a worst-score-dominant aggregation strategy. 
	Instead of reducing a high score based on lower scores (as in the best-score-dominant strategy), this strategy begins with the worst score in the set ($s_{a,h+1,n,base}$ in line \#19) and incrementally increases it by up to one score step ($\Delta D$), depending on the count $ct$ and values $val$ of the better scoring instances. The denominator term $ \sum \left( \mathbf{scoreCt} \right)^{(6 - val)}$ ensures that higher-scoring instances with value $val$ contribute larger increments, \eg a score of 5 provides the maximum uplift, while lower scores contribute exponentially diminishing increments. This prevents a few slightly better instances from overshadowing the dominance of the worst configuration.
	This worst-score-dominant logic is applied to attributes $f_{pp1,inst}$, $f_{pp2,inst}$, $f_{cp2,inst}$, $f_{dp1,inst}$, $f_{da1,inst}$, and $f_{da1,serv}$, as indicated in Fig.~\ref{fig:attribute}. As shown in our quantitative evaluation (Appendix~\ref{Appendix:aggregationEvaluation}), both the best-case–dominant and worst-case–dominant strategies in Algorithm~\ref{algo:1} exhibit the intended aggregation behavior. 
	
	\section{Assessment Results for Government Domain Names in the Six Studied Countries}\label{sec:insights}
	
	In this section, we present the assessment results and key insights obtained by applying our systematic methodology (developed in \S\ref{sec:assessment}) to the dataset of government domains structured using our multi-sourced data schema (\S\ref{sec:sectiondataSchema}). Our analysis covers six representative countries: Australia, Brazil, France, Indonesia, the U.K., and the U.S. We begin with a comparative overview of authoritative DNS resilience across countries (\S\ref{subsec:insightsCountries}), and then examine the assessment scores of all 18 infrastructure-level attributes for individual domain names. This detailed analysis highlights strengths and weaknesses across the three operational phases: infrastructure placement (\S\ref{subsec:infraPlacement}), service configuration (\S\ref{subsec:serviceConfig}), and DNS record dispatch (\S\ref{subsec:recordDisp}). 
	
	\begin{figure}[t!]
		\centering
		\begin{subfigure}{0.24\linewidth}
			\includegraphics[width=\linewidth]{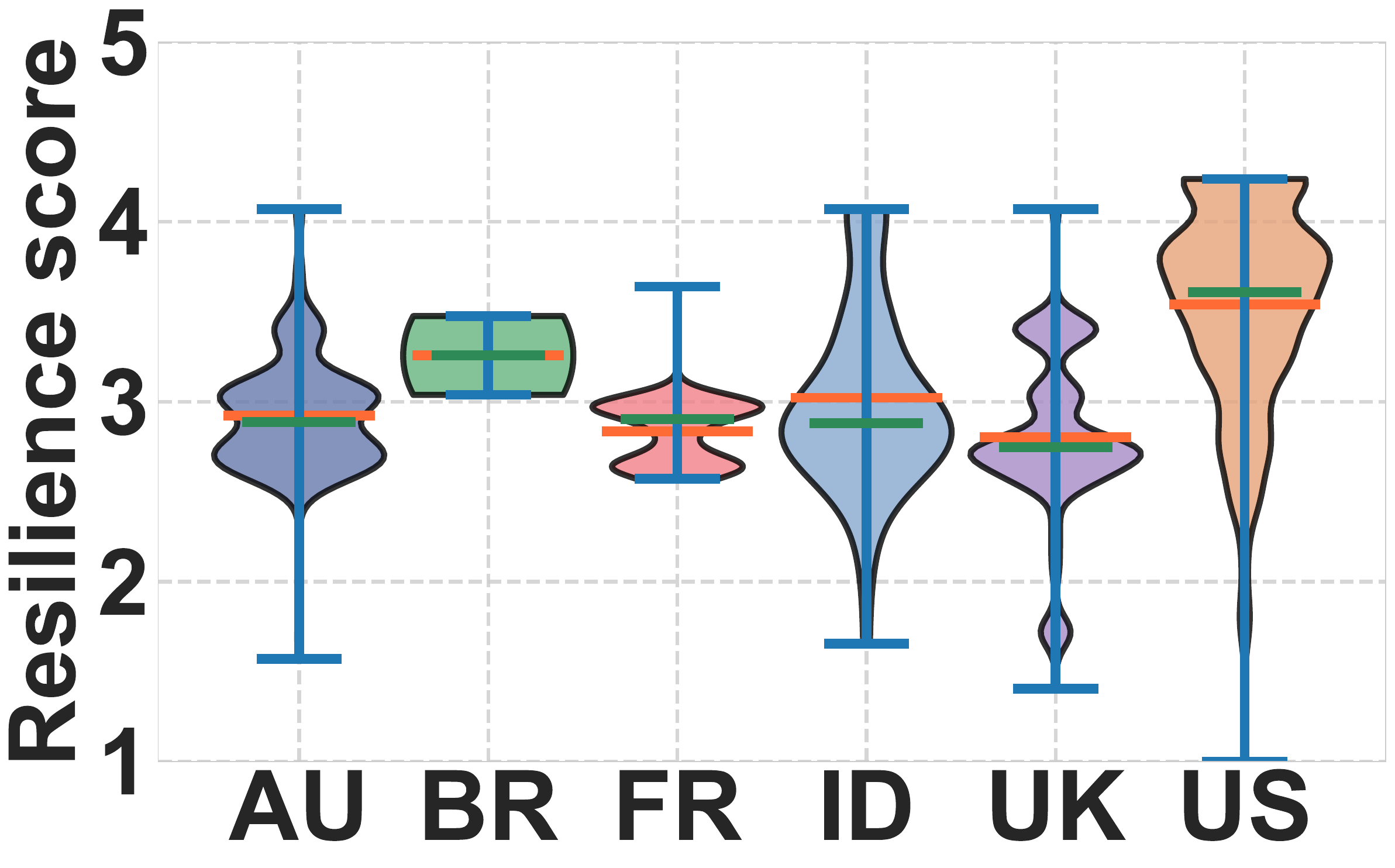}
			\caption{Overall}
			\label{fig:violin_mean}
		\end{subfigure}
		\begin{subfigure}{0.24\linewidth}
			\includegraphics[width=\linewidth]{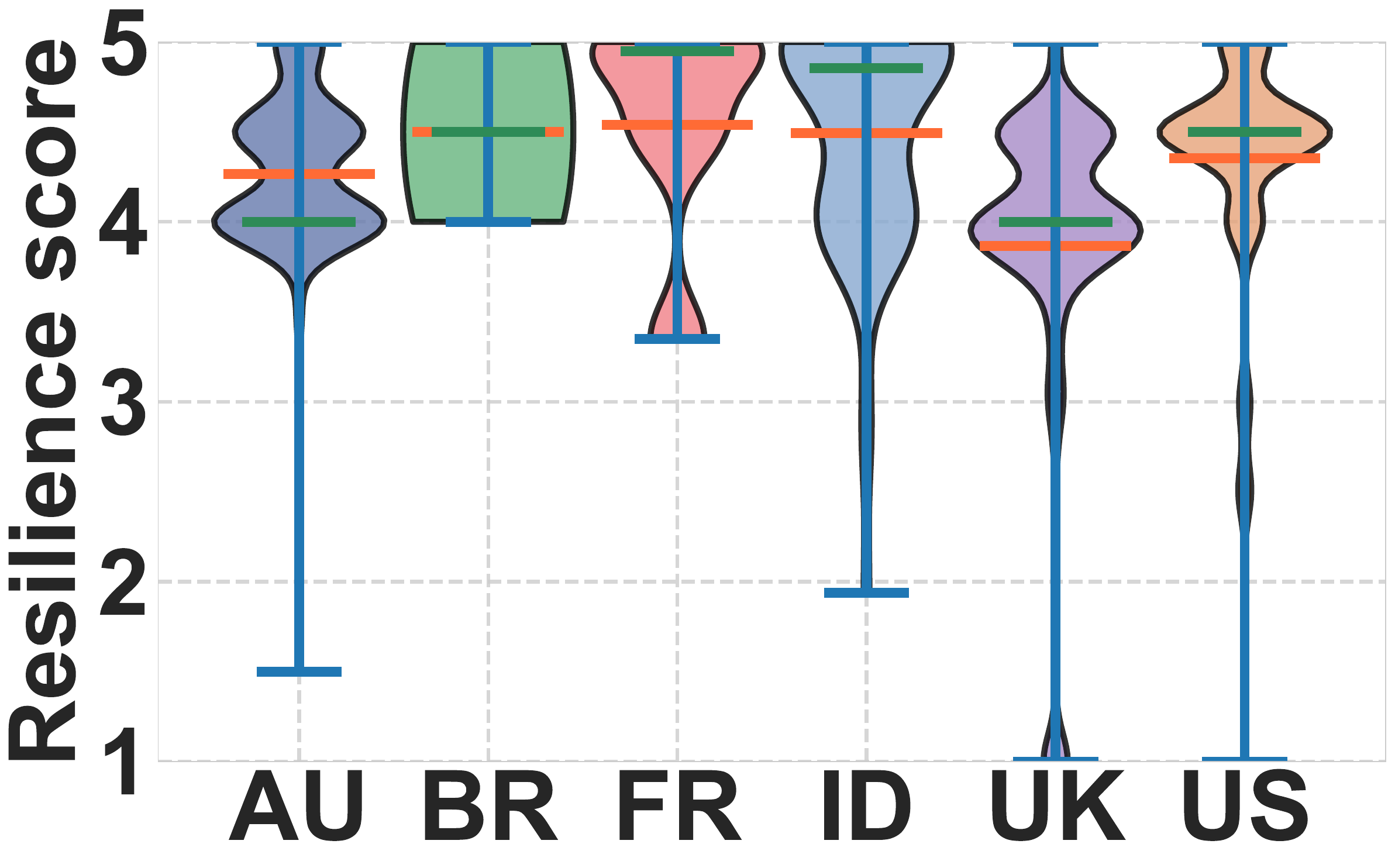}
			\caption{Infra. placement}
			\label{fig:violin_placement}
		\end{subfigure}
		\begin{subfigure}{0.24\linewidth}
			\includegraphics[width=\linewidth]{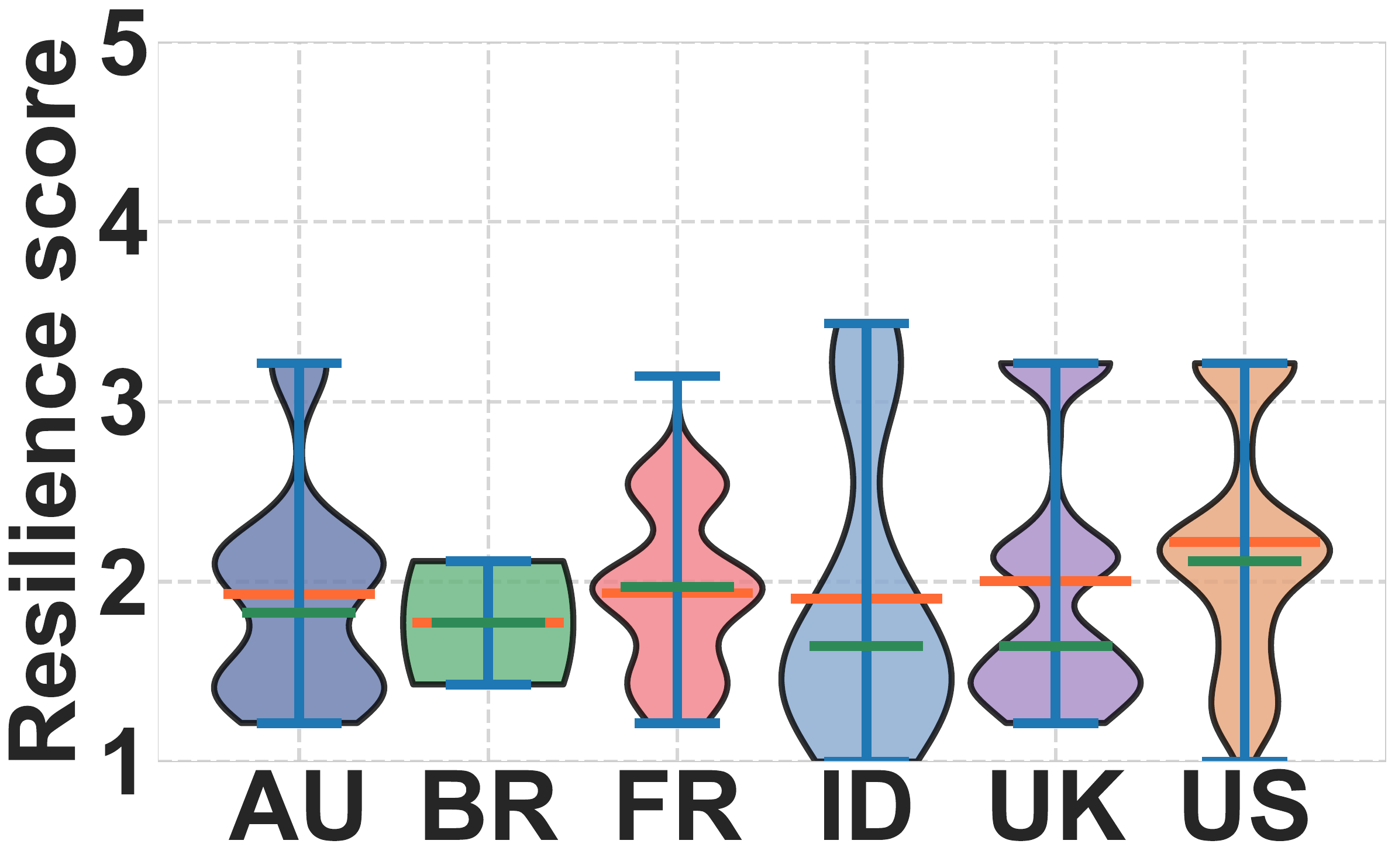}
			\caption{Serv. configuration}
			\label{fig:violin_configuration}
		\end{subfigure}
		\begin{subfigure}{0.24\linewidth}
			\includegraphics[width=\linewidth]{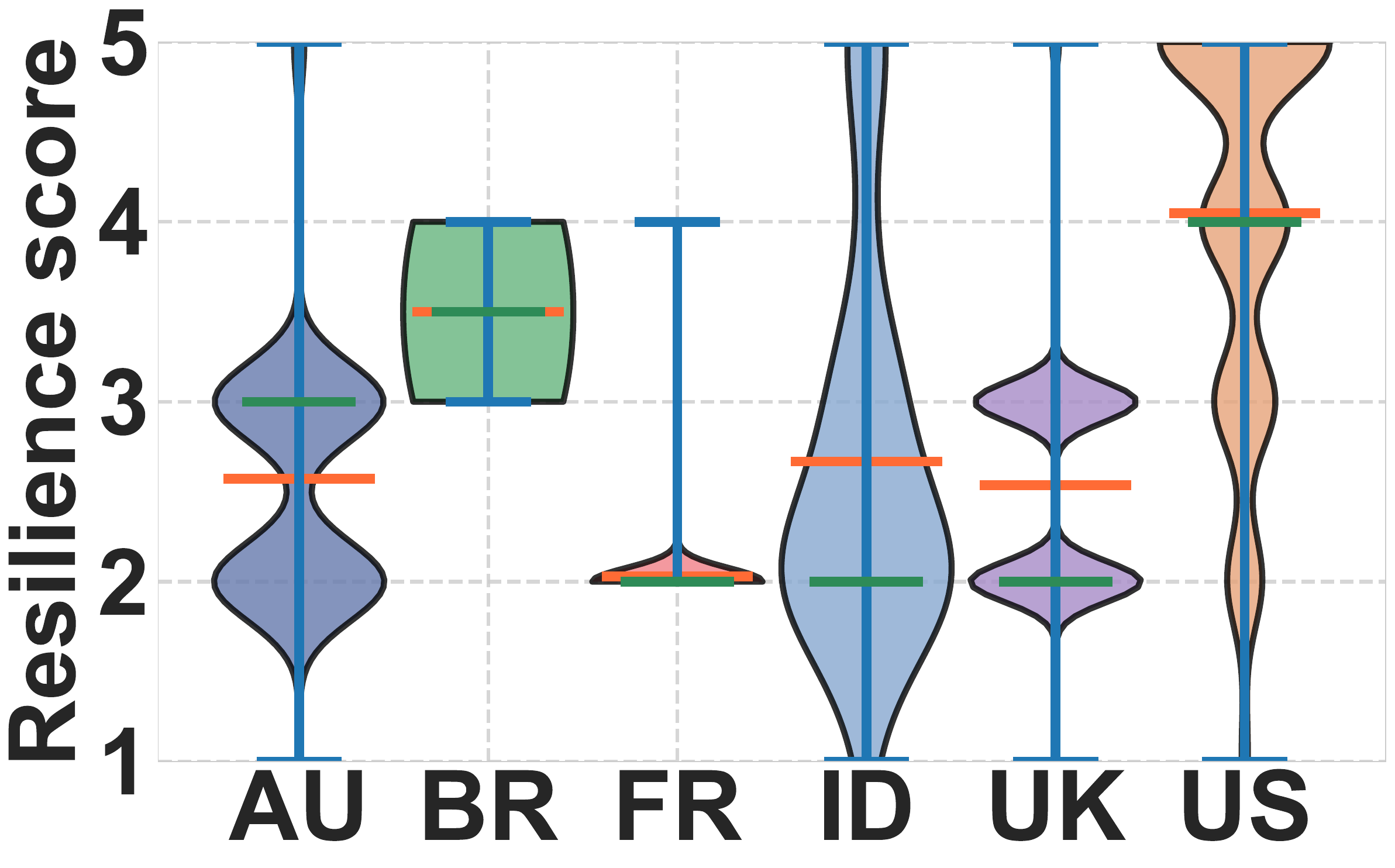}
			\caption{Record dispatch}
			\label{fig:violin_dispatch}
		\end{subfigure}
		\caption{Authoritative DNS resilience scores for government domain names in the six studied countries across: (a) overall operation, (b) infrastructure placement, (c) service configuration, and (d) record dispatch. The average values are represented as red horizontal lines.}
		\label{fig:violin_plot}
	\end{figure}
	
	\subsection{Overview of Government Authoritative DNS Resilience in the Six Countries}\label{subsec:insightsCountries}
	
	Before examining the 18 attribute scores aggregated at the infrastructure level for each domain, we first provide an executive-level overview of the operational resilience (across infrastructure placement, service configuration, DNS record dispatch, and overall dimensions), summarized using the averaging functions shown in the right region of Fig.~\ref{fig:attribute}. The distribution of the overall resilience scores per domain name in each country is visualized in Fig.~\ref{fig:violin_mean}, while Fig.~\ref{fig:violin_plot} presents the distributions for each operational phase, namely placement, configuration, and dispatch.
	
	Across all countries, the U.S. demonstrates the strongest overall operational resilience. 
	With 1,308 unique government domain names, the second largest among the six countries, the U.S. achieves an average score of 3.5/5 and a median of 3.6/5, driven by consistently higher resilience across all three operational phases compared to the others.
	
	\begin{figure}[t!]
		\centering
		\begin{subfigure}[b]{\linewidth}
			\centering
			\includegraphics[width=0.985\linewidth]{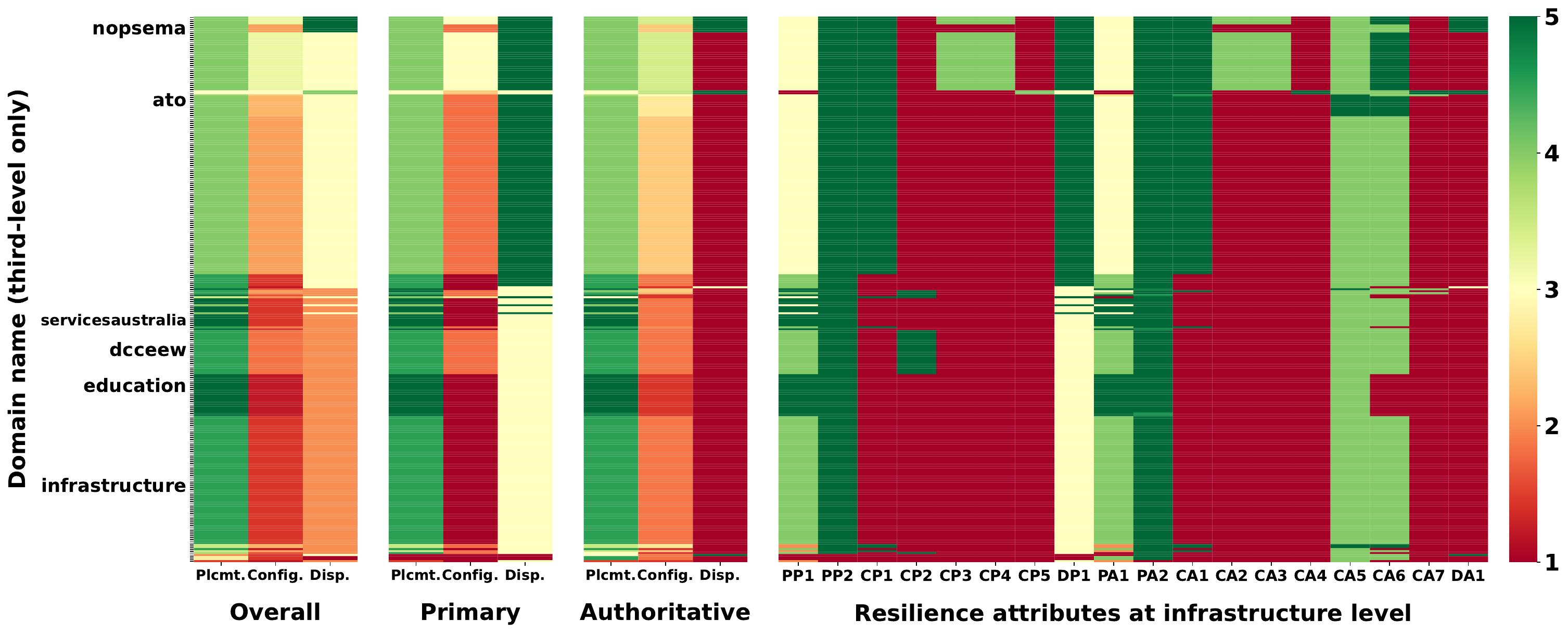}
			\caption{273 unique government domain names in \textbf{Australia}.}
			\label{fig:heatmap_au}
		\end{subfigure}
		
		\begin{subfigure}[b]{\linewidth}
			\centering
			\includegraphics[width=0.985\linewidth]{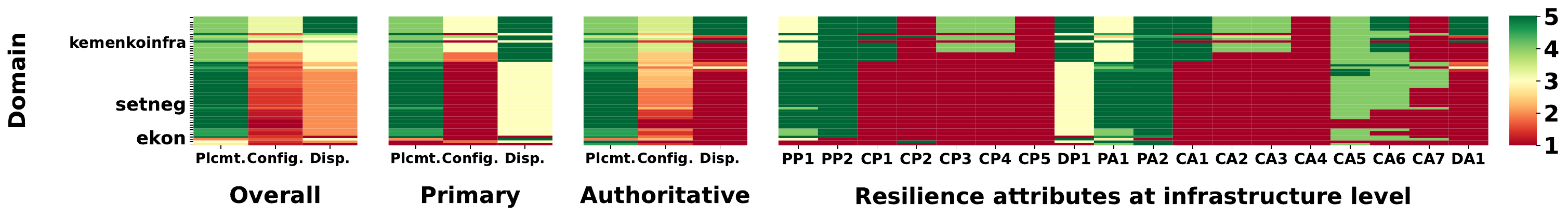}
			\caption{54 unique government domain names in \textbf{Indonesia}.}
			\label{fig:heatmap_id}
		\end{subfigure}
		
		\begin{subfigure}[b]{\linewidth}
			\centering
			\includegraphics[width=0.985\linewidth]{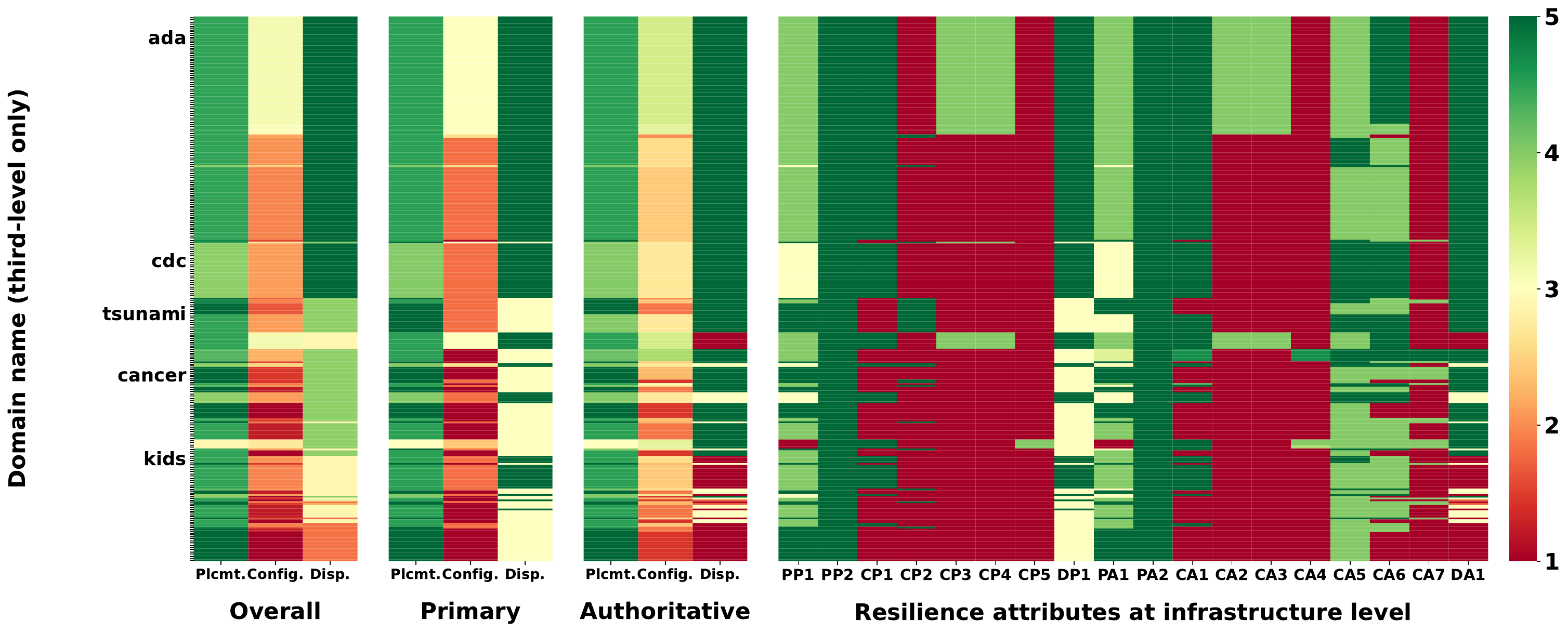}
			\caption{300 representative government domain names in the \textbf{U.S.}}
			\label{fig:heatmap_us}
		\end{subfigure}
		\vspace{-5mm}
		\caption{Resilience heatmaps for authoritative DNS infrastructures supporting government domains in (a) Australia, (b) Indonesia, and (c) the U.S. Each row corresponds to a domain name, and each column summarizes resilience scores across infrastructure placement, service configuration, and DNS record dispatch attributes.}
		\vspace{-3mm}
		\label{fig:heatmaps}
	\end{figure}

	\begin{figure}[t!]
		\centering
		\begin{subfigure}[b]{\linewidth}
			\centering
			\includegraphics[width=0.985\linewidth]{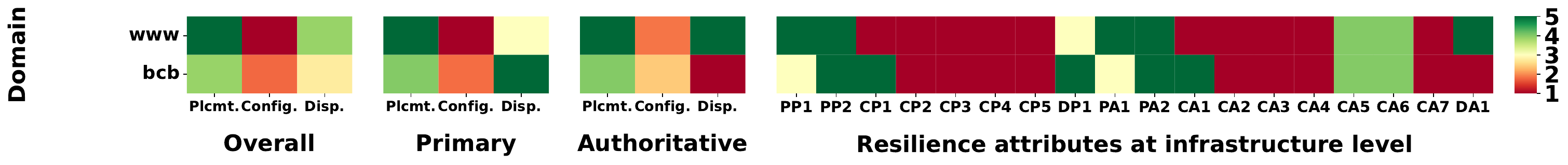}
			\caption{2 unique government domain names in \textbf{Brazil}.}
			\label{fig:heatmap_br}
		\end{subfigure}
		
		\begin{subfigure}[b]{\linewidth}
			\centering
			\includegraphics[width=0.985\linewidth]{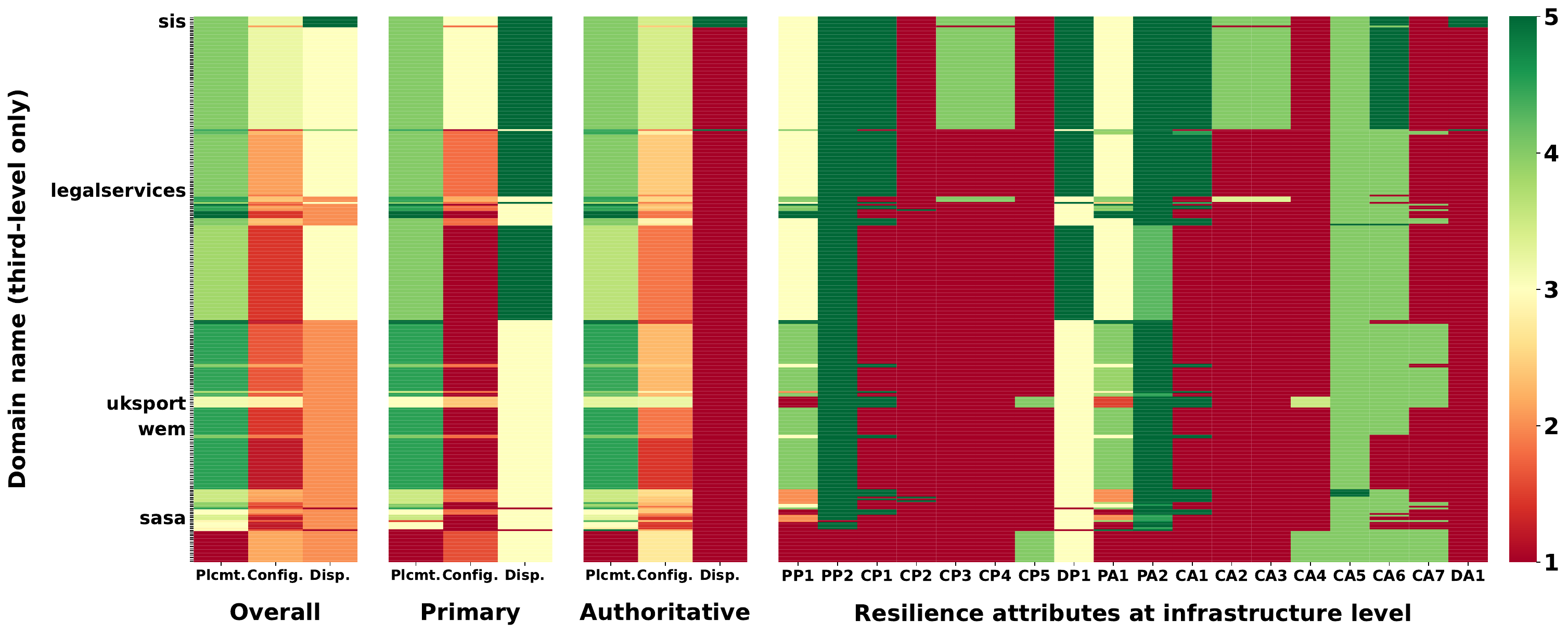}
			\caption{300 representative government domain names in the \textbf{U.K}.}
			\label{fig:heatmap_uk}
		\end{subfigure}
		
		\begin{subfigure}[b]{\linewidth}
			\centering
			\includegraphics[width=0.985\linewidth]{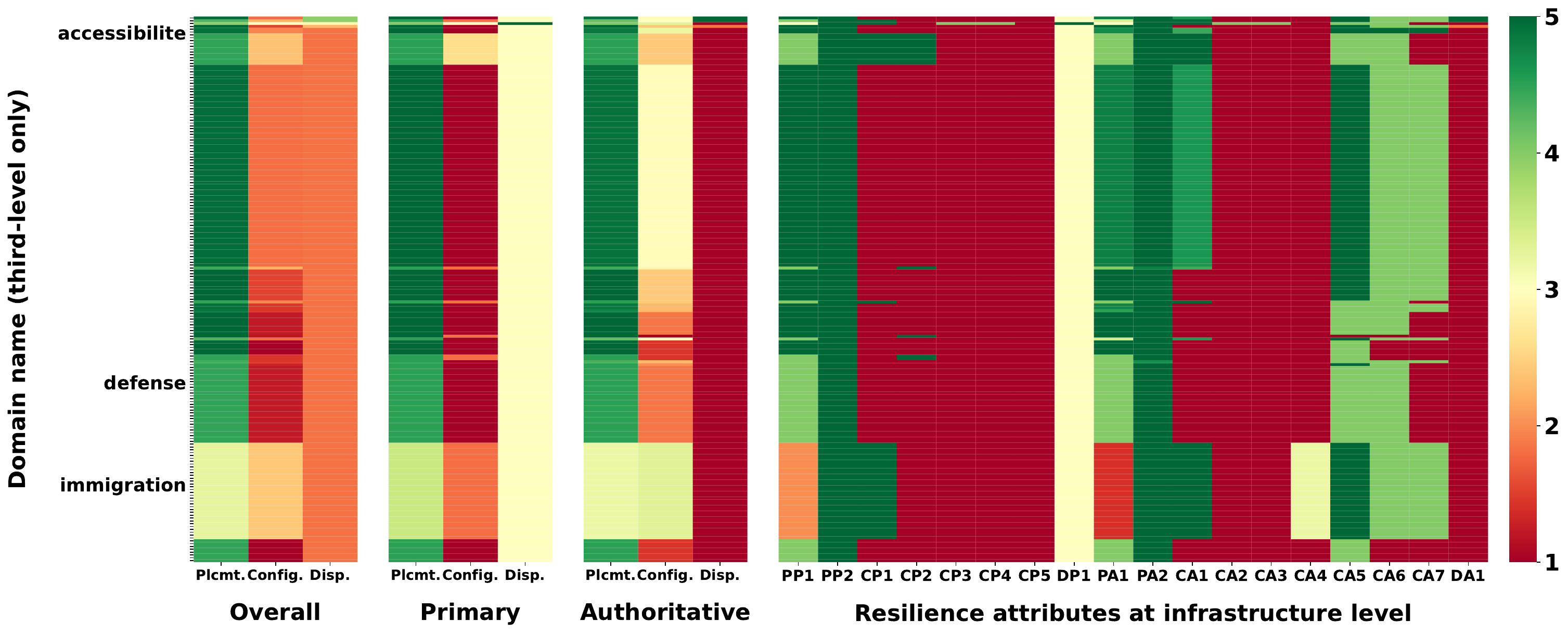}
			\caption{192 unique government domain names in \textbf{France}.}
			\label{fig:heatmap_fr}
		\end{subfigure}
		\vspace{-5mm}
		\caption{Resilience heatmaps for authoritative DNS infrastructures supporting government domain names in Brazil, the U.K., and France. Each row represents a domain name, and each column reflects resilience scores across placement, configuration, and dispatch attributes.}
		\label{fig:heatmaps_appendix}
		\vspace{-3mm}
	\end{figure}

	The U.K., which manages the largest number of unique government domains (5,475), exhibits mean and median scores of approximately 2.8/5, placing it near the cross-country average. Its distribution forms three distinct clusters, approximately 3.5/5 (strong), 3/5 (moderate), and 1.5/5 (weak). Although we cannot confirm administrative factors without access to confidential data, these clusters can reflect differing operational models for management practices across government agencies in the respective countries. 
	
	Australia, France, and Indonesia show relatively balanced distributions, with their average and median scores near 3/5 and without strongly separated clusters. However, France displays a small subset of domains with notably weaker resilience (below 2.9/5). Brazil, despite having only two unique government domains, each supported by a single authoritative DNS infrastructure, achieves the second-highest overall resilience, with both domains scoring 3.3/5 on average and at the median.
	
	We next analyze the attribute-level assessment results at the authoritative DNS infrastructure level, highlighting the distinct strengths and weaknesses in operational practices observed across the three phases for government domain names in each country.
	
	\subsection{Resilience Scores for Infrastructure Placement Attributes}
	\label{subsec:infraPlacement}
	
	We now examine the assessment results for each government domain name, shown as rows in the heatmaps in Fig.~\ref{fig:heatmaps} and Fig.~\ref{fig:heatmaps_appendix}.
	Across the {\myverb{Plcmt.}} column within the {\myverb{Overall}} blocks for each country, both \textbf{Indonesia} and \textbf{Brazil} demonstrate strong placement resilience, with nearly all domains scoring above 4/5 and almost no domains exhibiting amber or red cells (indicating scores below 3/5). This pattern is consistent across both primary and authoritative placement attributes (\ie {\myverb{PP1-2}} and {\myverb{PA1-2}}), with only  a few exceptions. These results likely reflect the relatively small number of government domains in both countries and their correspondingly centralized management. 
	
	For example, Brazil manages only two unique government domains {\myverb{www.gov.br}} and {\myverb{bcb.gov.br}}. All primary and authoritative name-server instances for {\myverb{www.gov.br}} are hosted by a Brazilian state-owned enterprise, and all instances for {\myverb{bcb.gov.br}} are hosted by a major cloud provider. In Indonesia, a national regulation \cite{presiden_peraturan_2018} requires government digital services to be operated by the national data center (\ie SOE), and most domains comply with this requirement. The one exception is \myverb{ekon.go.id},  which delegates its primary name server to a small foreign company (as shown in the last row of Fig.~\ref{fig:heatmap_id}).
	\textbf{France} also exhibits high placement resilience for most government domains, consistent with its strict national cybersecurity regulations \cite{anssi_french_2025}. However, a small subset of domains perform poorly, particularly for IP administration country attributes (\ie {\myverb{PP2}} and {\myverb{PA2}}), suggesting that some servers are still hosted or administered outside national borders.
	
	Government domains in \textbf{Australia} and the \textbf{U.K.} show more moderate placement resilience, with most domains scoring 3-4 out of 5 across the {\myverb{plcmt.}} columns in the {\myverb{Overall}}, {\myverb{Primary}}, and {\myverb{Authoritative}} blocks. This is largely driven by the attributes of the hosting provider (\ie {\myverb{PP1}} and {\myverb{PA1}}), as both countries rely extensively on US-headquartered cloud providers for authoritative DNS services.
	Finally, in the \textbf{U.S.}, most government domains are supported by cloud DNS infrastructures owned and operated by US entities with domestically administrated IP address space. A small number of domains rely on SOEs or private SMEs, but these exceptions do not significantly affect the overall resilience profile of placement.

	\begin{takeawaybox}
		Indonesia and Brazil which have relatively few government domains, and France, which enforces strict national cybersecurity regulations, demonstrate stronger resilience in infrastructure placement than the U.S., U.K. and Australia, where government services more heavily rely on foreign or local cloud providers.
	\end{takeawaybox}
	
	\subsection{Resilience Scores for Service Configuration Attributes}
	\label{subsec:serviceConfig}
	
	We now examine resilience scores for service-configuration practices, as reflected in the {\myverb{Config.}} column and the underlying attributes {\myverb{CP1-5}} and {\myverb{CA1-7}} across the six countries studied. While configurations vary widely between individual domain names, three broad patterns emerge.
	
	Overall, service-configuration is modest in all six countries, with most domains scoring between 2 and 3 out of 5. Among them, the \textbf{U.S.} (Fig.~\ref{fig:heatmap_us}) demonstrates the strongest performance, with the {\myverb{Config.}} columns dominated by amber-colored cells in, indicating comparatively better configuration practices than those observed in other countries, where orange and light-red cells are more common. Weak resilience is most pronounced for the primary functionality, especially in {\myverb{CP5}} (AS redundancy), a challenge shared across all domains. 
	Maintaining multiple autonomous systems for redundancy is operationally costly, making it difficult for domain operators to adopt this best practice. 
	Even so, 294 U.S. domains (22.48\% of 1,308) do exhibit strong redundancy in hosting IPs and subnets, as indicated by higher scores in {\myverb{CP3-4}} and {\myverb{CA2-3}}. 
	
	Surprisingly, co-hosting authoritative and primary name-server functionality (indicated by {\myverb{CP2}}) is common across all six countries. Although operationally convenient, co-hosting exposes the primary name server, a critical, non-public function, to unnecessary Internet-facing attack surfaces.
	
	Government domains in \textbf{Australia} (Fig.~\ref{fig:heatmap_au}) and the \textbf{U.K.} (Fig.~\ref{fig:heatmap_uk}) show slightly lower configuration resilience compared to the U.S. These domains frequently score poorly in regional accessibility (indicated by {\myverb{CP1}}) and in redundancy across IPs (by {\myverb{CP3}} and {\myverb{CA2}}) and subnets (by {\myverb{CP4}} and {\myverb{CA3}}), especially when compared to the U.S.
	
	Interestingly, the countries with the strongest infrastructure placement resilience, \textbf{Indonesia} (Fig.~\ref{fig:heatmap_id}), \textbf{Brazil} (Fig.~\ref{fig:heatmap_br}), and \textbf{France} (Fig.~\ref{fig:heatmap_fr}), often exhibit the weakest service configurations. This contrast may reflect limited operational capacity among local hosting enterprises, despite strong placement policies. Nonetheless, there are exceptions. For example, {\myverb{accessibilite.gouv.fr}} achieves strong configuration resilience, with Anycast-enabled regional accessibility (\ie {\myverb{CP1}} and {\myverb{CA1}}) and separation of primary and authoritative (\ie {\myverb{CP2}}) implemented by a French provider.
	
	\begin{takeawaybox}
		All studied countries show moderate configuration resilience and common weaknesses, including limited AS-level redundancy and co-hosting of primary and authoritative name servers on the same IP addresses. The U.S., U.K. and Australia demonstrate comparatively better redundancy and regional accessibility.
	\end{takeawaybox}
	
	\subsection{Resilience Scores for DNS Record Dispatch Attributes}
	\label{subsec:recordDisp}
	
	Reflected in {\myverb{Disp.}}, {\myverb{DP1}} and {\myverb{DA1}} columns, government domains in the \textbf{U.S.} demonstrate the strongest resilience in record dispatch among the six countries. Most US domains display amber or green cell (scores 3-5) for both primary and authoritative functions. A substantial majority, 1,042 domains (79.66\%), deploy DNSSEC on authoritative servers to protect DNS responses against hijacking or manipulations, with 953 domains (72.86\%) using algorithms recommended by RFC8624 \cite{rfc8624}. 
	
	For primary servers, almost all government domains across all countries, except for isolated cases such as {\myverb{fha.gov}}, have AXFR (\ie {\myverb{DP1}}) enforced either manually or automatically. Instances with a red {\myverb{DP1}} score (no AXFR enforcement), although rare, pose meaningful risks by permitting unauthorized retrieval of authoritative DNS records.
	
	Government domains in \textbf{Australia}, \textbf{Indonesia}, the \textbf{U.K.}, and \textbf{Brazil} exhibit weaker resilience in the record dispatch phase, primarily due to the absence of DNSSEC on their authoritative name servers (\ie {\myverb{DA1}}). Without DNSSEC, authoritative responses remain vulnerable to on-path manipulation even when AXFR is correctly enforced at the primary server.
	
	Among the six countries, \textbf{France} displays the most concerning resilience profile for DNS record dispatch. Except for {\myverb{dmp.gouv.fr}}, all French government domains rely on manual enforcement of AXFR for primary servers, introducing additional operational risk through human-dependent configuration. Only three domains (including {\myverb{finances.gouv.fr}}, {\myverb{culture.gouv.fr}}, and {\myverb{douanne.gouv.fr}}) implement DNSSEC, though all three correctly use recommended algorithms. The near absence of DNSSEC deployment reduces resilience against data-integrity attacks.
	
	\begin{takeawaybox}
		The U.S. shows the strongest DNS record dispatch resilience, driven by broader DNSSEC deployment, while France shows the weakest due to low DNSSEC adoption and greater reliance on manual AXFR enforcement.
	\end{takeawaybox}
	
	\section{Related Work}
	\textbf{Assessing resilience of government public services:} Given the criticality of public services for both taxpayers and governments, evaluating the resilience of digital infrastructures that support these services has been an active topic in the measurement community. Prior works typically study isolated components of the ecosystem. 
	Kumar \textit{et al.} \cite{kumar_choices_2024} analyze the resilience from the perspective of website-hosting infrastructures for government services worldwide, focusing on the organizational profile of hosting entities (\eg third-party providers vs. domestic providers).
	Habib \textit{et al.} \cite{habib_formalizing_2025} quantify the dependence of web infrastructures, including those supporting government portals, on geopolitical factors in 150 countries.
	Houser \textit{et al.} \cite{houser_comprehensive_2022} evaluate the availability and legitimacy of DNS records served by authoritative name servers for public-service domains.
	Sommese \textit{et al.} \cite{sommese_assessing_2022} assess redundancy in recursive DNS resolution paths for government domains in the U.S., Netherlands, Sweden, and Switzerland.
	Jonker \textit{et al.} \cite{jonker_where_2022} study longitudinal changes in Russia's digital-infrastructure ecosystem, including naming, hosting, and certificate issuance, under resilience and sovereignty pressures.
	To the best of our knowledge, our work is the first to evaluate the resilience of authoritative DNS infrastructures with a holistic view of the network operational process, spanning infrastructure placement, service configuration, and record dispatch.

	\textbf{Analyzing resilience and security of DNS infrastructure:} 
	Beyond government-specific studies, a substantial body of literature examines the resilience and security of DNS infrastructure more broadly. 
	The work in \cite{COMNET2022} provides a large-scale characterization of DNS behavior, query failures, and underlying causes, including those relevant to DNS infrastructure.
	Several works identify integrity risks and configuration vulnerabilities in resolver infrastructures \cite{xu_tsuking_2023, hilton_fourteen_2023, afek_nrdelegationattack_2023, randall_trufflehunter_2020, man_dns_2021, li_tudoor_2024, dahlberg_timeless_2023}. 
	Cross-country dependency analyses \cite{kumar_each_2023, kashaf_analyzing_2020} evaluate geopolitical intergovernmental factors that shape DNS operational resilience. 
	Another line of research studies confidentiality attacks, including domain hijacking, that exploit weaknesses in DNS service operation \cite{moura_characterizing_2024, zhang_silence_2023, sommese_investigating_2022, akiwate_retroactive_2022, deccio_behind_2020, alowaisheq_zombie_2020}. 
	Complementary efforts \cite{zhang_rethinking_2024, galloway_practical_2024, liu_dial_2023,lyu_classifying_2022, kosek_dns_2022, isobe_first_2022, chhabra_measuring_2021, moura_clouding_2020, izhikevich_zdns_2022} examine DNS end-host behavior, including deployment of encrypted DNS protocols \cite{sommese_darkdns_2024, nisenoff_user_2023, randall_home_2021, dong_exploring_2024}, enterprise DNS-query patterns \cite{lyu_enterprise_2023, lyu_hierarchical_2021}, and DNS-based attacks such as reflection, resource exhaustion, and query amplification \cite{duan_camp_2024, nawrocki_far_2021, li_dnsbomb_2024, moura_tsuname_2021}. 
	These works highlight the security and performance implications of resolver design, encryption, and attacker behavior.
	Complementing prior studies, our work is the first to systematically assess the \textbf{authoritative} DNS infrastructure of a domain through the lens of its full network operational process. By modeling placement decisions, configuration practices, and record-dispatch mechanisms, we surface resilience issues that are inherently distributed and often overlooked by domain owners.
	
	\section{Discussion on Limitations}
	We acknowledge the limitations identified in our proposed methodology.
	To assign assessment scores for each attribute, we adopt a five-point Likert scale that serves as a diagnostic heuristic. The score aggregation from server instances to the overall infrastructure level uses impact-driven strategies that reflect our heuristic design principles from best practices and expert judgment. Additionally, evaluating our assessment method against real-world outages caused by resilience-related practices is beyond the scope of this work. Therefore, further investigations such as systematic sensitivity analysis and empirical validation against real-world outcomes are valuable future works.
	For example, assessment scores of each domain can be correlated with observed real-world outages caused by various factors described by the resilience attributes. A more systematic sensitivity analysis over score aggregation weight (beyond the analysis in Appendix~\ref{Appendix:weightAnalysis}) can help clarify the intended use of our assessment framework and reduce the risk of over-interpreting the quantitative scores. Also, additional signals such as routing security can be further incorporated in our assessment framework to provide a more comprehensive coverage of resilience aspects.

	\section{Conclusion}
	We presented a systematic framework to assess the network operational resilience of authoritative DNS infrastructures supporting public-service domains operated by government agencies. Designed to serve both strategic policymakers (who require an aggregated understanding of national DNS resilience) and the individual departments responsible for operating these infrastructures, our scoring-based methodology evaluates a comprehensive set of attributes derived from a multi-sourced data schema. These attributes capture operational practices at four hierarchical levels of the authoritative DNS infrastructure and across three phases of the network operational process. 
	We applied our framework to government domain names published by six representative countries and analyzed the resulting dataset. The assessment revealed previously unknown insights into the resilience of authoritative DNS infrastructures at both the country and domain levels. Our findings demonstrate the value of a structured, data-driven approach for governance agencies seeking to understand systematic resilience, and for domain owners to identify weaknesses and improve operational practices in their DNS configurations.
	
	\section*{Acknowledgment}
	We thank our anonymous shepherd and the anonymous reviewers for their insightful feedback and suggestions that helped us further improve the clarity and quality of this paper.

	\bibliographystyle{ACM-Reference-Format}
	\bibliography{bibliodns}

@misc{dataset,
	author = {Lyu, Minzhao},
	title = {{Critical Infrastructure Measurement Dataset}},
	year = {2026},
	url = {https://minzhaolyu.github.io/dataset/InternetInfrastructureDataset},
	note = {Accessed: 2026-03-19}
}

@article{ASeptiadiSIGMETRICS26,
	author = {Septiadi, Agung and Lyu, Minzhao and Habibi Gharakheili, Hassan and Sivaraman, Vijay},
	title = {{Assessing Resilience in Authoritative DNS Infrastructure
	Supporting Government Services}},
	journal = {Proceedings of the ACM on Measurement and Analysis of Computing Systems},
	year = {2026},
	month = {Jun},
}

@inproceedings {afek_nrdelegationattack_2023,
author = {{Yehuda Afek and Anat Bremler-Barr and Shani Stajnrod}},
title = {{NRDelegationAttack}: Complexity {DDoS} attack on {DNS} Recursive Resolvers},
booktitle = {Proc. USENIX Security},
year = {2023},
month = aug,
address = {Anaheim, CA, USA}
}

@inproceedings{akiwate_retroactive_2022,
author = {Akiwate, Gautam and Sommese, Raffaele and Jonker, Mattijs and Durumeric, Zakir and Claffy, KC and Voelker, Geoffrey M. and Savage, Stefan},
title = {{Retroactive identification of targeted DNS infrastructure hijacking}},
year = {2022},
month = oct,
booktitle = {Proc. ACM IMC},
address = {Nice, France}
}

@inproceedings{alowaisheq_zombie_2020,
author = {Alowaisheq, Eihal and Tang, Siyuan and Wang, Zhihao and Alharbi, Fatemah and Liao, Xiaojing and Wang, XiaoFeng},
title = {{Zombie Awakening: Stealthy Hijacking of Active Domains through DNS Hosting Referral}},
year = {2020},
booktitle = {Proc. ACM CCS},
address = {Virtual Event}
}

@inproceedings{arghavani_suss_2024,
author = {Arghavani, Mahdi and Zhang, Haibo and Eyers, David and Arghavani, Abbas},
title = {{SUSS: Improving TCP Performance by Speeding Up Slow-Start}},
year = {2024},
booktitle = {Proc. ACM SIGCOMM},
address = {Sydney, NSW, Australia}
}

@inproceedings{chhabra_measuring_2021,
author = {Chhabra, Rishabh and Murley, Paul and Kumar, Deepak and Bailey, Michael and Wang, Gang},
title = {{Measuring DNS-over-HTTPS performance around the world}},
year = {2021},
booktitle = {Proc. ACM IMC},
address = {Virtual Event},
}

@inproceedings {dahlberg_timeless_2023,
author = {Rasmus Dahlberg and Tobias Pulls},
title = {{Timeless Timing Attacks and Preload Defenses in Tor{\textquoteright}s {DNS} Cache}},
booktitle = {Proc. USENIX Security},
year = {2023},
address = {Anaheim, CA, USA},
month = aug,
}

@inproceedings{deccio_behind_2020,
author = {Deccio, Casey and Hilton, Alden and Briggs, Michael and Avery, Trevin and Richardson, Robert},
title = {{Behind Closed Doors: A Network Tale of Spoofing, Intrusion, and False DNS Security}},
year = {2020},
booktitle = {Proc. ACM IMC},
address = {Virtual Event, USA}
}

@inproceedings{dong_exploring_2024,
author = {Dong, Hongying and Zhang, Yizhe and Lee, Hyeonmin and Huque, Shumon and Sun, Yixin},
title = {{Exploring the Ecosystem of DNS HTTPS Resource Records: An End-to-End Perspective}},
year = {2024},
booktitle = {Proc. ACM IMC},
address = {Madrid, Spain}
}

@inproceedings{duan_camp_2024,
author = {Huayi Duan and Marco Bearzi and Jodok Vieli and David Basin and Adrian Perrig and Si Liu and Bernhard Tellenbach},
title = {{{CAMP}: Compositional Amplification Attacks against {DNS}}},
booktitle = {Proc. USENIX Security},
year = {2024},
address = {Philadelphia, PA, USA},
month = aug
}

@inproceedings{galloway_practical_2024,
author={Galloway, Tillson and Karakolios, Kleanthis and Ma, Zane and Perdisci, Roberto and Keromytis, Angelos and Antonakakis, Manos},
booktitle={Proc. IEEE S\&P}, 
title={{Practical Attacks Against DNS Reputation Systems}}, 
year={2024},
}

@inproceedings{griffioen_have_2024,
author = {Griffioen, Harm and Koursiounis, Georgios and Smaragdakis, Georgios and Doerr, Christian},
title = {{Have you SYN me? Characterizing Ten Years of Internet Scanning}},
year = {2024},
booktitle = {Proc. ACM IMC},
address = {Madrid, Spain}
}

@inproceedings{habib_formalizing_2025,
author = {Habib, Rumaisa and Ruth, Kimberly and Akiwate, Gautam and Durumeric, Zakir},
title = {{Formalizing Dependence of Web Infrastructure}},
year = {2025},
booktitle = {Proc. ACM SIGCOMM},
address = {S\~{a}o Francisco Convent, Coimbra, Portugal}
}

@inproceedings{heftrig_harder_2024,
author = {Heftrig, Elias and Schulmann, Haya and Vogel, Niklas and Waidner, Michael},
title = {{The Harder You Try, The Harder You Fail: The KeyTrap Denial-of-Service Algorithmic Complexity Attacks on DNSSEC}},
year = {2024},
booktitle = {Proc. ACM CCS},
}

@inproceedings{hiesgen_age_2024,
author = {Hiesgen, Raphael and Nawrocki, Marcin and Barcellos, Marinho and Kopp, Daniel and Hohlfeld, Oliver and Chan, Echo and Dobbins, Roland and Doerr, Christian and Rossow, Christian and Thomas, Daniel R. and Jonker, Mattijs and Mok, Ricky and Luo, Xiapu and Kristoff, John and Schmidt, Thomas C. and W\"{a}hlisch, Matthias and claffy, kc},
title = {{The Age of DDoScovery: An Empirical Comparison of Industry and Academic DDoS Assessments}},
year = {2024},
booktitle = {Proc. ACM IMC},
address = {Madrid, Spain}
}

@inproceedings{hilton_fourteen_2023,
author = {Alden Hilton and Casey Deccio and Jacob Davis},
title = {{Fourteen Years in the Life: A Root {Server{\textquoteright}s} Perspective on {DNS} Resolver Security}},
booktitle = {Proc. USENIX Security},
year = {2023},
address = {Anaheim, CA, USA},
month = aug
}

@inproceedings{houser_comprehensive_2022,
author={Houser, Rebekah and Hao, Shuai and Cotton, Chase and Wang, Haining},
booktitle={Proc. IEEE/IFIP DSN}, 
title={{A Comprehensive, Longitudinal Study of Government DNS Deployment at Global Scale}}, 
year={2022},
}

@inproceedings{isobe_first_2022,
author = {Isobe, Katsuki and Kondo, Daishi and Tode, Hideki},
title = {{A first look at the name resolution latency on handshake}},
year = {2022},
booktitle = {Proc. ACM IMC},
address = {Nice, France}
}

@inproceedings{izhikevich_zdns_2022,
author = {Izhikevich, Liz and Akiwate, Gautam and Berger, Briana and Drakontaidis, Spencer and Ascheman, Anna and Pearce, Paul and Adrian, David and Durumeric, Zakir},
title = {{ZDNS: a fast DNS toolkit for internet measurement}},
year = {2022},
booktitle = {Proc. ACM IMC},
address = {Nice, France}
}

@inproceedings{jain_ukrainian_2022,
author = {Jain, Akshath and Patra, Deepayan and Xu, Peijing and Sherry, Justine and Gill, Phillipa},
title = {{The ukrainian internet under attack: an NDT perspective}},
year = {2022},
booktitle = {Proc. ACM IMC},
address = {Nice, France}
}

@inproceedings{jonker_where_2022,
author = {Jonker, Mattijs and Akiwate, Gautam and Affinito, Antonia and Claffy, kc and Botta, Alessio and Voelker, Geoffrey M. and van Rijswijk-Deij, Roland and Savage, Stefan},
title = {{Where .ru? Assessing the Impact of Conflict on Russian Domain Infrastructure}},
year = {2022},
booktitle = {Proc. ACM IMC},
address = {Nice, France}
}

@inproceedings{kashaf_analyzing_2020,
author = {Kashaf, Aqsa and Sekar, Vyas and Agarwal, Yuvraj},
title = {{Analyzing Third Party Service Dependencies in Modern Web Services: Have We Learned from the Mirai-Dyn Incident?}},
year = {2020},
booktitle = {Proc. ACM IMC},
address = {Virtual Event}
}

@inproceedings{kosek_dns_2022,
author = {Kosek, Mike and Schumann, Luca and Marx, Robin and Doan, Trinh Viet and Bajpai, Vaibhav},
title = {{DNS privacy with speed? evaluating DNS over QUIC and its impact on web performance}},
year = {2022},
booktitle = {Proc. ACM IMC},
address = {Nice, France}
}

@inproceedings{kumar_choices_2024,
author = {Kumar, Rashna and Carisimo, Esteban and Riva, Lukas De Angelis and Buzzone, Mauricio and Bustamante, Fabi\'{a}n E. and Qazi, Ihsan Ayyub and Beir\'{o}, Mariano G.},
title = {{Of Choices and Control - A Comparative Analysis of Government Hosting}},
year = {2024},
booktitle = {Proc. ACM IMC},
address = {Madrid, Spain}
}

@inproceedings{lee_chatfive_2024,
author = {Lee, Jungjae and Choi, Yubin and Song, Minhyuk and Park, Sanghyun},
title = {{ChatFive: Enhancing User Experience in Likert Scale Personality Test through Interactive Conversation with LLM Agents}},
year = {2024},
booktitle = {Proc. ACM CUI},
address = {Luxembourg, Luxembourg}
}

@inproceedings{li_dnsbomb_2024,
author={Li, Xiang and Wu, Dashuai and Duan, Haixin and Li, Qi},
booktitle={IEEE Symposium on Security and Privacy}, 
title={{DNSBomb: A New Practical-and-Powerful Pulsing DoS Attack Exploiting DNS Queries-and-Responses}}, 
year={2024},
}

@inproceedings{li_tudoor_2024,
author={Li, Xiang and Xu, Wei and Liu, Baojun and Zhang, Mingming and Li, Zhou and Zhang, Jia and Chang, Deliang and Zheng, Xiaofeng and Wang, Chuhan and Chen, Jianjun and Duan, Haixin and Li, Qi},
booktitle={IEEE Symposium on Security and Privacy}, 
title={{TuDoor Attack: Systematically Exploring and Exploiting Logic Vulnerabilities in DNS Response Pre-processing with Malformed Packets}}, 
year={2024},
}

@inproceedings{liu_dial_2023,
author = {Liu, Guannan and Jin, Lin and Hao, Shuai and Zhang, Yubao and Liu, Daiping and Stavrou, Angelos and Wang, Haining},
title = {{Dial "N" for NXDomain: The Scale, Origin, and Security Implications of DNS Queries to Non-Existent Domains}},
year = {2023},
booktitle = {Proc. ACM IMC},
address = {Montreal QC, Canada},
}

@inproceedings{man_dns_2021,
author = {Man, Keyu and Zhou, Xin'an and Qian, Zhiyun},
title = {{DNS Cache Poisoning Attack: Resurrections with Side Channels}},
year = {2021},
booktitle = {Proc. ACM CCS},
address = {Virtual Event},
}

@inproceedings{moura_clouding_2020,
author = {Moura, Giovane C. M. and Castro, Sebastian and Hardaker, Wes and Wullink, Maarten and Hesselman, Cristian},
title = {{Clouding up the Internet: how centralized is DNS traffic becoming?}},
year = {2020},
booktitle = {Proc. ACM IMC},
address = {Virtual Event},
}

@inproceedings{moura_tsuname_2021,
author = {Moura, Giovane C. M. and Castro, Sebastian and Heidemann, John and Hardaker, Wes},
title = {{TsuNAME: exploiting misconfiguration and vulnerability to DDoS DNS}},
year = {2021},
booktitle = {Proc. ACM IMC},
location = {Virtual Event},
}

@inproceedings{moura_characterizing_2024,
author = {Moura, Giovane C. M. and Daniels, Thomas and Bosteels, Maarten and Castro, Sebastian and M\"{u}ller, Moritz and Wabeke, Thymen and van den Hout, Thijs and Korczy\'{n}ski, Maciej and Smaragdakis, Georgios},
title = {{Characterizing and Mitigating Phishing Attacks at ccTLD Scale}},
year = {2024},
booktitle = {Proc. ACM CCS},
}

@inproceedings{nawrocki_far_2021,
author = {Nawrocki, Marcin and Jonker, Mattijs and Schmidt, Thomas C. and W\"{a}hlisch, Matthias},
title = {{The far side of DNS amplification: tracing the DDoS attack ecosystem from the internet core}},
year = {2021},
booktitle = {Proc. ACM IMC},
location = {Virtual Event},
}

@inproceedings {nisenoff_user_2023,
author = {Alexandra Nisenoff and Ranya Sharma and Nick Feamster},
title = {{User Awareness and Behaviors Concerning Encrypted {DNS} Settings in Web Browsers}},
booktitle = {Proc. USENIX Security},
year = {2023},
address = {Anaheim, CA, USA},
month = aug
}

@inproceedings{osali_sibling_2025,
author = {Osali, Fariba and Sediqi, Khwaja Zubair and Gasser, Oliver},
title = {{Sibling Prefixes: Identifying Similarities in IPv4 and IPv6 Prefixes}},
year = {2025},
booktitle = {Proc. ACM IMC},
address = {USA},
}

@inproceedings{randall_trufflehunter_2020,
author = {Randall, Audrey and Liu, Enze and Akiwate, Gautam and Padmanabhan, Ramakrishna and Voelker, Geoffrey M. and Savage, Stefan and Schulman, Aaron},
title = {{Trufflehunter: Cache Snooping Rare Domains at Large Public DNS Resolvers}},
year = {2020},
booktitle = {Proc. ACM IMC},
location = {Virtual Event},
}

@inproceedings{randall_home_2021,
author = {Randall, Audrey and Liu, Enze and Padmanabhan, Ramakrishna and Akiwate, Gautam and Voelker, Geoffrey M. and Savage, Stefan and Schulman, Aaron},
title = {{Home is where the hijacking is: understanding DNS interception by residential routers}},
year = {2021},
booktitle = {Proc. ACM IMC},
address = {Virtual Event},
}

@inproceedings{sommese_darkdns_2024,
author = {Sommese, Raffaele and Akiwate, Gautam and Affinito, Antonia and Muller, Moritz and Jonker, Mattijs and claffy, kc},
title = {{DarkDNS: Revisiting the Value of Rapid Zone Update}},
year = {2024},
booktitle = {Proc. ACM on IMC},
address = {Madrid, Spain},
}

@inproceedings{sommese_investigating_2022,
author = {Sommese, Raffaele and Claffy, KC and van Rijswijk-Deij, Roland and Chattopadhyay, Arnab and Dainotti, Alberto and Sperotto, Anna and Jonker, Mattijs},
title = {{Investigating the impact of DDoS attacks on DNS infrastructure}},
year = {2022},
booktitle = {Proc. ACM IMC},
address = {Nice, France},
}

@INPROCEEDINGS{sommese_assessing_2022,
author={Sommese, Raffaele and Jonker, Mattijs and van der Ham, Jeroen and Moura, Giovane C. M.},
booktitle={Proc. International Conference on Network and Service Management}, 
title={{Assessing e-Government DNS Resilience}}, 
year={2022},
}

@inproceedings{steurer_measuring_2025,
author = {Steurer, Florian and Phokeer, Amreesh and Paeps, Philip and Izhikevich, Liz},
title = {{Measuring Resilience of Authoritative DNS}},
year = {2025},
booktitle = {Proc. ACM SIGCOMM Posters and Demos},
address = {Coimbra, Portugal},
}

@inproceedings{xing_yesterday_2024,
author = {Xing, Yunpeng and Lu, Chaoyi and Liu, Baojun and Duan, Haixin and Sun, Junzhe and Li, Zhou},
title = {{Yesterday Once More: Global Measurement of Internet Traffic Shadowing Behaviors}},
year = {2024},
booktitle = {Proc. ACM IMC},
address = {Madrid, Spain},
}

@inproceedings{xu_tsuking_2023,
author = {Xu, Wei and Li, Xiang and Lu, Chaoyi and Liu, Baojun and Duan, Haixin and Zhang, Jia and Chen, Jianjun and Wan, Tao},
title = {{TsuKing: Coordinating DNS Resolvers and Queries into Potent DoS Amplifiers}},
year = {2023},
booktitle = {Proc. ACM CCS},
address = {Copenhagen, Denmark},
}

@inproceedings{zhang_silence_2023,
author = {Zhang, Fenglu and Liu, Baojun and Alowaisheq, Eihal and Chen, Jianjun and Lu, Chaoyi and Song, Linjian and Ma, Yong and Liu, Ying and Duan, Haixin and Yang, Min},
title = {{Silence is not Golden: Disrupting the Load Balancing of Authoritative DNS Servers}},
year = {2023},
booktitle = {Proc. ACM CCS},
address = {Copenhagen, Denmark},
}

@inproceedings {zhang_rethinking_2024,
author = {Yunyi Zhang and Baojun Liu and Haixin Duan and Min Zhang and Xiang Li and Fan Shi and Chengxi Xu and Eihal Alowaisheq},
title = {{Rethinking the Security Threats of Stale {DNS} Glue Records}},
booktitle = {Proc. USENIX Security},
year = {2024},
address = {Philadelphia, PA, USA},
month = aug
}

@inproceedings{zhou_regional_2023,
author = {Zhou, Minyuan and Zhang, Xiao and Hao, Shuai and Yang, Xiaowei and Zheng, Jiaqi and Chen, Guihai and Dou, Wanchun},
title = {{Regional IP Anycast: Deployments, Performance, and Potentials}},
year = {2023},
booktitle = {Proc. ACM SIGCOMM},
address = {New York, NY, USA},
}

@article{COMNET2022,
  title   = {{A Deep Dive into {DNS} Behavior and Query Failures}},
  author  = {Yang, Donghui and Li, Zhenyu and Jiang, Haiyang and Tyson, Gareth and Li, Hongtao and Xie, Gaogang and Zeng, Yu},
  journal = {Computer Networks},
  volume  = {214},
  pages   = {109131},
  year    = {2022},
}

@article{englbrecht_towards_2020,
title = {{Towards a Capability Maturity Model for Digital Forensic Readiness}},
author = {Englbrecht, Ludwig and Meier, Stefan and Pernul, G{\"u}nther},
year = 2020,
month = oct,
journal = {Wireless Networks},
volume = {26},
number = {7},
address = {Berlin, Heidelberg}
}

@article{kumar_each_2023,
author = {Kumar, Rashna and Asif, Sana and Lee, Elise and Bustamante, Fabian E.},
title = {{Each at its Own Pace: Third-Party Dependency and Centralization Around the World}},
year = {2023},
volume = {7},
number = {1},
journal = {Proc. ACM Meas. Anal. Comput. Syst.},
month = mar,
}

@article{lyu_hierarchical_2021,
author={Lyu, Minzhao and Gharakheili, Hassan Habibi and Russell, Craig and Sivaraman, Vijay},
journal={IEEE TNSM}, 
title={{Hierarchical Anomaly-Based Detection of Distributed DNS Attacks on Enterprise Networks}}, 
year={2021},
volume={18},
number={1},
}

@article{lyu_enterprise_2023,
author={Lyu, Minzhao and Habibi Gharakheili, Hassan and Russell, Craig and Sivaraman, Vijay},
journal={IEEE TNSM}, 
title={{Enterprise DNS Asset Mapping and Cyber-Health Tracking via Passive Traffic Analysis}}, 
year={2023},
volume={20},
number={3},
}

@article{lyu_classifying_2022,
title = {{Classifying and tracking enterprise assets via dual-grained network behavioral analysis}},
journal = {Computer Networks},
volume = {218},
year = {2022},
author = {Minzhao Lyu and Hassan {Habibi Gharakheili} and Vijay Sivaraman},
}

@misc{anssi_french_2025,
author = {{ANSSI}},
title = {{French Cybersecurity Agency}},
year = 2025,
month = nov,
url = {https://cyber.gouv.fr/},
note = {Accessed: 2025-11-14}
}

@misc{agor_quarterly_2025,
author = {{AGOR}},
title = {{AGOR Quarterly Report}},
year = 2025,
month = nov,
url = {https://www.finance.gov.au/government/managing-commonwealth-resources/structure-australian-government-public-sector},
note = {Accessed: 2025-07-21}
}

@misc{acsc_annual_2025,
author = {{ACSC}},
title = {{Annual Cyber Threat Report 2024-2025}},
year = 2025,
month = oct,
url = {https://www.cyber.gov.au/about-us/view-all-content/reports-and-statistics/annual-cyber-threat-report-2024-2025},
note = {Accessed: 2025-11-11}
}

@misc{european_digital_2025,
author = {{European Commision}},
title = {{Digital Tracking}},
year = 2025,
month = jun,
url = {https://commission.europa.eu/strategy-and-policy/eu-budget/performance-and-reporting/horizontal-priorities/digital-tracking_en},
note = {Accessed: 2025-11-11}
}

@misc{congress_information_2024,
author = {{U.S. Congress}},
title = {{Information Technology Spending in the President’s Budget Submission for FY2025: In Brief}},
year = 2024,
month = apr,
url = {https://www.congress.gov/crs-product/R48049},
note = {Accessed: 2025-11-11}
}

@misc{ipinfo_ipinfo_2025,
author = {{IPinfo}},
title = {{IPinfo: The Trusted Source For IP Address Data}},
year = 2025,
month = nov,
url = {https://ipinfo.io},
note = {Accessed: 2025-11-14}
}

@misc{presiden_peraturan_2018,
author = {{Presiden Republik Indonesia}},
title = {{Peraturan Presiden Nomor 95 Tahun 2018 tentang Sistem Pemerintahan Berbasis Elektronik}},
year = 2018,
month = okt,
url = {https://peraturan.bpk.go.id/Details/96913/perpres-no-95-tahun-2018},
note = {Accessed: 2025-11-12}
}

@misc{setkab_kabinet_2924,
author = {{Sekretaris Kabinet Republik Indonesia}},
title = {{Kabinet Pemerintahan Indonesia}},
year = 2024,
month = okt,
url = {https://setkab.go.id/profil-kabinet/},
note = {Accessed: 2025-11-12}
}

@misc{rdap_registration_2025,
author = {{RDAP.org}},
title = {{The Registration Data Access Protocol}},
year = 2025,
month = nov,
url = {https://about.rdap.org},
note = {Accessed: 2025-11-01}
}

@misc{csis_significant_2025,
author = {{CSIS}},
title = {{Significant Cyber Incidents}},
year = 2025,
month = nov,
url = {https://www.csis.org/programs/strategic-technologies-program/significant-cyber-incidents},
note = {Accessed: 2025-11-11}
}

@misc{robinson_single_2025,
author = {{Robinson, Dan}},
title = {{A single DNS race condition brought Amazon's cloud empire to its knees}},
year = 2025,
month = oct,
url = {https://www.theregister.com/2025/10/23/amazon_outage_postmortem/},
note = {Accessed: 2025-11-11}
}

@misc{united_e-gov_2024,
author = {{United Nation}},
title = {{E-Government Development Index (EGDI)}},
year = 2024,
month = sep,
url = {https://publicadministration.un.org/egovkb/en-us/About/Overview/-E-Government-Development-Index},
note = {Accessed: 2025-08-22}
}

@misc{brazil_orgaos_2025,
author = {{Brazil Government}},
title = {{Órgãos do Governo}},
year = 2025,
month = aug,
url = {https://www.gov.br/pt-br/orgaos-do-governo/ministerios-e-orgaos-com-status-de-ministerios},
note = {Accessed: 2025-11-03}
}

@misc{datagouv_list_2022,
author = {{datagouv}},
title = {{Lists of gouv.fr sites}},
year = 2022,
month = feb,
url = {https://www.data.gouv.fr/datasets/listes-des-sites-gouv-fr/},
note = {Accessed: 2022-07-22}
}

@misc{pallarito_cloudflare_2022,
author = {{Pallarito, Ash and Abley, Joe}},
title = {{Cloudflare 1.1.1.1 incident on July 14, 2025}},
year = 2022,
month = jul,
url = {https://blog.cloudflare.com/cloudflare-1-1-1-1-incident-on-july-14-2025/},
note = {Accessed: 2025-11-11}
}

@misc{ukgov_list_2025,
author = {{Government Digital Service and Central Digital and Data Office}},
title = {{List of .gov.uk domain names}},
year = 2025,
month = jan,
url = {https://www.gov.uk/government/publications/list-of-gov-uk-domain-names},
note = {Accessed: 2025-07-22}
}

@misc{uksecretary_state_2025,
author = {{Secretary of State Science}},
title = {{State of digital government review}},
year = 2025,
month = jan,
url = {https://www.gov.uk/government/publications/state-of-digital-government-review/state-of-digital-government-review},
note = {Accessed: 2025-08-11}
}

@misc{cisa_getgov_2025,
author = {{Cybersecurity and Infrastructure Security Agency}},
title = {{get.gov}},
year = 2025,
url = {https://get.gov/about/data/},
note = {Accessed: 2025-04-20}
}

@misc{mitchell_what_2025,
author = {{Mitchell, Robert}},
title = {{What Is Internet Resilience?}},
year = 2025,
month = aug,
url = {https://www.internetsociety.org/blog/2025/08/what-is-internet-resilience/},
note = {Accessed: 2025-11-11}
}

@misc{moss_magnitude_2025,
author = {{Moss, Sebastian}},
title = {{8.8 magnitude earthquake causes Internet outages in Kamchatka, Russia}},
year = 2025,
month = jul,
url = {https://www.datacenterdynamics.com/en/news/88-magnitude-earthquake-causes-internet-outages-in-kamchatka-russia/},
note = {Accessed: 2025-10-30}
}

@misc{akamai_designing_2025,
author = {{Akamai}},
title = {{Designing DNS for Availability and Resilience Against DDoS Attacks}},
year = 2025,
month = nov,
url = {https://www.akamai.com/resources/white-paper/designing-dns-for-availability-and-resilience-against-ddos-attacks},
note = {Accessed: 2025-11-28}
}

@misc{cloudflare_improving_2020,
author = {{Cloudflare}},
title = {{Improving DNS security, performance, and reliability}},
year = 2020,
url = {https://www.cloudflare.com/en-au/improving-dns-security-performance-reliability/},
note = {Accessed: 2025-11-28}
}

@misc{examlabs_understanding_2025,
author = {{ExamLabs}},
title = {{Understanding the Role of DNS Authoritative Name Servers}},
year = 2025,
url = {https://www.exam-labs.com/blog/understanding-the-role-of-dns-authoritative-name-servers},
note = {Accessed: 2025-11-28}
}

@misc{evans_why_2022,
author = {{Evans, Julia}},
title = {{Why might you run your own DNS server?}},
year = 2022,
month = jan,
url = {https://jvns.ca/blog/2022/01/05/why-might-you-run-your-own-dns-server-/},
note = {Accessed: 2025-11-28}
}

@misc{mcleod_likert_2025,
author = {{McLeod, Saul}},
title = {{Likert Scale Questionnaire: Examples \& Analysis}},
year = 2025,
month = nov,
url = {https://www.simplypsychology.org/likert-scale.html},
note = {Accessed: 2025-12-01}
}

@misc{joshi_biggest_2025,
author = {{Joshi, Sagar}},
title = {{10 Biggest IT Outages in History: Who Pulled the Plug?}},
year = 2025,
month = jul,
url = {https://learn.g2.com/biggest-it-outages-in-history},
note = {Accessed: 2025-12-01}
}

@misc{google_cloud_2025,
author = {{Google}},
title = {{Cloud DNS}},
year = 2025,
month = dec,
url = {https://cloud.google.com/dns/},
note = {Accessed: 2025-12-01}
}

@misc{cloudflare_cloudflare_2025,
author = {{Cloudflare}},
title = {{Cloudflare DNS}},
year = 2025,
month = dec,
url = {https://www.cloudflare.com/application-services/products/dns/},
note = {Accessed: 2025-12-01}
}

@misc{akamai_edge_2025,
author = {{Akamai}},
title = {{Edge DNS}},
year = 2025,
month = dec,
url = {https://www.akamai.com/products/edge-dns},
note = {Accessed: 2025-12-01}
}

@misc{rfc2182,
series =    {Request for Comments},
number =    2182,
howpublished =  {RFC 2182},
publisher = {RFC Editor},
author =    {Michael A. Patton and Scott O. Bradner and Robert Elz and Randy Bush},
title =     {{Selection and Operation of Secondary DNS Servers}},
pagetotal = 11,
year =      1997,
month =     jul,
}

@misc{rfc4786,
series =    {Request for Comments},
number =    4786,
howpublished =  {RFC 4786},
publisher = {RFC Editor},
author =    {Kurt Erik Lindqvist and Joe Abley},
title =     {{Operation of Anycast Services}},
pagetotal = 24,
year =      2006,
month =     dec,
}

@misc{rfc5936,
series =    {Request for Comments},
number =    5936,
howpublished =  {RFC 5936},
publisher = {RFC Editor},
author =    {Edward P. Lewis and Alfred Hoenes},
title =     {{DNS Zone Transfer Protocol (AXFR)}},
pagetotal = 29,
year =      2010,
month =     jun,
}

@misc{rfc8624,
author =    {Paul Wouters and Ondřej Surý},
title =     {{Algorithm Implementation Requirements and Usage Guidance for DNSSEC}},
url =       {https://www.rfc-editor.org/info/rfc8624},
series =    {Request for Comments},
number =    8624,
howpublished =  {RFC 8624},
publisher = {RFC Editor},
pagetotal = 11,
year =      2019,
month =     jun,
}
	
	\appendix
	
	\section{Source Code and Data Availability}
	\label{appendix:dataAvailability}
	We make publicly available the source code of our data collection and assessment pipeline, along with the datasets collected and generated in this study. These include the authoritative DNS resilience dataset for government domains in the six studied countries (\S\ref{subsec:dataCollection}), local configurations and catalogs used in the assessment process, and the full set of assessment scores reported in \S\ref{sec:assessment}. All resources  are accessible via a permanent public link \cite{dataset} on our academic website under the ``Authoritative DNS Infrastructure'' category.
	
	\section{Data Collection and Resilience Assessment Pipeline}
	\label{appendix:dataCollection}
	
	\begin{figure}[h]
		\centering
		\includegraphics[width=0.99\linewidth]{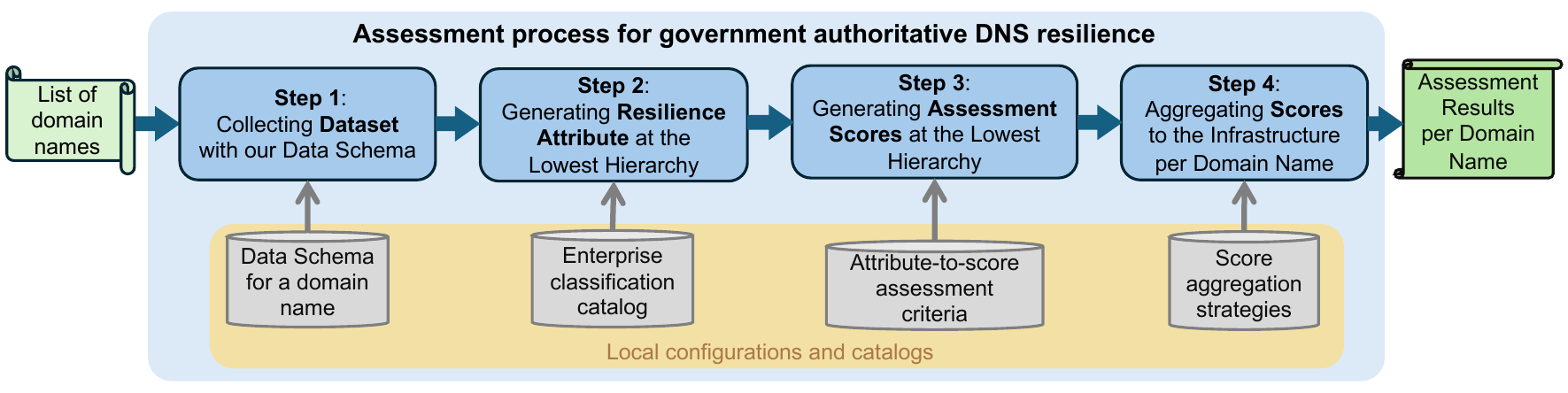}
		\vspace{-2mm}
		\caption{Overview of our data collection and assessment pipeline for evaluating the operational resilience of authoritative DNS infrastructures supporting government services.} 
		\label{fig:fig_assessment_process}
	\end{figure}
	
	As shown in Fig.~\ref{fig:fig_assessment_process}, our automated assessment pipeline comprises four stages: data collection using our data schema, generation of resilience attributes, assignment of scores on a five-point scale, and aggregation of scores to produce domain-level assessments. The pipeline accepts a CSV file listing domain names to evaluate and processes each domain name individually. 
	
	In the \textbf{first} step, the pipeline validates the existence of each domain name via DNS lookups using the Linux {\myverb{dig}} tool. A hierarchical JSON object is then instantiated to store all data fields defined in our schema (Fig.~\ref{fig:thedataschema}). 
	For each authoritative name server identified in the NS record, we query relevant DNS records: {\myverb{SOA}}, {\myverb{NS}}, {\myverb{A}}, {\myverb{AAAA}}, {\myverb{DNSSEC}}, and {\myverb{AXFR}}, from our measurement vantage points. By design, authoritative DNS records are retrieved from authoritative name servers as the only source of truth, and variation of measurement vantage points has negligible impact under normal circumstances in the absence of data integrity attacks (\eg hijacking). As an additional validation, we collect data from two vantage points: one with a public IP address in our lab and one located in the studied countries via VPN services. We also configure the DNS resolution using three options: the local ISP resolver and public resolvers operated by Google and Cloudflare. No inconsistencies are observed across these collection settings.
	To obtain registration information, the pipeline uses the Python IPWhois API to retrieve RDAP records (\myverb{rdap.org}) for every IP address associated with primary and authoritative name-server instances. The IP operational attributes (\eg Anycast status) are collected using the IPInfo Python API. Similarly, no inconsistency are observed from across vantage points, as both RDAP and IPInfo rely on centrally maintained databases.

	\begin{table}[t!]
		\tiny
		\caption{Resilience attributes at their finest hierarchical levels, together with the criteria and five-point scores used in our assessment.}
		\vspace{-2mm}
		\label{tab:resilience_framework}
		\begin{tabular}{|llcllc|}
			\hline
			\rowcolor[HTML]{C0C0C0} 
			\multicolumn{1}{|l|}{\cellcolor[HTML]{C0C0C0}\textbf{NS Function}} & \multicolumn{1}{l|}{\cellcolor[HTML]{C0C0C0}\textbf{Assessment aspect}} & \multicolumn{1}{l|}{\cellcolor[HTML]{C0C0C0}\textbf{Attribute}} & \multicolumn{1}{l|}{\cellcolor[HTML]{C0C0C0}\textbf{Hierarchy}} & \multicolumn{1}{l|}{\cellcolor[HTML]{C0C0C0}\textbf{Assessment criteria}} & \multicolumn{1}{l|}{\cellcolor[HTML]{C0C0C0}\textbf{Score}} \\ \hline
			\rowcolor[HTML]{EFEFEF} 
			\multicolumn{6}{|c|}{\cellcolor[HTML]{EFEFEF}\textbf{Infrastructure placement}} \\ \hline
			\multicolumn{1}{|l|}{} & \multicolumn{1}{l|}{} & \multicolumn{1}{c|}{} & \multicolumn{1}{l|}{\cellcolor[HTML]{F2DCDB}{\color[HTML]{C00000} }} & \multicolumn{1}{l|}{State-owned enterprise} & 5 \\ \cline{5-6} 
			\multicolumn{1}{|l|}{} & \multicolumn{1}{l|}{} & \multicolumn{1}{c|}{} & \multicolumn{1}{l|}{\cellcolor[HTML]{F2DCDB}{\color[HTML]{C00000} }} & \multicolumn{1}{l|}{Local private enterprise} & 4 \\ \cline{5-6} 
			\multicolumn{1}{|l|}{} & \multicolumn{1}{l|}{} & \multicolumn{1}{c|}{} & \multicolumn{1}{l|}{\cellcolor[HTML]{F2DCDB}{\color[HTML]{C00000} }} & \multicolumn{1}{l|}{Foreign large enterprise} & 3 \\ \cline{5-6} 
			\multicolumn{1}{|l|}{} & \multicolumn{1}{l|}{} & \multicolumn{1}{c|}{} & \multicolumn{1}{l|}{\cellcolor[HTML]{F2DCDB}{\color[HTML]{C00000} }} & \multicolumn{1}{l|}{Foreign SME enterprise} & 2 \\ \cline{5-6} 
			\multicolumn{1}{|l|}{} & \multicolumn{1}{l|}{\multirow{-5}{*}{Hosting enterprise type}} & \multicolumn{1}{c|}{\multirow{-5}{*}{\textbf{PP1}}} & \multicolumn{1}{l|}{\multirow{-5}{*}{\cellcolor[HTML]{F2DCDB}{\color[HTML]{C00000} \textit{\textbf{Instance}}}}} & \multicolumn{1}{l|}{Unregistered enterprise} & 1 \\ \cline{2-6} 
			\multicolumn{1}{|l|}{} & \multicolumn{1}{l|}{} & \multicolumn{1}{c|}{} & \multicolumn{1}{l|}{\cellcolor[HTML]{F2DCDB}{\color[HTML]{C00000} }} & \multicolumn{1}{l|}{Inside the country} & 5 \\ \cline{5-6} 
			\multicolumn{1}{|l|}{\multirow{-7}{*}{\textbf{Primary}}} & \multicolumn{1}{l|}{\multirow{-2}{*}{Hosting geolocation}} & \multicolumn{1}{c|}{\multirow{-2}{*}{\textbf{PP2}}} & \multicolumn{1}{l|}{\multirow{-2}{*}{\cellcolor[HTML]{F2DCDB}{\color[HTML]{C00000} \textit{\textbf{Instance}}}}} & \multicolumn{1}{l|}{Outside the country} & 1 \\ \hline
			\multicolumn{1}{|l|}{} & \multicolumn{1}{l|}{} & \multicolumn{1}{c|}{} & \multicolumn{1}{l|}{\cellcolor[HTML]{F2DCDB}{\color[HTML]{C00000} }} & \multicolumn{1}{l|}{State-owned enterprise} & 5 \\ \cline{5-6} 
			\multicolumn{1}{|l|}{} & \multicolumn{1}{l|}{} & \multicolumn{1}{c|}{} & \multicolumn{1}{l|}{\cellcolor[HTML]{F2DCDB}{\color[HTML]{C00000} }} & \multicolumn{1}{l|}{Local private enterprise} & 4 \\ \cline{5-6} 
			\multicolumn{1}{|l|}{} & \multicolumn{1}{l|}{} & \multicolumn{1}{c|}{} & \multicolumn{1}{l|}{\cellcolor[HTML]{F2DCDB}{\color[HTML]{C00000} }} & \multicolumn{1}{l|}{Foreign large enterprise} & 3 \\ \cline{5-6} 
			\multicolumn{1}{|l|}{} & \multicolumn{1}{l|}{} & \multicolumn{1}{c|}{} & \multicolumn{1}{l|}{\cellcolor[HTML]{F2DCDB}{\color[HTML]{C00000} }} & \multicolumn{1}{l|}{Foreign SME enterprise} & 2 \\ \cline{5-6} 
			\multicolumn{1}{|l|}{} & \multicolumn{1}{l|}{\multirow{-5}{*}{Hosting enterprise type}} & \multicolumn{1}{c|}{\multirow{-5}{*}{\textbf{PA1}}} & \multicolumn{1}{l|}{\multirow{-5}{*}{\cellcolor[HTML]{F2DCDB}{\color[HTML]{C00000} \textit{\textbf{Instance}}}}} & \multicolumn{1}{l|}{Unregistered enterprise} & 1 \\ \cline{2-6} 
			\multicolumn{1}{|l|}{} & \multicolumn{1}{l|}{} & \multicolumn{1}{c|}{} & \multicolumn{1}{l|}{\cellcolor[HTML]{F2DCDB}{\color[HTML]{C00000} }} & \multicolumn{1}{l|}{Inside the country} & 5 \\ \cline{5-6} 
			\multicolumn{1}{|l|}{\multirow{-7}{*}{\textbf{Authoritative}}} & \multicolumn{1}{l|}{\multirow{-2}{*}{Hosting geolocation}} & \multicolumn{1}{c|}{\multirow{-2}{*}{\textbf{PA2}}} & \multicolumn{1}{l|}{\multirow{-2}{*}{\cellcolor[HTML]{F2DCDB}{\color[HTML]{C00000} \textit{\textbf{Instance}}}}} & \multicolumn{1}{l|}{Outside the country} & 1 \\ \hline
			\rowcolor[HTML]{EFEFEF} 
			\multicolumn{6}{|c|}{\cellcolor[HTML]{EFEFEF}\textbf{Service configuration}} \\ \hline
			\multicolumn{1}{|l|}{} & \multicolumn{1}{l|}{} & \multicolumn{1}{c|}{} & \multicolumn{1}{l|}{\cellcolor[HTML]{F2DCDB}{\color[HTML]{C00000} }} & \multicolumn{1}{l|}{Anycast implemented} & 5 \\ \cline{5-6} 
			\multicolumn{1}{|l|}{} & \multicolumn{1}{l|}{\multirow{-2}{*}{Regional accessibility}} & \multicolumn{1}{c|}{\multirow{-2}{*}{\textbf{CP1}}} & \multicolumn{1}{l|}{\multirow{-2}{*}{\cellcolor[HTML]{F2DCDB}{\color[HTML]{C00000} \textit{\textbf{Instance}}}}} & \multicolumn{1}{l|}{Anycast not implemented} & 1 \\ \cline{2-6} 
			\multicolumn{1}{|l|}{} & \multicolumn{1}{l|}{} & \multicolumn{1}{c|}{} & \multicolumn{1}{l|}{\cellcolor[HTML]{F2DCDB}{\color[HTML]{C00000} }} & \multicolumn{1}{l|}{Instance is listed in authoritative NS} & 5 \\ \cline{5-6} 
			\multicolumn{1}{|l|}{} & \multicolumn{1}{l|}{\multirow{-2}{*}{Public exposure}} & \multicolumn{1}{c|}{\multirow{-2}{*}{\textbf{CP2}}} & \multicolumn{1}{l|}{\multirow{-2}{*}{\cellcolor[HTML]{F2DCDB}{\color[HTML]{C00000} \textit{\textbf{Instance}}}}} & \multicolumn{1}{l|}{IP is not listed in authoritative NS} & 1 \\ \cline{2-6} 
			\multicolumn{1}{|l|}{} & \multicolumn{1}{l|}{} & \multicolumn{1}{c|}{} & \multicolumn{1}{l|}{\cellcolor[HTML]{FDEADA}{\color[HTML]{984807} }} & \multicolumn{1}{l|}{More than 4 IPs used for the name server} & 5 \\ \cline{5-6} 
			\multicolumn{1}{|l|}{} & \multicolumn{1}{l|}{} & \multicolumn{1}{c|}{} & \multicolumn{1}{l|}{\cellcolor[HTML]{FDEADA}{\color[HTML]{984807} }} & \multicolumn{1}{l|}{2, 3, or 4 IPs used for the name server} & 4 \\ \cline{5-6} 
			\multicolumn{1}{|l|}{} & \multicolumn{1}{l|}{\multirow{-3}{*}{Instance redundancy}} & \multicolumn{1}{c|}{\multirow{-3}{*}{\textbf{CP3}}} & \multicolumn{1}{l|}{\multirow{-3}{*}{\cellcolor[HTML]{FDEADA}{\color[HTML]{984807} \textit{\textbf{Name server}}}}} & \multicolumn{1}{l|}{1 IP used for the name server} & 1 \\ \cline{2-6} 
			\multicolumn{1}{|l|}{} & \multicolumn{1}{l|}{} & \multicolumn{1}{c|}{} & \multicolumn{1}{l|}{\cellcolor[HTML]{FDEADA}{\color[HTML]{984807} }} & \multicolumn{1}{l|}{More than 4 subnets used for the name server} & 5 \\ \cline{5-6} 
			\multicolumn{1}{|l|}{} & \multicolumn{1}{l|}{} & \multicolumn{1}{c|}{} & \multicolumn{1}{l|}{\cellcolor[HTML]{FDEADA}{\color[HTML]{984807} }} & \multicolumn{1}{l|}{2, 3, or 4 subnets used for the name server} & 4 \\ \cline{5-6} 
			\multicolumn{1}{|l|}{} & \multicolumn{1}{l|}{\multirow{-3}{*}{Subnet redundancy - name server}} & \multicolumn{1}{c|}{\multirow{-3}{*}{\textbf{CP4}}} & \multicolumn{1}{l|}{\multirow{-3}{*}{\cellcolor[HTML]{FDEADA}{\color[HTML]{984807} \textit{\textbf{Name server}}}}} & \multicolumn{1}{l|}{1 subnet used for the name server} & 1 \\ \cline{2-6} 
			\multicolumn{1}{|l|}{} & \multicolumn{1}{l|}{} & \multicolumn{1}{c|}{} & \multicolumn{1}{l|}{\cellcolor[HTML]{FDEADA}{\color[HTML]{984807} }} & \multicolumn{1}{l|}{More than 4 ASes used for the name server} & 5 \\ \cline{5-6} 
			\multicolumn{1}{|l|}{} & \multicolumn{1}{l|}{} & \multicolumn{1}{c|}{} & \multicolumn{1}{l|}{\cellcolor[HTML]{FDEADA}{\color[HTML]{984807} }} & \multicolumn{1}{l|}{2, 3, or 4 ASes used for the name server} & 4 \\ \cline{5-6} 
			\multicolumn{1}{|l|}{\multirow{-13}{*}{\textbf{Primary}}} & \multicolumn{1}{l|}{\multirow{-3}{*}{AS redundancy - name server}} & \multicolumn{1}{c|}{\multirow{-3}{*}{\textbf{CP5}}} & \multicolumn{1}{l|}{\multirow{-3}{*}{\cellcolor[HTML]{FDEADA}{\color[HTML]{984807} \textit{\textbf{Name server}}}}} & \multicolumn{1}{l|}{1 AS used for the name server} & 1 \\ \hline
			\multicolumn{1}{|l|}{} & \multicolumn{1}{l|}{} & \multicolumn{1}{c|}{} & \multicolumn{1}{l|}{\cellcolor[HTML]{F2DCDB}{\color[HTML]{C00000} }} & \multicolumn{1}{l|}{Anycast implemented} & 5 \\ \cline{5-6} 
			\multicolumn{1}{|l|}{} & \multicolumn{1}{l|}{\multirow{-2}{*}{Regional accessibility}} & \multicolumn{1}{c|}{\multirow{-2}{*}{\textbf{CA1}}} & \multicolumn{1}{l|}{\multirow{-2}{*}{\cellcolor[HTML]{F2DCDB}{\color[HTML]{C00000} \textit{\textbf{Instance}}}}} & \multicolumn{1}{l|}{Anycast not impemented} & 1 \\ \cline{2-6} 
			\multicolumn{1}{|l|}{} & \multicolumn{1}{l|}{} & \multicolumn{1}{c|}{} & \multicolumn{1}{l|}{\cellcolor[HTML]{FDEADA}{\color[HTML]{984807} }} & \multicolumn{1}{l|}{More than 4 IPs used for the name server} & 5 \\ \cline{5-6} 
			\multicolumn{1}{|l|}{} & \multicolumn{1}{l|}{} & \multicolumn{1}{c|}{} & \multicolumn{1}{l|}{\cellcolor[HTML]{FDEADA}{\color[HTML]{984807} }} & \multicolumn{1}{l|}{2, 3, or 4 IPs used for the name server} & 4 \\ \cline{5-6} 
			\multicolumn{1}{|l|}{} & \multicolumn{1}{l|}{\multirow{-3}{*}{Instance redundancy}} & \multicolumn{1}{c|}{\multirow{-3}{*}{\textbf{CA2}}} & \multicolumn{1}{l|}{\multirow{-3}{*}{\cellcolor[HTML]{FDEADA}{\color[HTML]{984807} \textit{\textbf{Name server}}}}} & \multicolumn{1}{l|}{1 IP used for the name server} & 1 \\ \cline{2-6} 
			\multicolumn{1}{|l|}{} & \multicolumn{1}{l|}{} & \multicolumn{1}{c|}{} & \multicolumn{1}{l|}{\cellcolor[HTML]{FDEADA}{\color[HTML]{984807} }} & \multicolumn{1}{l|}{More than 4 subnets used for the name server} & 5 \\ \cline{5-6} 
			\multicolumn{1}{|l|}{} & \multicolumn{1}{l|}{} & \multicolumn{1}{c|}{} & \multicolumn{1}{l|}{\cellcolor[HTML]{FDEADA}{\color[HTML]{984807} }} & \multicolumn{1}{l|}{2, 3, or 4 subnets used for the name server} & 4 \\ \cline{5-6} 
			\multicolumn{1}{|l|}{} & \multicolumn{1}{l|}{\multirow{-3}{*}{Subnet redundancy - name server}} & \multicolumn{1}{c|}{\multirow{-3}{*}{\textbf{CA3}}} & \multicolumn{1}{l|}{\multirow{-3}{*}{\cellcolor[HTML]{FDEADA}{\color[HTML]{984807} \textit{\textbf{Name server}}}}} & \multicolumn{1}{l|}{1 subnet used for the name server} & 1 \\ \cline{2-6} 
			\multicolumn{1}{|l|}{} & \multicolumn{1}{l|}{} & \multicolumn{1}{c|}{} & \multicolumn{1}{l|}{\cellcolor[HTML]{FDEADA}{\color[HTML]{984807} }} & \multicolumn{1}{l|}{More than 4 ASes used for the name server} & 5 \\ \cline{5-6} 
			\multicolumn{1}{|l|}{} & \multicolumn{1}{l|}{} & \multicolumn{1}{c|}{} & \multicolumn{1}{l|}{\cellcolor[HTML]{FDEADA}{\color[HTML]{984807} }} & \multicolumn{1}{l|}{2, 3, or 4 ASes used for the name server} & 4 \\ \cline{5-6} 
			\multicolumn{1}{|l|}{} & \multicolumn{1}{l|}{\multirow{-3}{*}{AS redundancy - name server}} & \multicolumn{1}{c|}{\multirow{-3}{*}{\textbf{CA4}}} & \multicolumn{1}{l|}{\multirow{-3}{*}{\cellcolor[HTML]{FDEADA}{\color[HTML]{984807} \textit{\textbf{Name server}}}}} & \multicolumn{1}{l|}{1 AS used for the name server} & 1 \\ \cline{2-6} 
			\multicolumn{1}{|l|}{} & \multicolumn{1}{l|}{} & \multicolumn{1}{c|}{} & \multicolumn{1}{l|}{\cellcolor[HTML]{E6E0EC}{\color[HTML]{604A7B} }} & \multicolumn{1}{l|}{More than 4 authoritative name servers used} & 5 \\ \cline{5-6} 
			\multicolumn{1}{|l|}{} & \multicolumn{1}{l|}{} & \multicolumn{1}{c|}{} & \multicolumn{1}{l|}{\cellcolor[HTML]{E6E0EC}{\color[HTML]{604A7B} }} & \multicolumn{1}{l|}{2, 3, or 4 name servers used} & 4 \\ \cline{5-6} 
			\multicolumn{1}{|l|}{} & \multicolumn{1}{l|}{\multirow{-3}{*}{Name server redundancy}} & \multicolumn{1}{c|}{\multirow{-3}{*}{\textbf{CA5}}} & \multicolumn{1}{l|}{\multirow{-3}{*}{\cellcolor[HTML]{E6E0EC}{\color[HTML]{604A7B} \textit{\textbf{NS functionality}}}}} & \multicolumn{1}{l|}{1 name server used} & 1 \\ \cline{2-6} 
			\multicolumn{1}{|l|}{} & \multicolumn{1}{l|}{} & \multicolumn{1}{c|}{} & \multicolumn{1}{l|}{\cellcolor[HTML]{E6E0EC}{\color[HTML]{604A7B} }} & \multicolumn{1}{l|}{More than 4 subnets used for the NS functionality} & 5 \\ \cline{5-6} 
			\multicolumn{1}{|l|}{} & \multicolumn{1}{l|}{} & \multicolumn{1}{c|}{} & \multicolumn{1}{l|}{\cellcolor[HTML]{E6E0EC}{\color[HTML]{604A7B} }} & \multicolumn{1}{l|}{2, 3, or 4 subnets used for the NS functionality} & 4 \\ \cline{5-6} 
			\multicolumn{1}{|l|}{} & \multicolumn{1}{l|}{\multirow{-3}{*}{Subnet redundancy - NS function}} & \multicolumn{1}{c|}{\multirow{-3}{*}{\textbf{CA6}}} & \multicolumn{1}{l|}{\multirow{-3}{*}{\cellcolor[HTML]{E6E0EC}{\color[HTML]{604A7B} \textit{\textbf{NS functionality}}}}} & \multicolumn{1}{l|}{1 subnet used for the NS functionality} & 1 \\ \cline{2-6} 
			\multicolumn{1}{|l|}{} & \multicolumn{1}{l|}{} & \multicolumn{1}{c|}{} & \multicolumn{1}{l|}{\cellcolor[HTML]{E6E0EC}{\color[HTML]{604A7B} }} & \multicolumn{1}{l|}{More than 4 ASes used for the NS functionality} & 5 \\ \cline{5-6} 
			\multicolumn{1}{|l|}{} & \multicolumn{1}{l|}{} & \multicolumn{1}{c|}{} & \multicolumn{1}{l|}{\cellcolor[HTML]{E6E0EC}{\color[HTML]{604A7B} }} & \multicolumn{1}{l|}{2, 3, or 4 ASes used for the NS functionality} & 4 \\ \cline{5-6} 
			\multicolumn{1}{|l|}{\multirow{-20}{*}{\textbf{Authoritative}}} & \multicolumn{1}{l|}{\multirow{-3}{*}{AS redundancy - NS function}} & \multicolumn{1}{c|}{\multirow{-3}{*}{\textbf{CA7}}} & \multicolumn{1}{l|}{\multirow{-3}{*}{\cellcolor[HTML]{E6E0EC}{\color[HTML]{604A7B} \textit{\textbf{NS functionality}}}}} & \multicolumn{1}{l|}{1 AS used for the NS functionality} & 1 \\ \hline
			\rowcolor[HTML]{EFEFEF} 
			\multicolumn{6}{|c|}{\cellcolor[HTML]{EFEFEF}\textbf{DNS record dispatch}} \\ \hline
			\multicolumn{1}{|l|}{} & \multicolumn{1}{l|}{} & \multicolumn{1}{c|}{} & \multicolumn{1}{l|}{\cellcolor[HTML]{F2DCDB}{\color[HTML]{C00000} }} & \multicolumn{1}{l|}{Automatic AXFR enforcement} & 5 \\ \cline{5-6} 
			\multicolumn{1}{|l|}{} & \multicolumn{1}{l|}{} & \multicolumn{1}{c|}{} & \multicolumn{1}{l|}{\cellcolor[HTML]{F2DCDB}{\color[HTML]{C00000} }} & \multicolumn{1}{l|}{Manual AXFR enforcement} & 3 \\ \cline{5-6} 
			\multicolumn{1}{|l|}{\multirow{-3}{*}{\textbf{Primary}}} & \multicolumn{1}{l|}{\multirow{-3}{*}{Zone file delivery}} & \multicolumn{1}{c|}{\multirow{-3}{*}{\textbf{DP1}}} & \multicolumn{1}{l|}{\multirow{-3}{*}{\cellcolor[HTML]{F2DCDB}{\color[HTML]{C00000} \textit{\textbf{Instance}}}}} & \multicolumn{1}{l|}{AXFR not enforced} & 1 \\ \hline
			\multicolumn{1}{|l|}{} & \multicolumn{1}{l|}{} & \multicolumn{1}{c|}{} & \multicolumn{1}{l|}{\cellcolor[HTML]{F2DCDB}{\color[HTML]{C00000} }} & \multicolumn{1}{l|}{Recommended DNSSEC algorithm} & 5 \\ \cline{5-6} 
			\multicolumn{1}{|l|}{} & \multicolumn{1}{l|}{} & \multicolumn{1}{c|}{} & \multicolumn{1}{l|}{\cellcolor[HTML]{F2DCDB}{\color[HTML]{C00000} }} & \multicolumn{1}{l|}{Not recommended DNSSEC algorithm} & 3 \\ \cline{5-6} 
			\multicolumn{1}{|l|}{\multirow{-3}{*}{\textbf{Authoritative}}} & \multicolumn{1}{l|}{\multirow{-3}{*}{DNS response delivery}} & \multicolumn{1}{c|}{\multirow{-3}{*}{\textbf{DA1}}} & \multicolumn{1}{l|}{\multirow{-3}{*}{\cellcolor[HTML]{F2DCDB}{\color[HTML]{C00000} \textit{\textbf{Instance}}}}} & \multicolumn{1}{l|}{DNSSEC not implemented} & 1 \\ \hline
		\end{tabular}
	\end{table}
	
	In the \textbf{second} step, all collected fields are converted into the 18 resilience attributes defined in \S\ref{subsec:attribute}. This includes mapping hosting-enterprise names to their corresponding categories using a curated enterprise-classification catalog (stored in JSON), converting geolocations to country codes, enumerating string-valued fields, counting server instances and name servers, and identifying unique administrative subnets and ASes.
	In the \textbf{third} step, each attribute is evaluated according to its scoring criteria in Table~\ref{tab:resilience_framework}. These criteria are implemented as modular scoring functions configured through a YAML file, allowing assessors to modify the scoring logic without altering the core pipeline.
	Lastly, in the \textbf{fourth} and final step, the pipeline aggregates attribute scores from their lowest hierarchical level upward, following the strategies shown in Fig.~\ref{fig:attribute}. Depending on the attribute, aggregation uses direct mapping, the worst-score-dominant strategy, or the best-score-dominant strategy, as specified in a YAML configuration file. The resulting JSON output contains all attributes and their scores throughout the hierarchy for the assessed domain. These aggregated results for government domains in the six studied countries are analyzed in \S\ref{sec:insights}.

	\section{List of Authoritative DNS Resilience Attributes and Assessment Criteria}\label{assessmentcriteria}
	
	Table~\ref{tab:resilience_framework} summarizes all resilience attributes used in our assessment of authoritative DNS infrastructures for government domains. Each attribute captures a unique aspect of operational practice, categorized by the operational phase, DNS functionality, and assessment dimension. All attributes are defined at their finest applicable hierarchy level, as described in \S\ref{subsec:attribute}. 
	The last two columns of Table~\ref{tab:resilience_framework} present the assessment criteria and corresponding five-point scores assigned to each attribute, following the scoring methodology discussed in \S\ref{subsec:scores}.
	
	\section{Simulation Results of Our Score Aggregation Strategies}
	\label{Appendix:aggregationEvaluation}
	
	To evaluate whether the two score-aggregation strategies in Algorithm~\ref{algo:1} behave as intended, as  described in \S\ref{subsec:scoreAggregation}, we conducted simulations using extensive combinations of score arrays $\mathbf{s}_{a,h,n}$ comprising different counts of values from the set $\{1,\ldots,5\}$. The simulations cover realistic configurations (reflecting the small  numbers of instances and name servers commonly observed in our dataset) and theoretical upper bounds, with array lengths increasing up to 1000 elements to demonstrate  convergence behavior.

	\begin{figure}[h]
		\centering
		\begin{subfigure}[h]{0.19\linewidth}
			\centering
			\includegraphics[width=\linewidth]{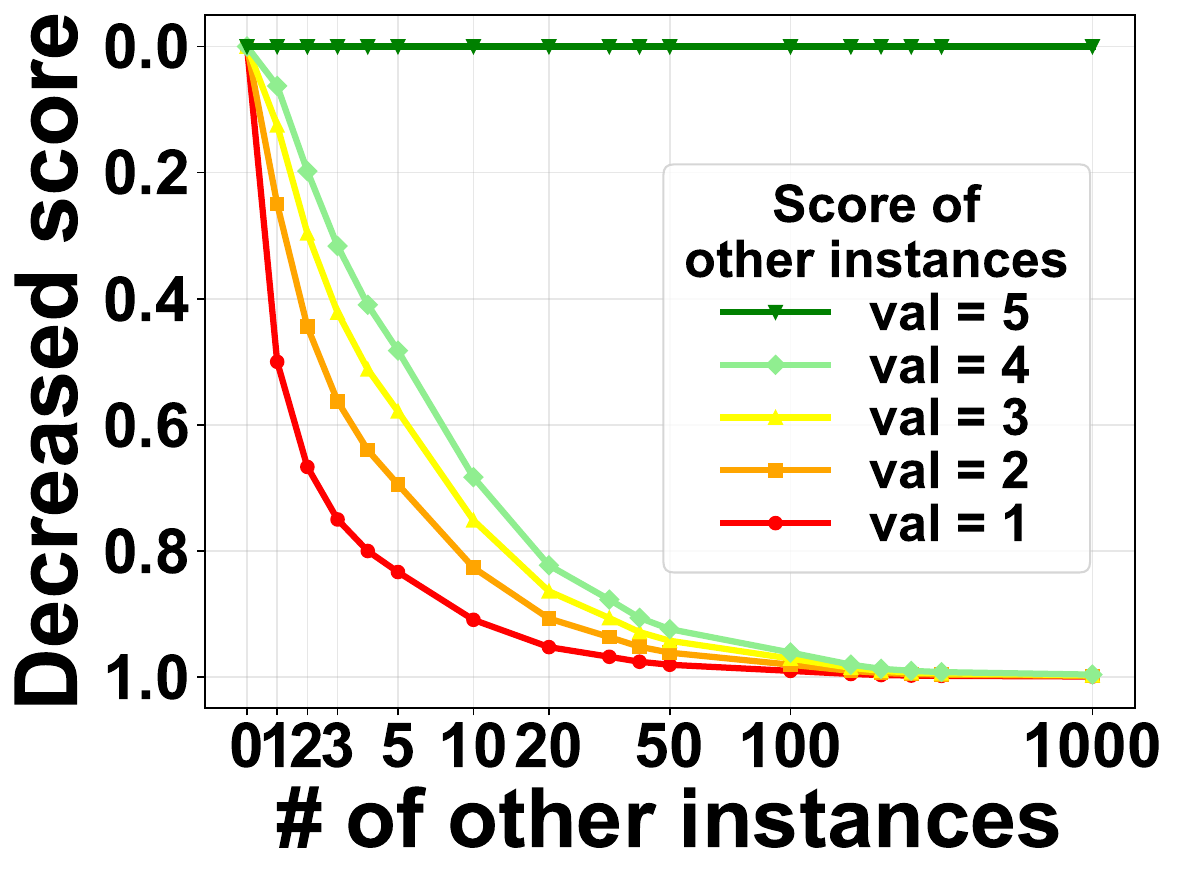}
			\caption{$s_{a,h+1,base}=5$.}
			\label{fig:aggregation_simulation_best_5}
		\end{subfigure}
		\hfill
		\begin{subfigure}[h]{0.19\linewidth}
			\centering
			\includegraphics[width=\linewidth]{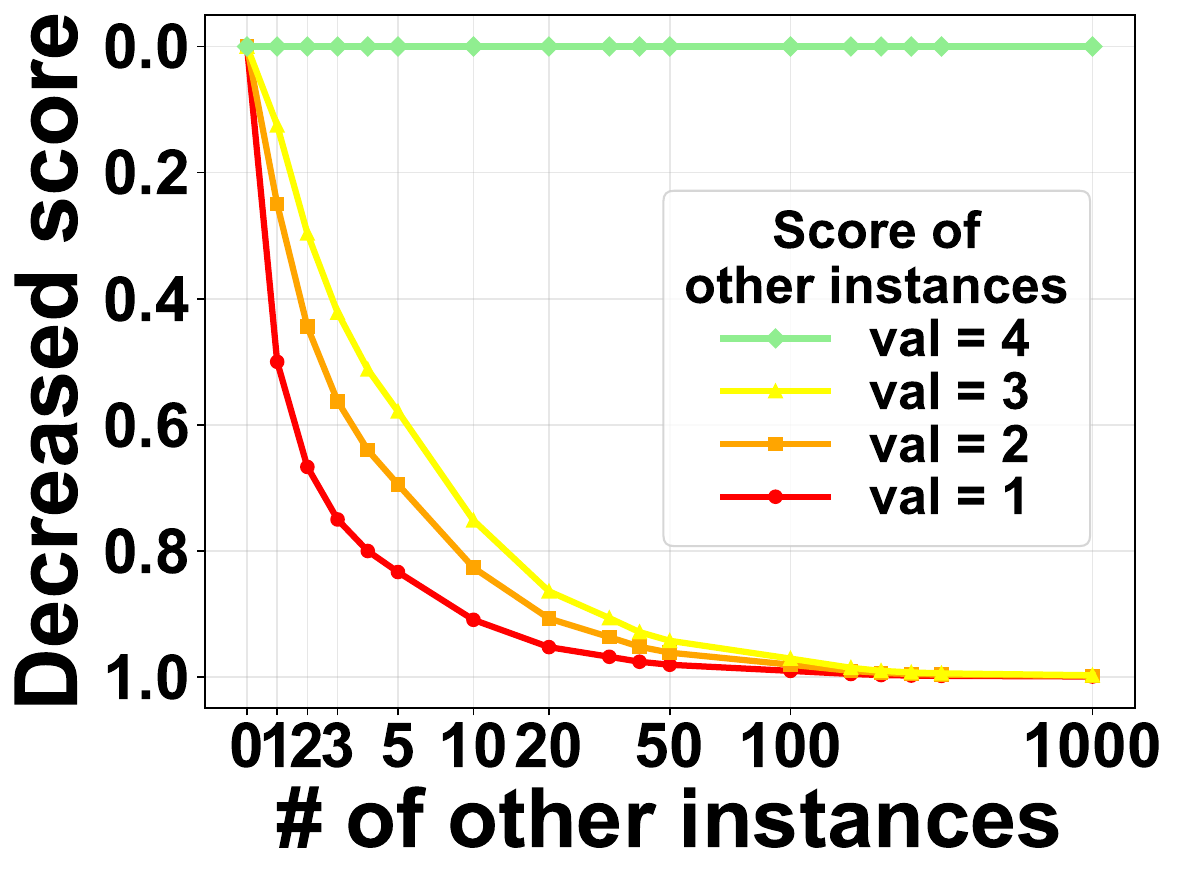}
			\caption{$s_{a,h+1,base}=4$.}
			\label{fig:aggregation_simulation_best_4}
		\end{subfigure}
		\hfill
		\begin{subfigure}[h]{0.19\linewidth}
			\centering
			\includegraphics[width=\linewidth]{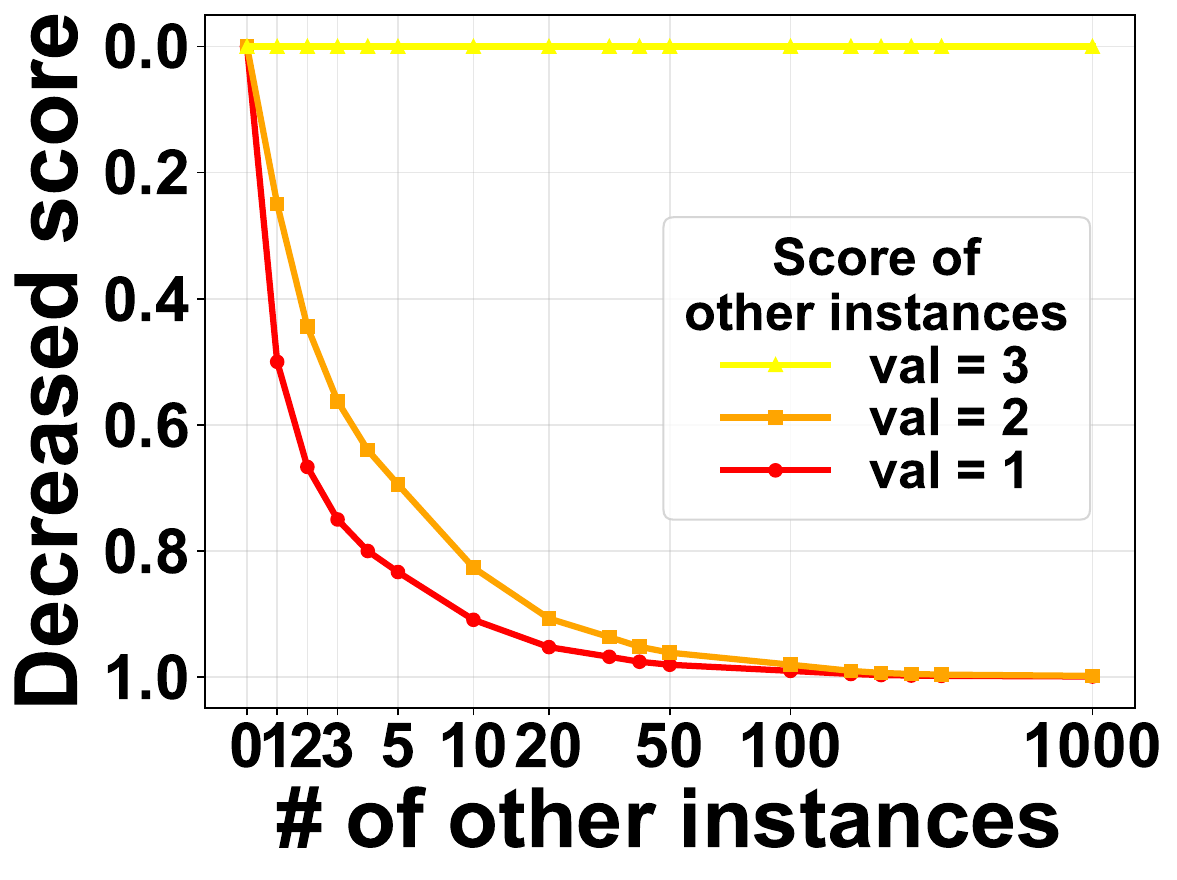}
			\caption{$s_{a,h+1,base}=3$.}
			\label{fig:aggregation_simulation_best_3}
		\end{subfigure}
		\hfill
		\begin{subfigure}[h]{0.19\linewidth}
			\centering
			\includegraphics[width=\linewidth]{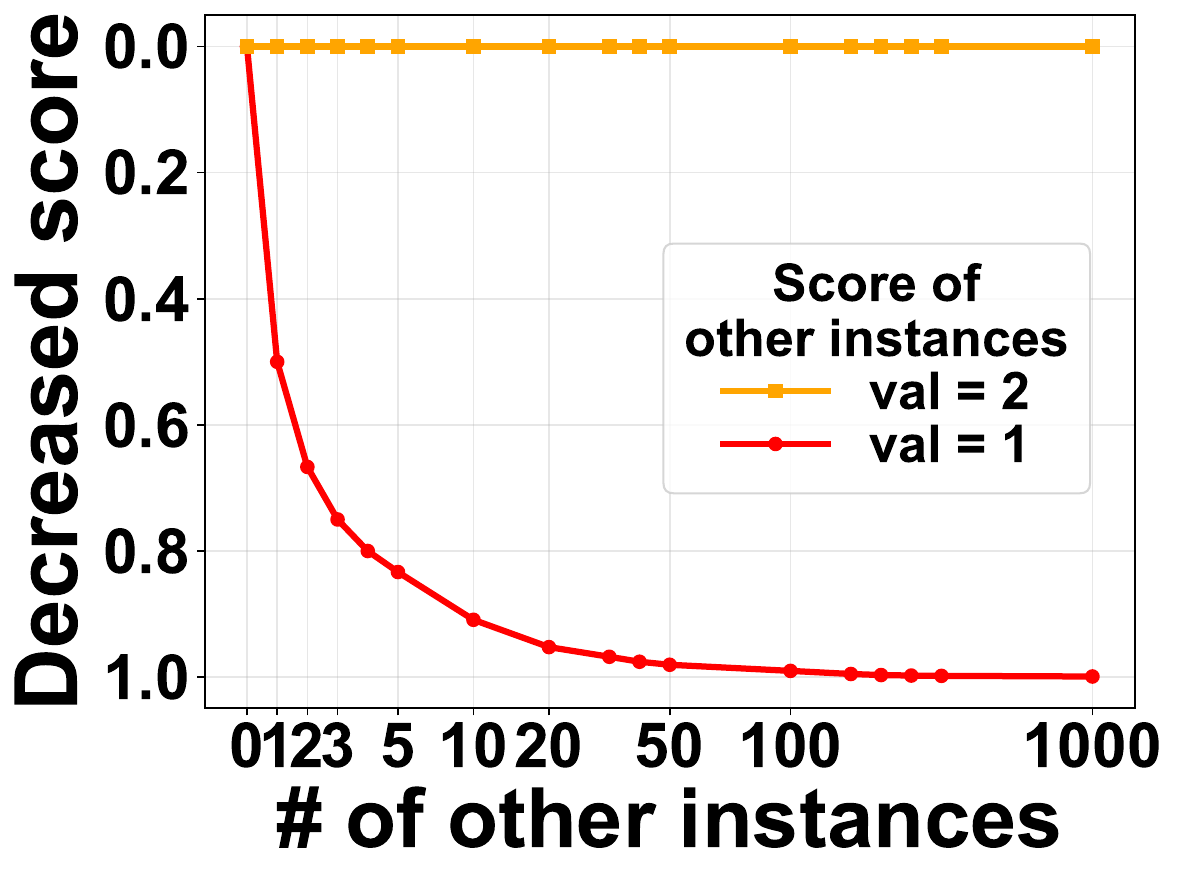}
			\caption{$s_{a,h+1,base}=2$.}
			\label{fig:aggregation_simulation_best_2}
		\end{subfigure}
		\hfill
		\begin{subfigure}[h]{0.19\linewidth}
			\centering
			\includegraphics[width=\linewidth]{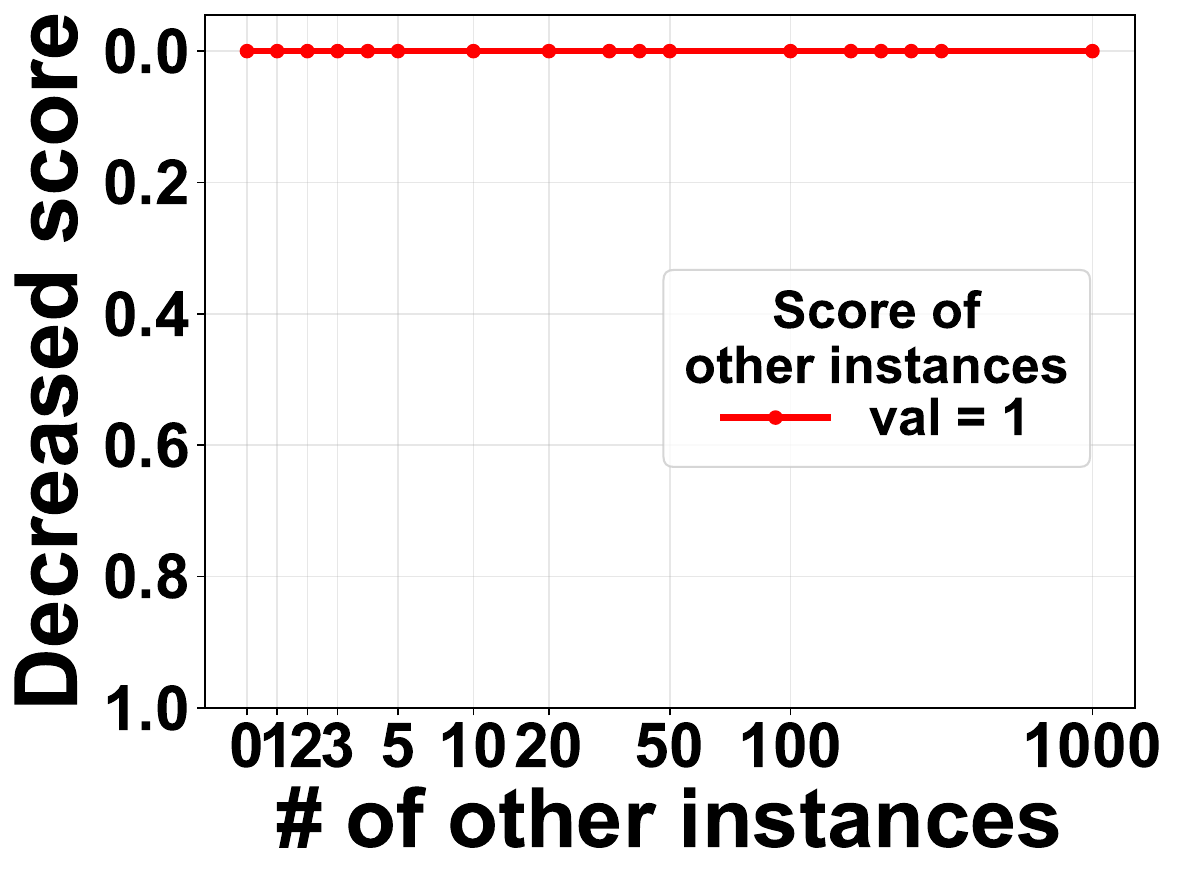}
			\caption{$s_{a,h+1,base}=1$.}
			\label{fig:aggregation_simulation_best_1}
		\end{subfigure}
		\vspace{-2mm}
		\caption{Simulation of the \textbf{best}-score-dominant strategy in Algorithm~\ref{algo:1}, showing aggregated outcomes for $s_{a,h+1,base}$ set to: (a) 5, (b) 4, (c) 3, (d) 2, and (e) 1. Each subfigure shows boundary cases with one best-scoring instance and all remaining instances sharing the same lower score.}
		\label{fig:aggregation_simulation_best}
	\end{figure}

	\begin{figure}[h]
		\centering
		\begin{subfigure}[h]{0.19\linewidth}
			\centering
			\includegraphics[width=\linewidth]{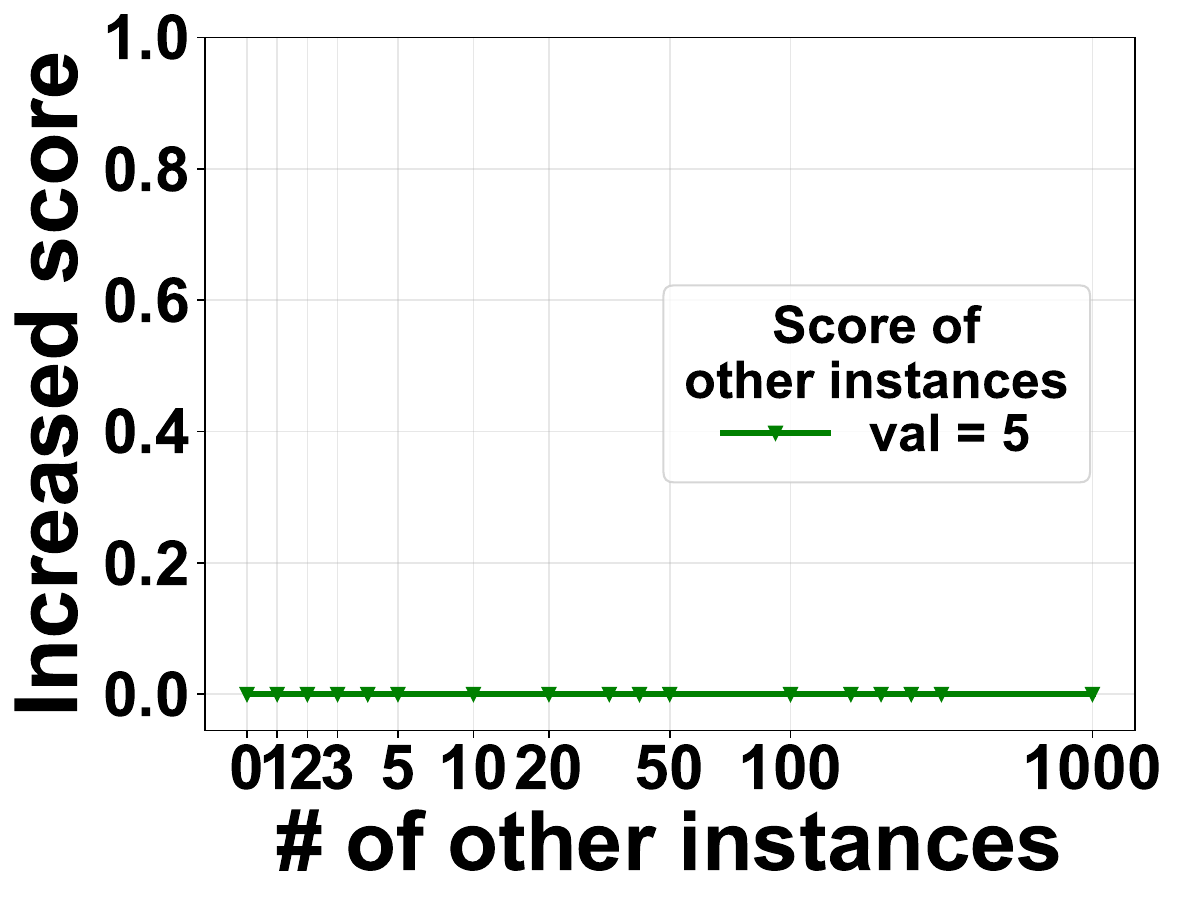}
			\caption{$s_{a,h+1,base}=5$.}
			\label{fig:aggregation_simulation_worst_5}
		\end{subfigure}
		\hfill
		\begin{subfigure}[h]{0.19\linewidth}
			\centering
			\includegraphics[width=\linewidth]{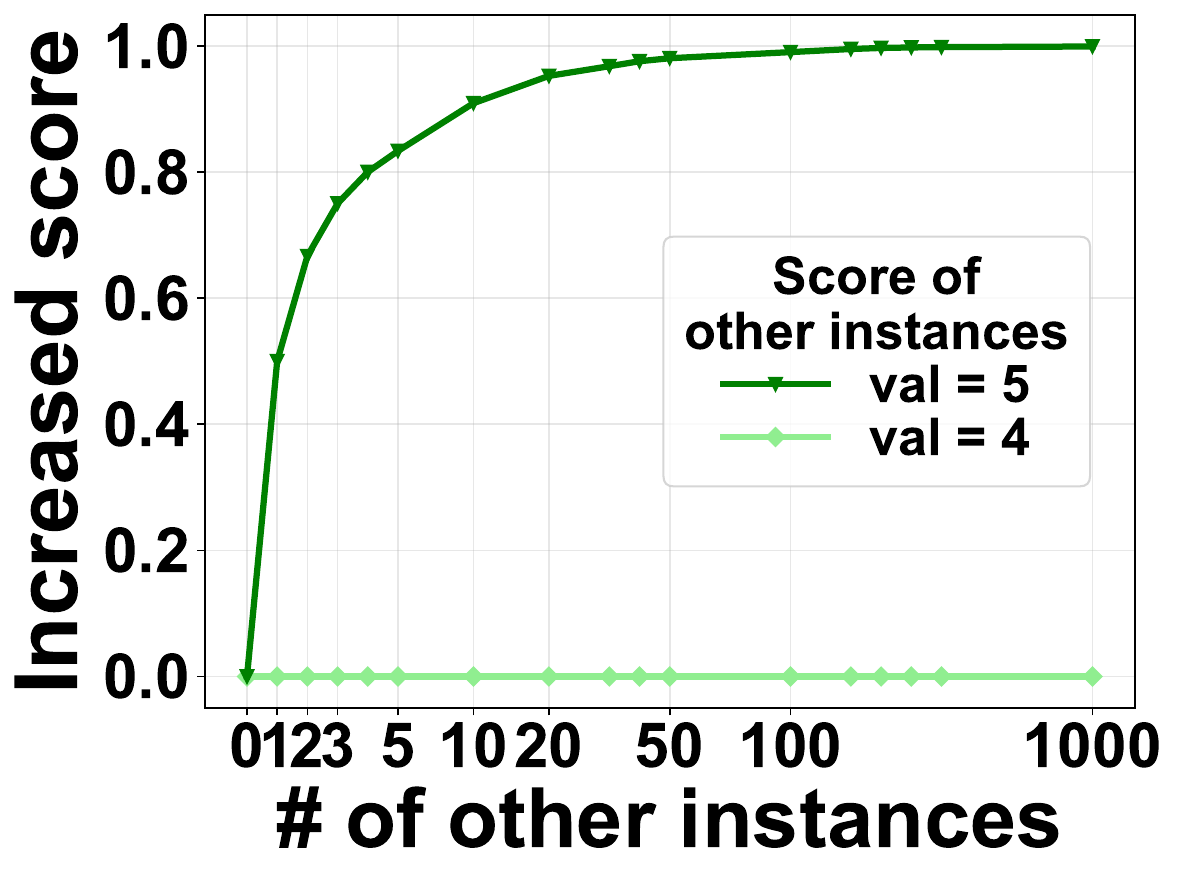}
			\caption{$s_{a,h+1,base}=4$.}
			\label{fig:aggregation_simulation_worst_4}
		\end{subfigure}
		\hfill
		\begin{subfigure}[h]{0.19\linewidth}
			\centering
			\includegraphics[width=\linewidth]{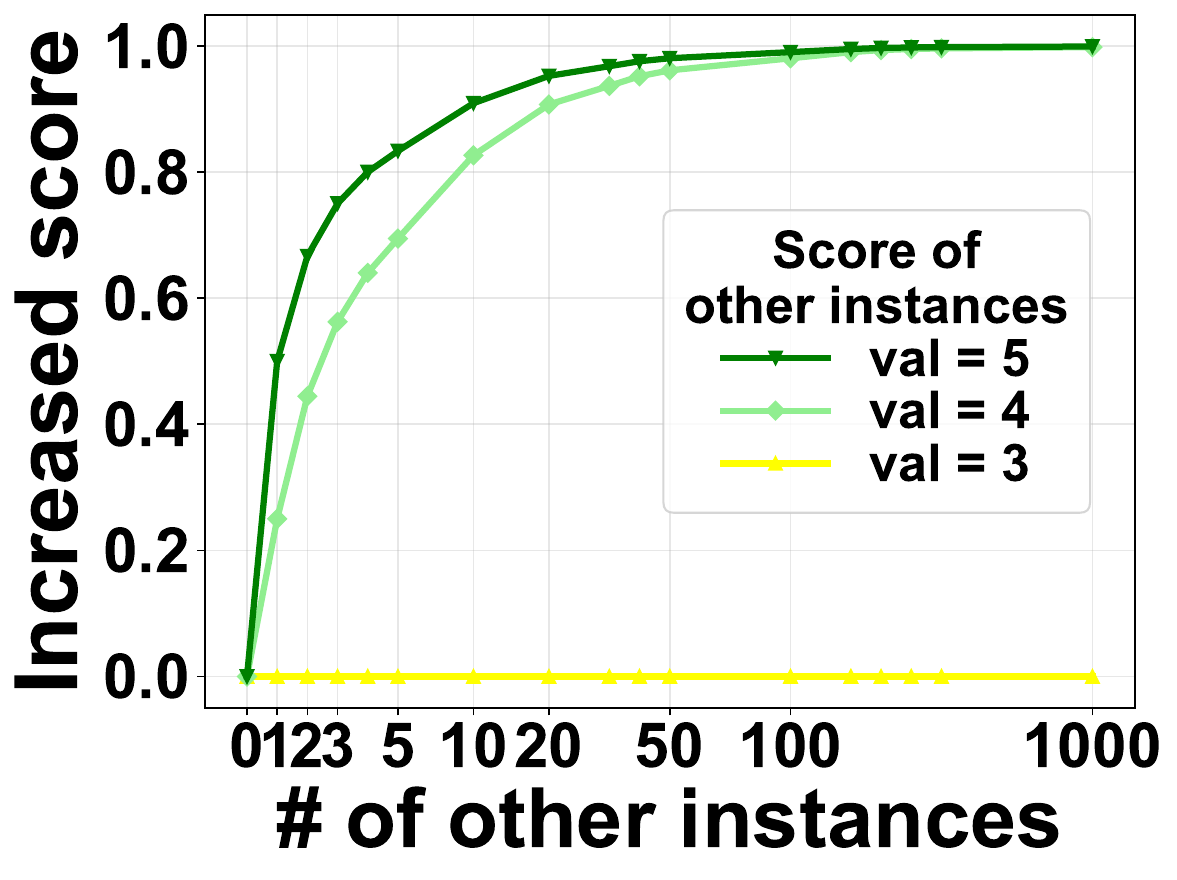}
			\caption{$s_{a,h+1,base}=3$.}
			\label{fig:aggregation_simulation_worst_3}
		\end{subfigure}
		\hfill
		\begin{subfigure}[h]{0.19\linewidth}
			\centering
			\includegraphics[width=\linewidth]{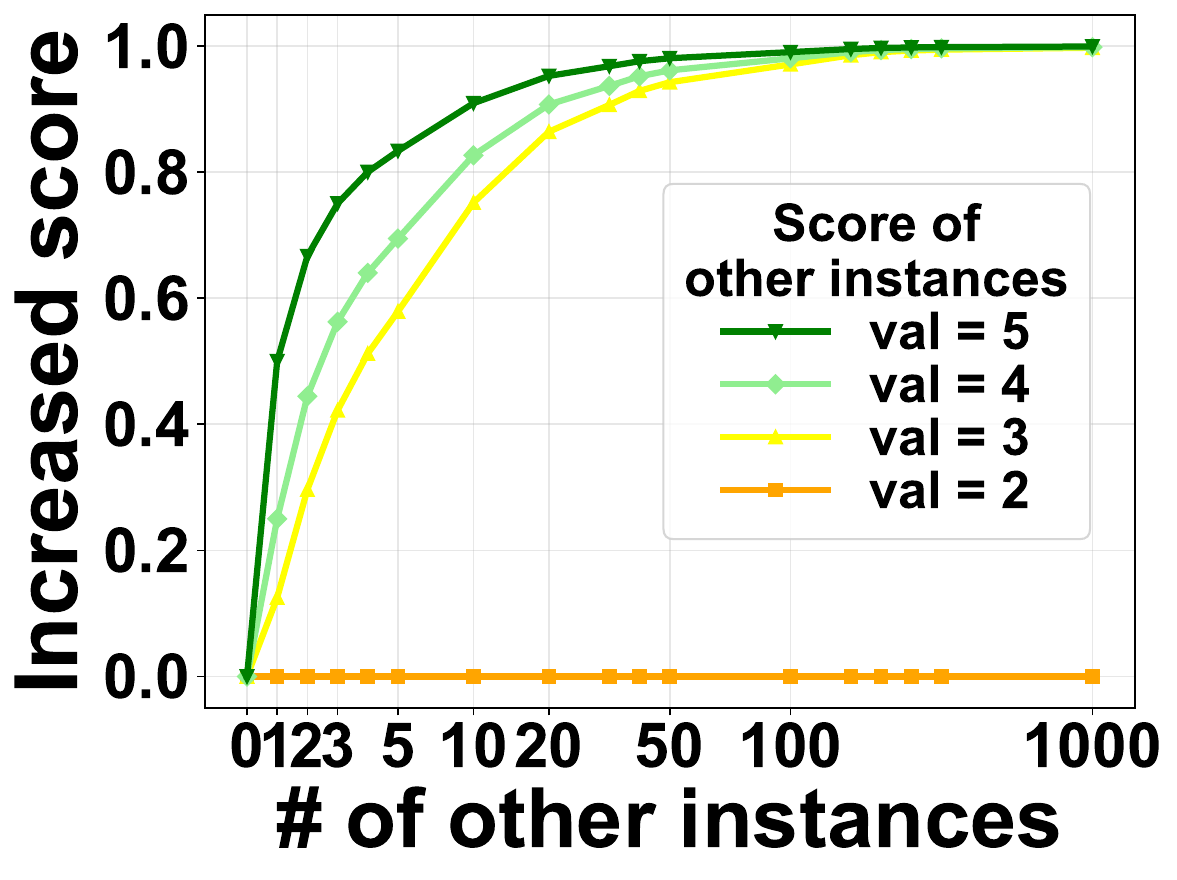}
			\caption{$s_{a,h+1,base}=2$.}
			\label{fig:aggregation_simulation_worst_2}
		\end{subfigure}
		\hfill
		\begin{subfigure}[h]{0.19\linewidth}
			\centering
			\includegraphics[width=\linewidth]{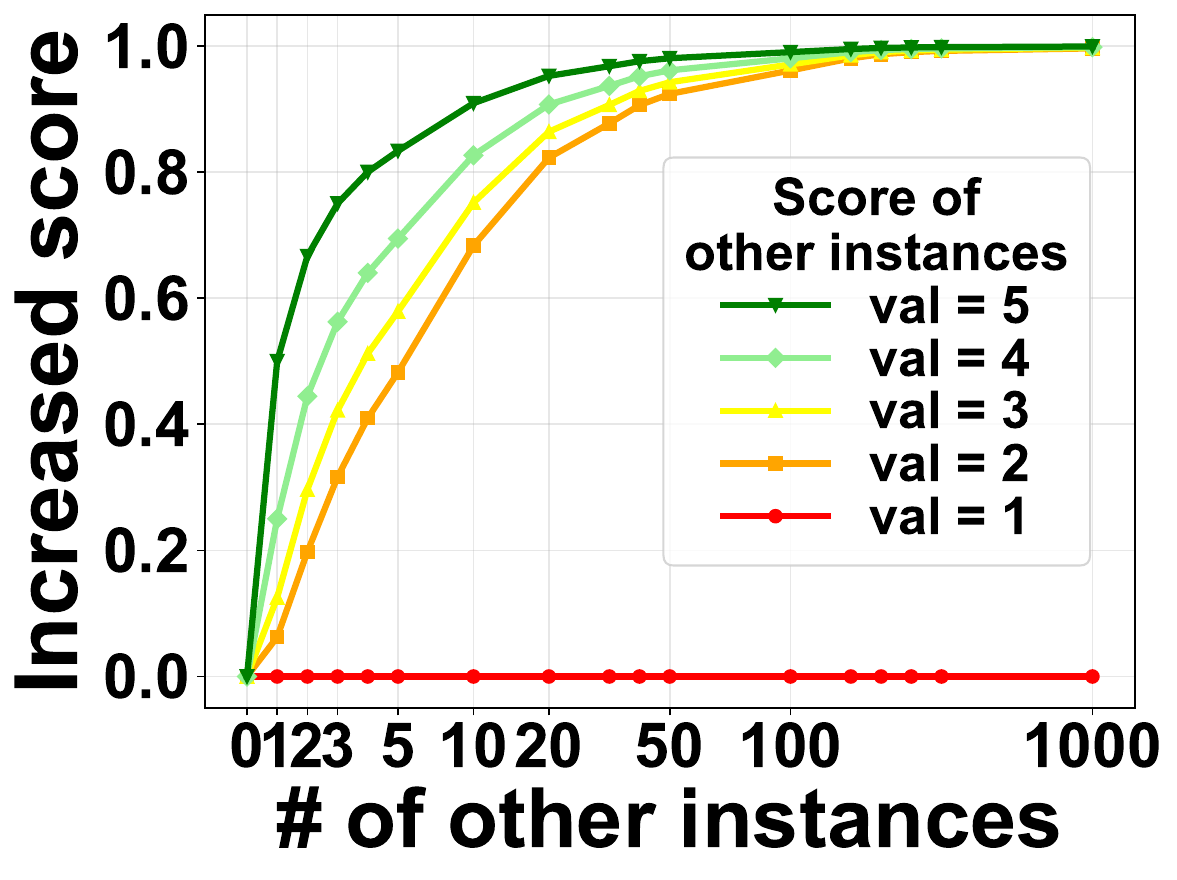}
			\caption{$s_{a,h+1,base}=1$.}
			\label{fig:aggregation_simulation_worst_1}
		\end{subfigure}
		\vspace{-2mm}
		\caption{Simulation of the \textbf{worst}-score-dominant strategy in Algorithm~\ref{algo:1}, showing aggregated outcomes for  $s_{a,h+1,base}$ set to: (a) 5, (b) 4, (c) 3, (d) 2, and (e) 1. Each subfigure shows boundary cases with one worst-scoring instance and all remaining instances sharing the same higher score.}
		\label{fig:aggregation_simulation_worst}
	\end{figure}

	For the best-score-dominant aggregation strategy, Fig~\ref{fig:aggregation_simulation_best} presents the boundary cases in which the array contains exactly one score instance with the best score $\mathbf{s}_{a,h+1,base}$ while all other instances share the same lower score. 
	As established in our design principles (\S\ref{subsec:scoreAggregation}), any number of instances with scores worse than the best score should never reduce the aggregated value by more than a single step $\Delta D=1$. This property is confirmed across all five possible best values in Fig.~\ref{fig:aggregation_simulation_best_5}--\ref{fig:aggregation_simulation_best_1}, where boundary curves converge to a reduction of exactly one step. 
	The figures also show that (i) the penalty introduced by a worse score grows exponentially as the score decreases, as seen when comparing vertical drops across values within the same plot, and (ii) the penalty is evenly distributed across, as demonstrated by horizontal shifts observed when varying the lower-scoring instances along the same curve. Aggregated scores for all other score arrays lie between the boundary curves, consistent with the expected behavior of the strategy.  
	
	For the worst-score-dominant aggregation strategy, simulation results in Fig.~\ref{fig:aggregation_simulation_worst} similarly validate our design objectives. In this case, the aggregated value starts at the lowest score in the array and increases by at most one step $\Delta D=1$, with increments contributed by higher-scoring instances decreasing exponentially according to Algorithm~\ref{algo:1}. The convergence trends observed in the simulated configurations confirm that the strategy precisely captures the intended dominance of the weakest configuration in determining resilience.

	\section{Weight Sensitivity Analysis for Country-level Results}
	\label{Appendix:weightAnalysis}
	\begin{figure}[h]
		\centering
		\begin{subfigure}{0.24\linewidth}
			\includegraphics[width=\linewidth]{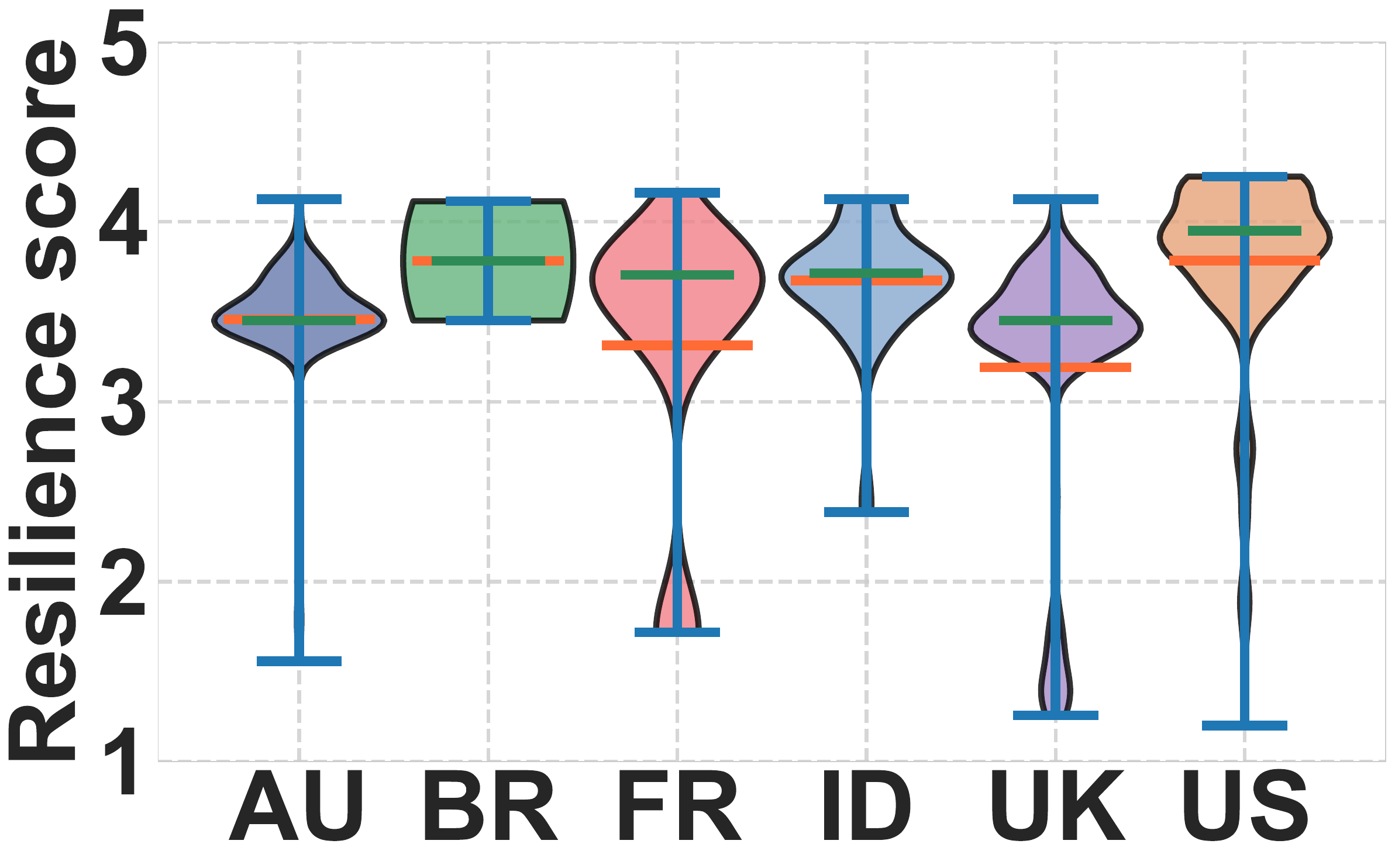}
			\caption{Placement prioritized}
			\label{fig:placement_prioritised}
		\end{subfigure}
		\begin{subfigure}{0.24\linewidth}
			\includegraphics[width=\linewidth]{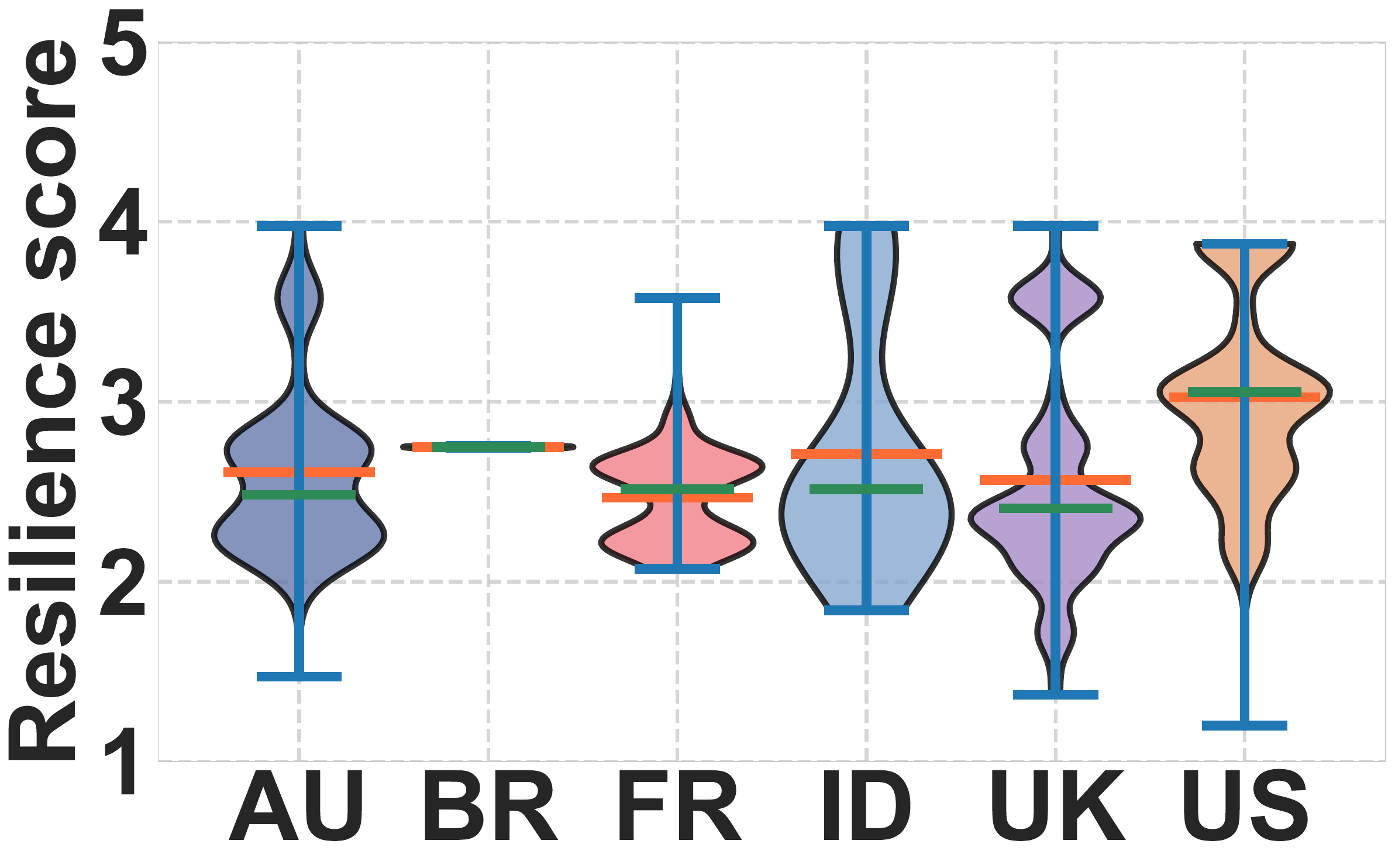}
			\caption{Config. prioritized}
			\label{fig:config_prioritised}
		\end{subfigure}
		\begin{subfigure}{0.24\linewidth}
			\includegraphics[width=\linewidth]{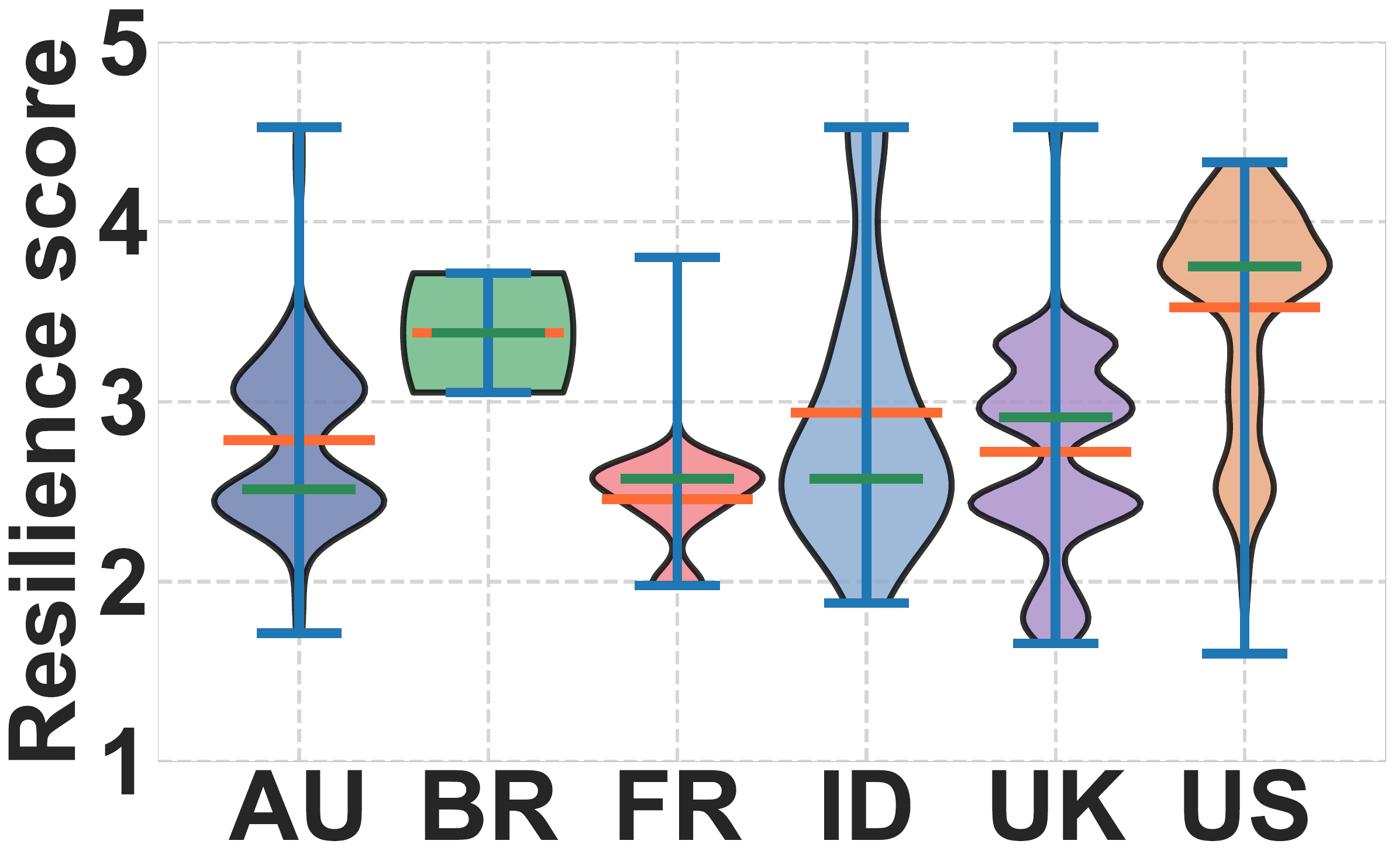}
			\caption{Dispatch prioritized}
			\label{fig:dispatch_prioritised}
		\end{subfigure}
		\begin{subfigure}{0.24\linewidth}
			\includegraphics[width=\linewidth]{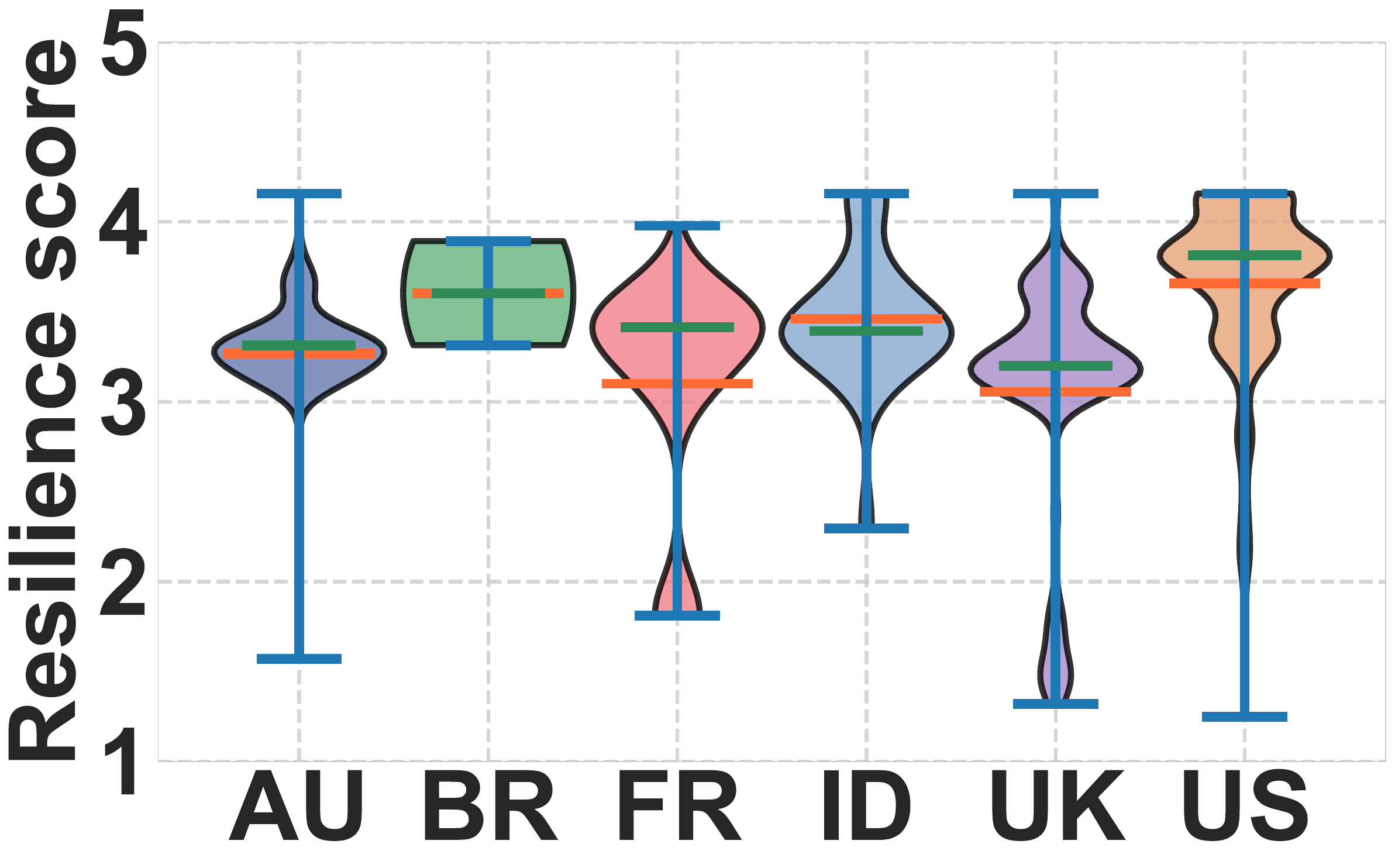}
			\caption{Placement focused}
			\label{fig:placement_focused}
		\end{subfigure}
		\vspace{-2mm}
		\caption{Overall authoritative DNS resilience scores for government domain names in the six studied countries with the weights of ``placement'', ``config'', and ``dispatch'' scores set to (a) high, low and low, (b) low, high and low (c) low, low and high, and (d) medium, low and low.}
		\label{fig:violin_plot_overall_variation}
	\end{figure}
	
	\begin{figure}[h]
		\centering
		\begin{subfigure}{0.24\linewidth}
			\includegraphics[width=\linewidth]{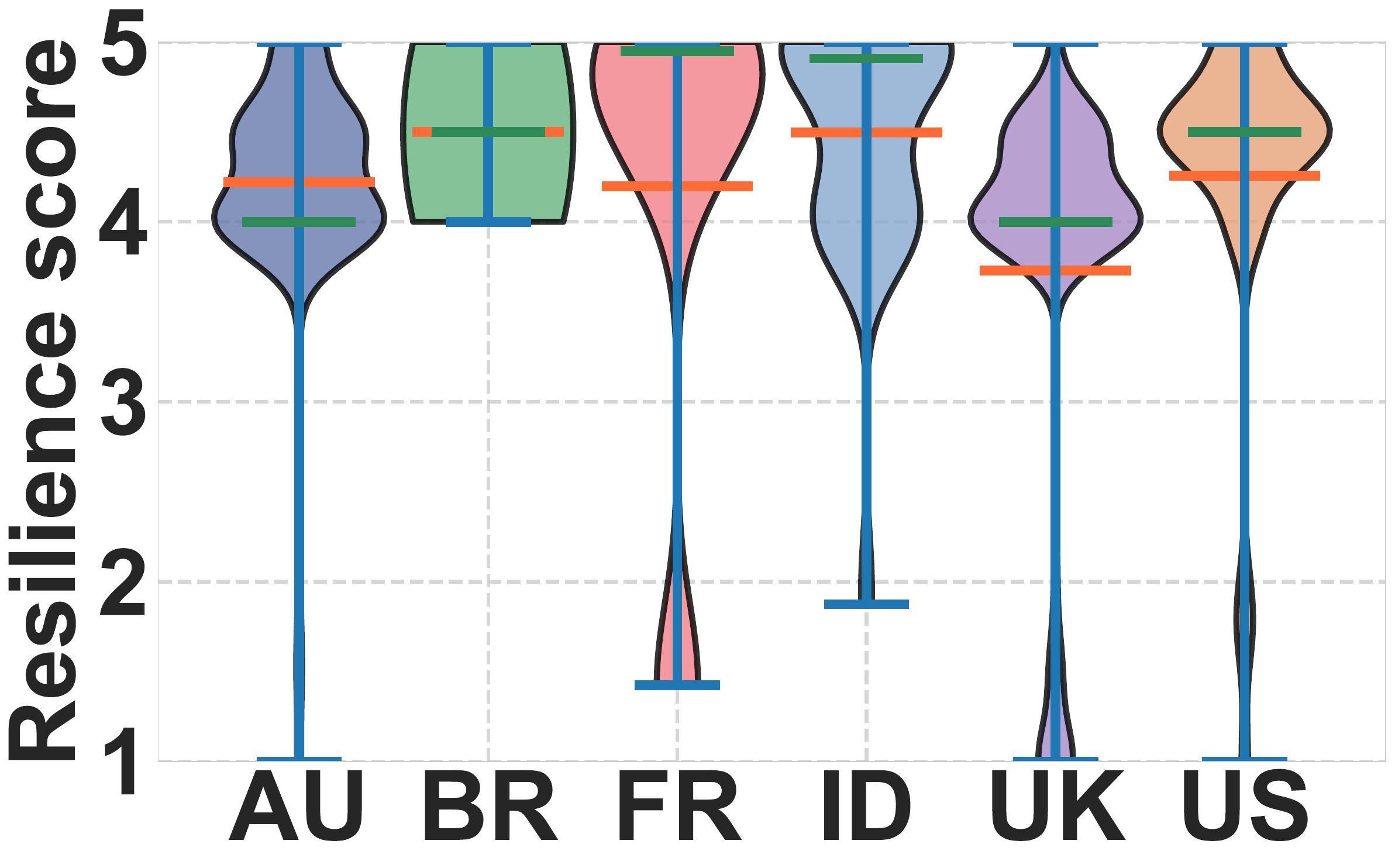}
			\caption{Primary prioritized}
			\label{fig:placement_primary_prioritised}
		\end{subfigure}
		\begin{subfigure}{0.24\linewidth}
			\includegraphics[width=\linewidth]{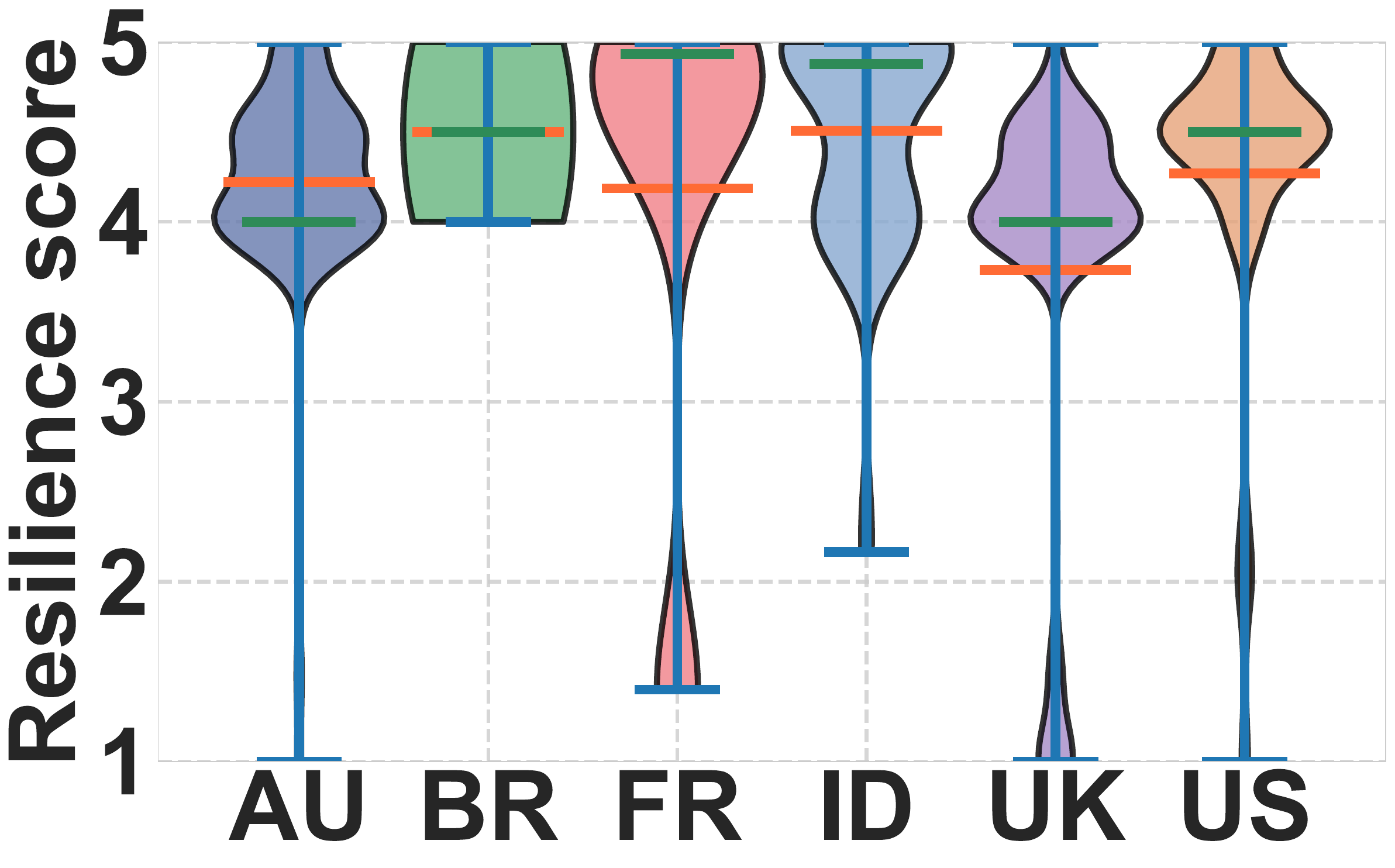}
			\caption{Primary focused}
			\label{fig:placement_primary_prioritised}
		\end{subfigure}
		\begin{subfigure}{0.24\linewidth}
			\includegraphics[width=\linewidth]{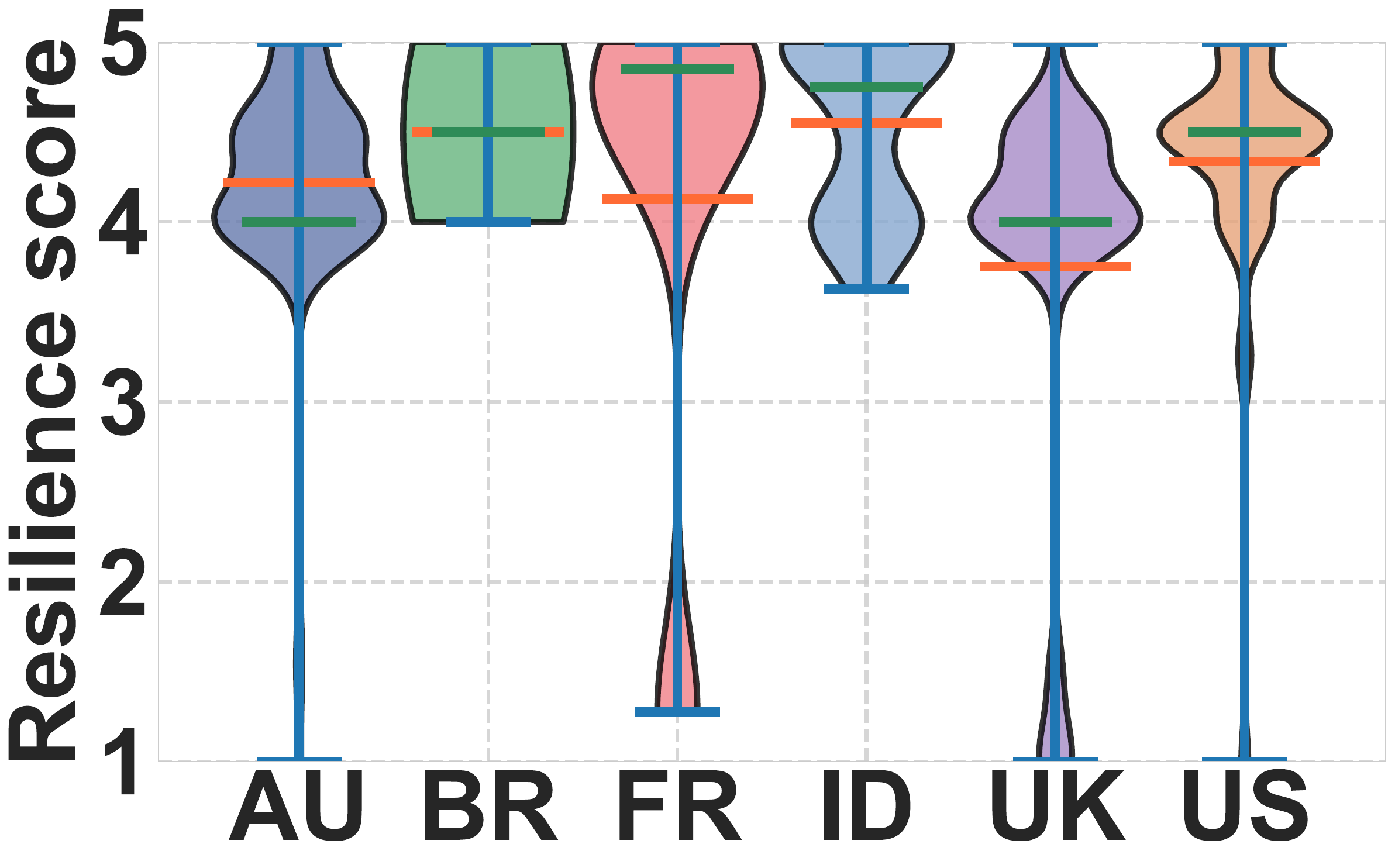}
			\caption{Authori. prioritized}
			\label{fig:placement_authoritative_prioritised}
		\end{subfigure}
		\begin{subfigure}{0.24\linewidth}
			\includegraphics[width=\linewidth]{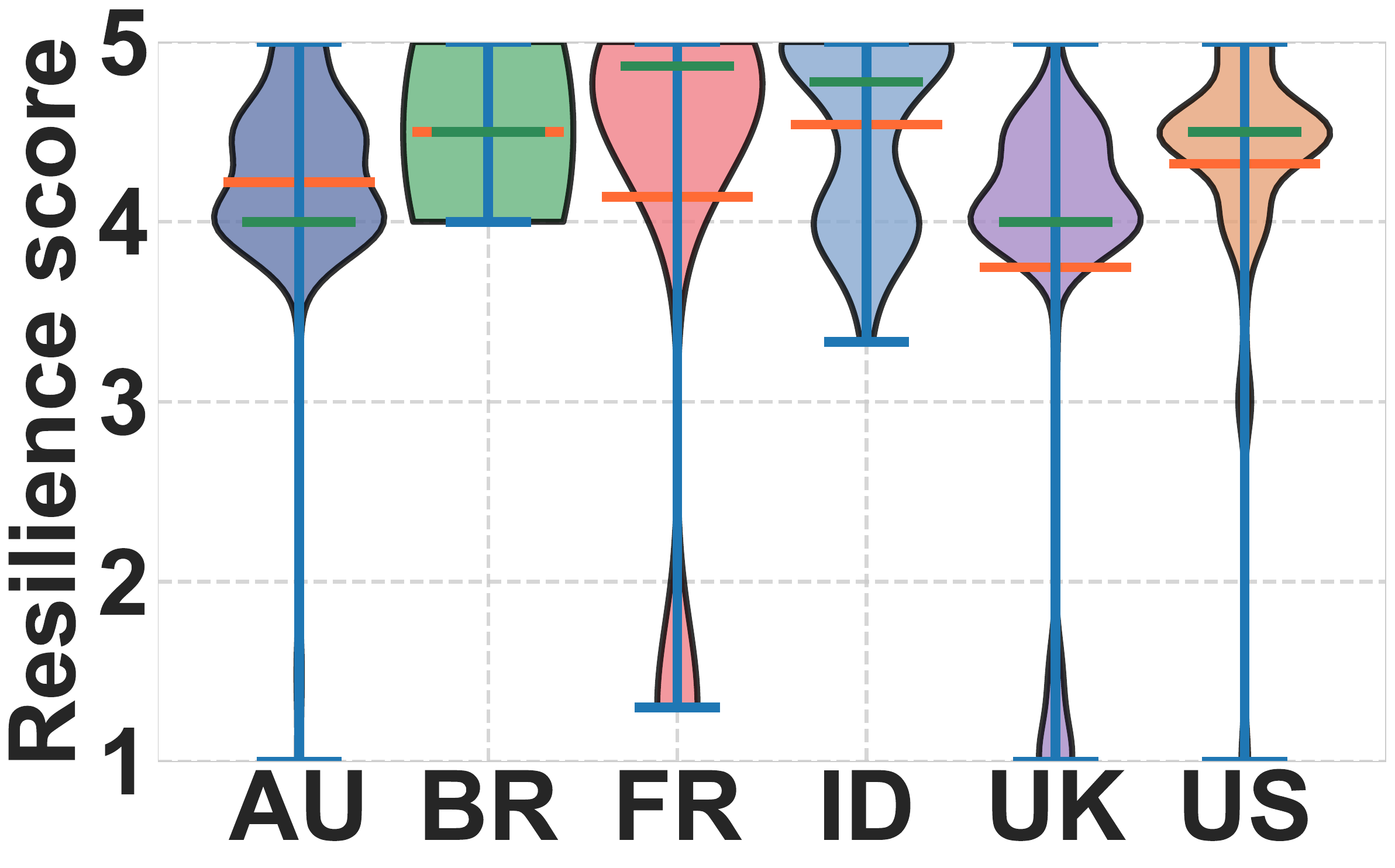}
			\caption{Authori. focused}
			\label{fig:placement_authoritative_focused}
		\end{subfigure}
		\vspace{-2mm}
		\caption{Authoritative DNS resilience scores in ``placement'' for government domain names in the six studied countries with the weights of ``prim\_place'' and ``auth\_place'' scores set to (a) high and low, (b) medium and low (c) low and high, and (d) low and medium.}
		\label{fig:violin_plot_placement_variation}
	\end{figure}
	
	\begin{figure}[h]
		\centering
		\begin{subfigure}{0.24\linewidth}
			\includegraphics[width=\linewidth]{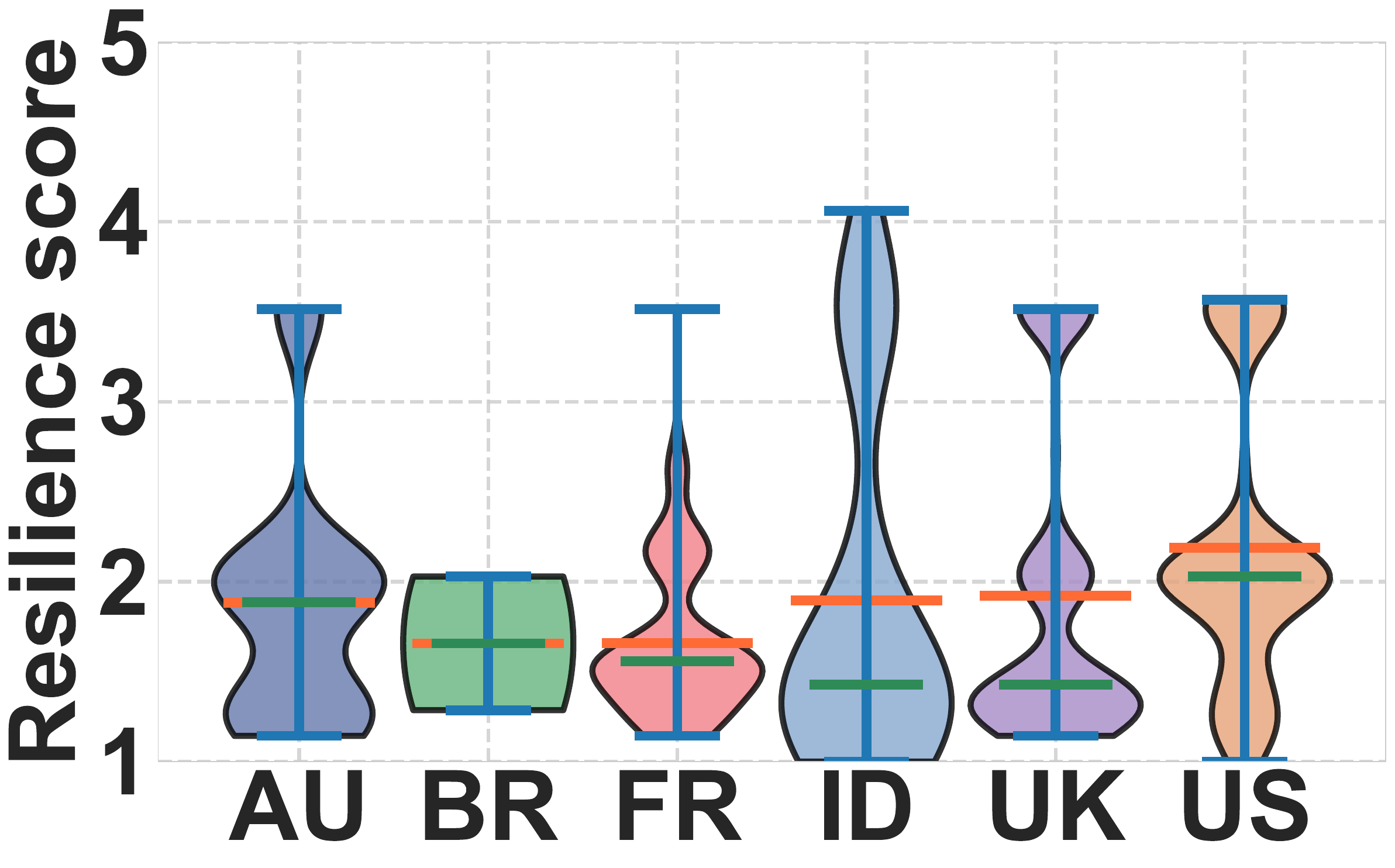}
			\caption{Primary prioritized}
			\label{fig:configuration_primary_prioritised}
		\end{subfigure}
		\begin{subfigure}{0.24\linewidth}
			\includegraphics[width=\linewidth]{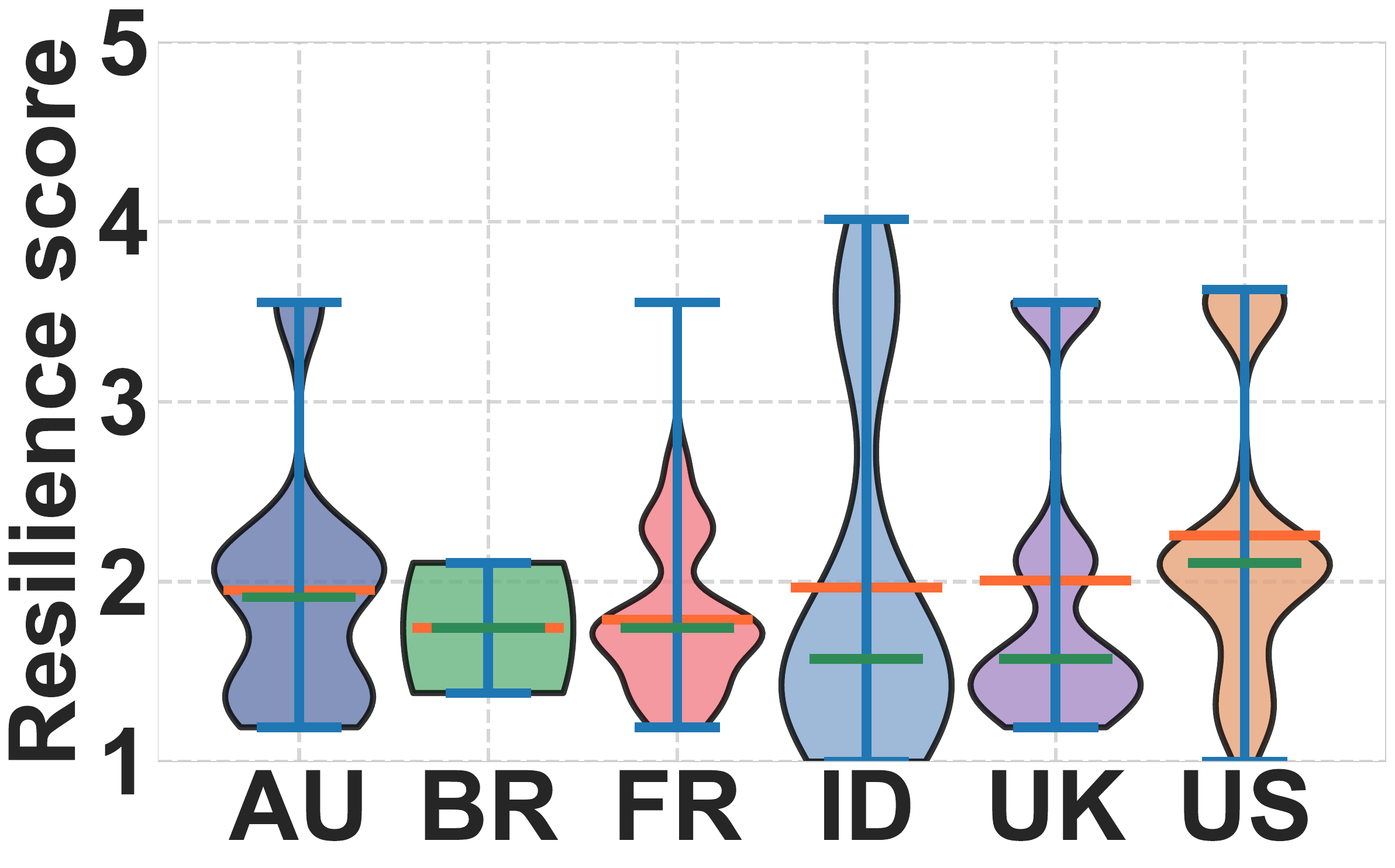}
			\caption{Primary focused}
			\label{fig:configuration_primary_prioritised}
		\end{subfigure}
		\begin{subfigure}{0.24\linewidth}
			\includegraphics[width=\linewidth]{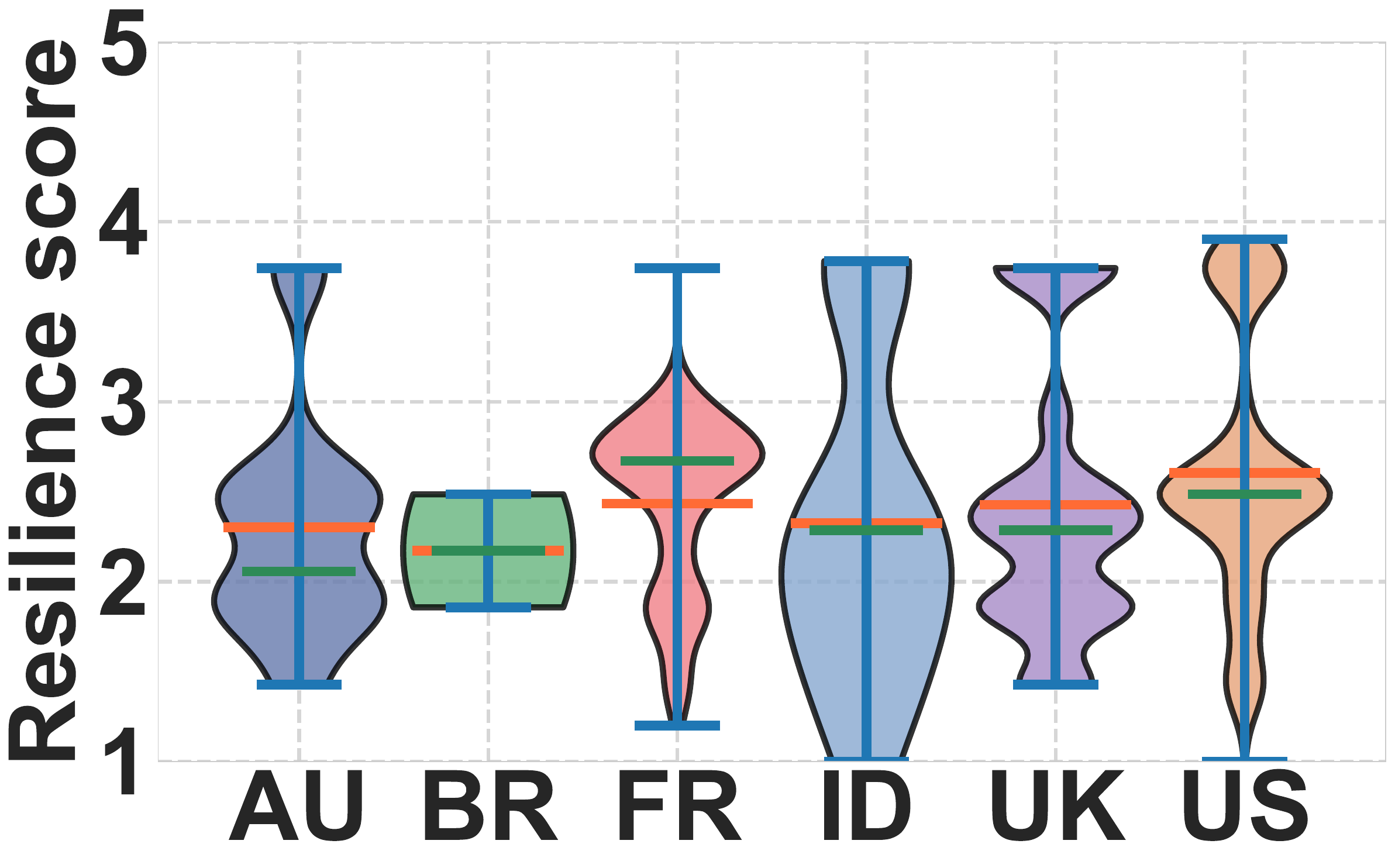}
			\caption{Authori. prioritized}
			\label{fig:configuration_authoritative_prioritised}
		\end{subfigure}
		\begin{subfigure}{0.24\linewidth}
			\includegraphics[width=\linewidth]{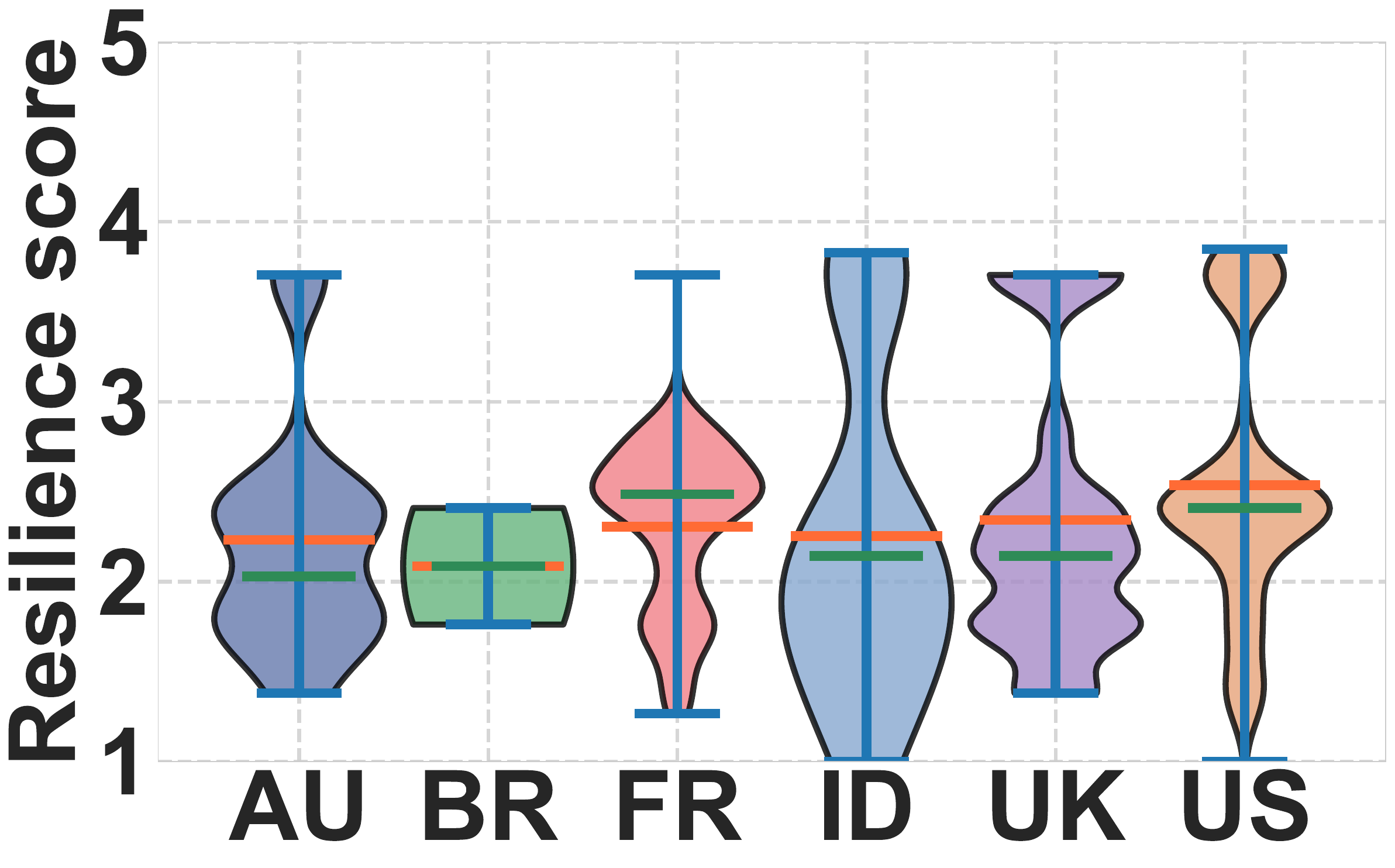}
			\caption{Authori. focused}
			\label{fig:configuration_authoritative_focused}
		\end{subfigure}
		\caption{Authoritative DNS resilience scores in ``configuration'' for government domain names in the six studied countries with the weights of ``prim\_config'' and ``auth\_config'' scores set to (a) high and low, (b) medium and low (c) low and high, and (d) low and medium.}
		\label{fig:violin_plot_configuration_variation}
	\end{figure}
	
	\begin{figure}[h]
		\centering
		\begin{subfigure}{0.24\linewidth}
			\includegraphics[width=\linewidth]{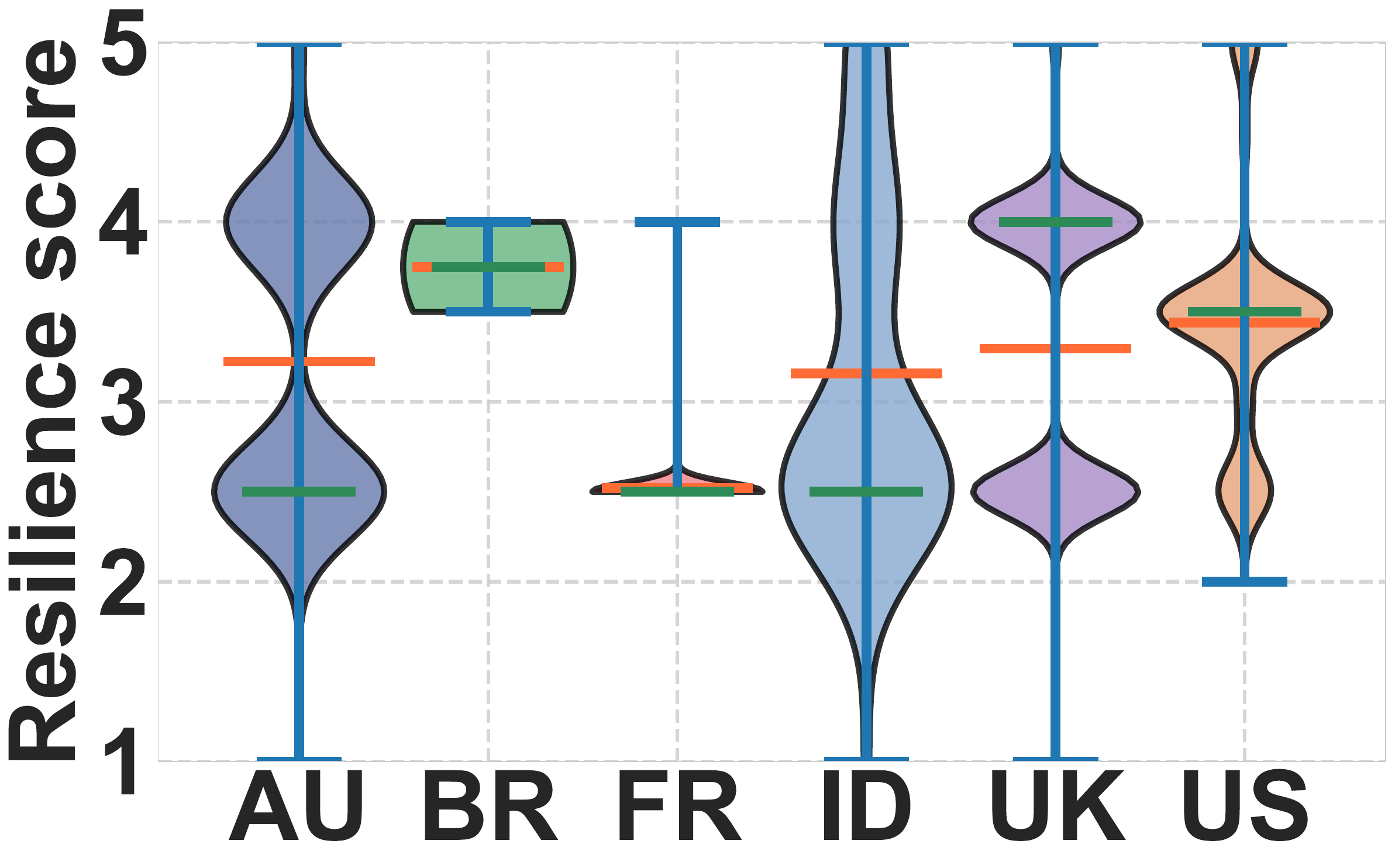}
			\caption{Primary prioritized}
			\label{fig:dispatch_primary_prioritised}
		\end{subfigure}
		\begin{subfigure}{0.24\linewidth}
			\includegraphics[width=\linewidth]{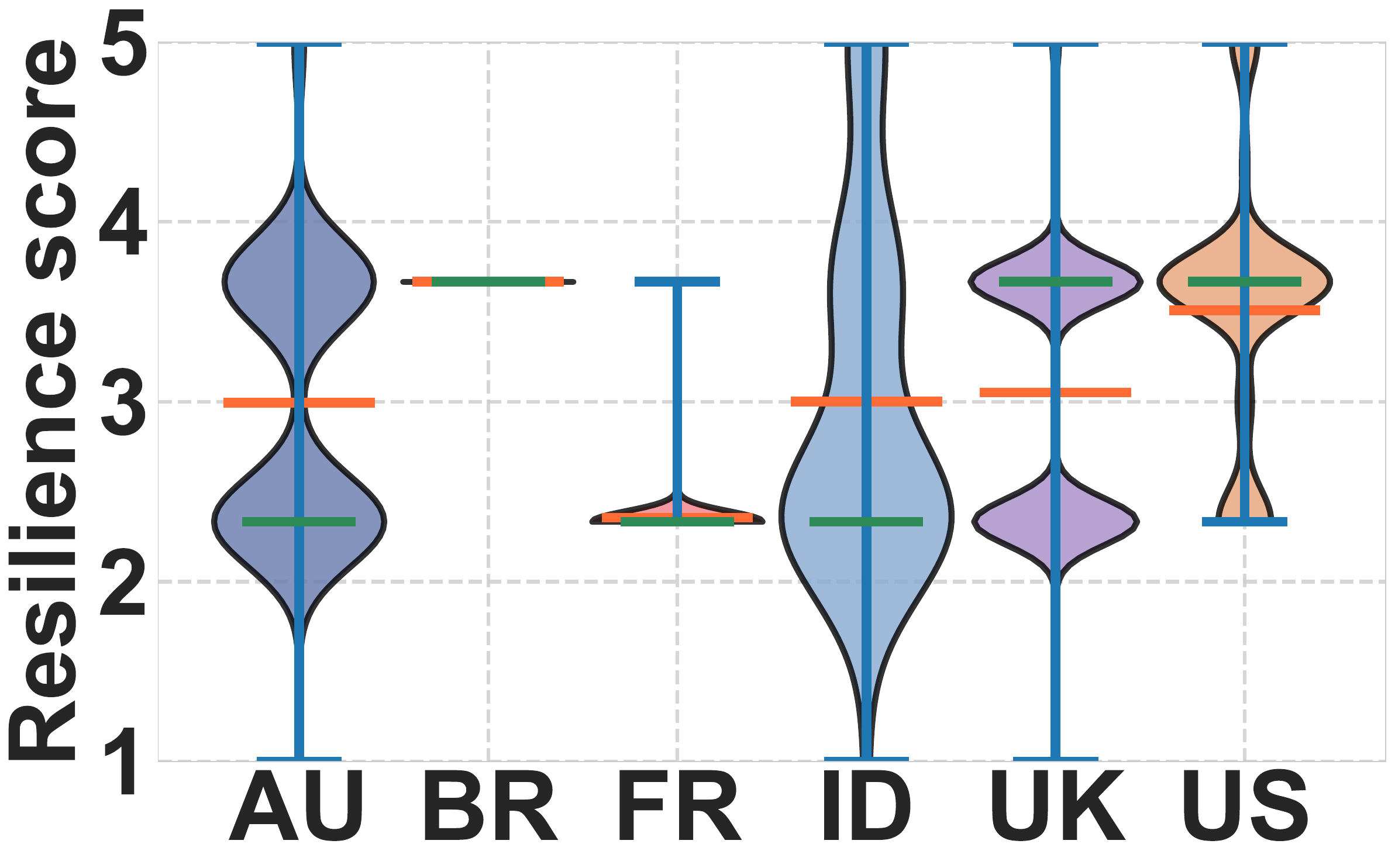}
			\caption{Primary focused}
			\label{fig:dispatch_primary_prioritised}
		\end{subfigure}
		\begin{subfigure}{0.24\linewidth}
			\includegraphics[width=\linewidth]{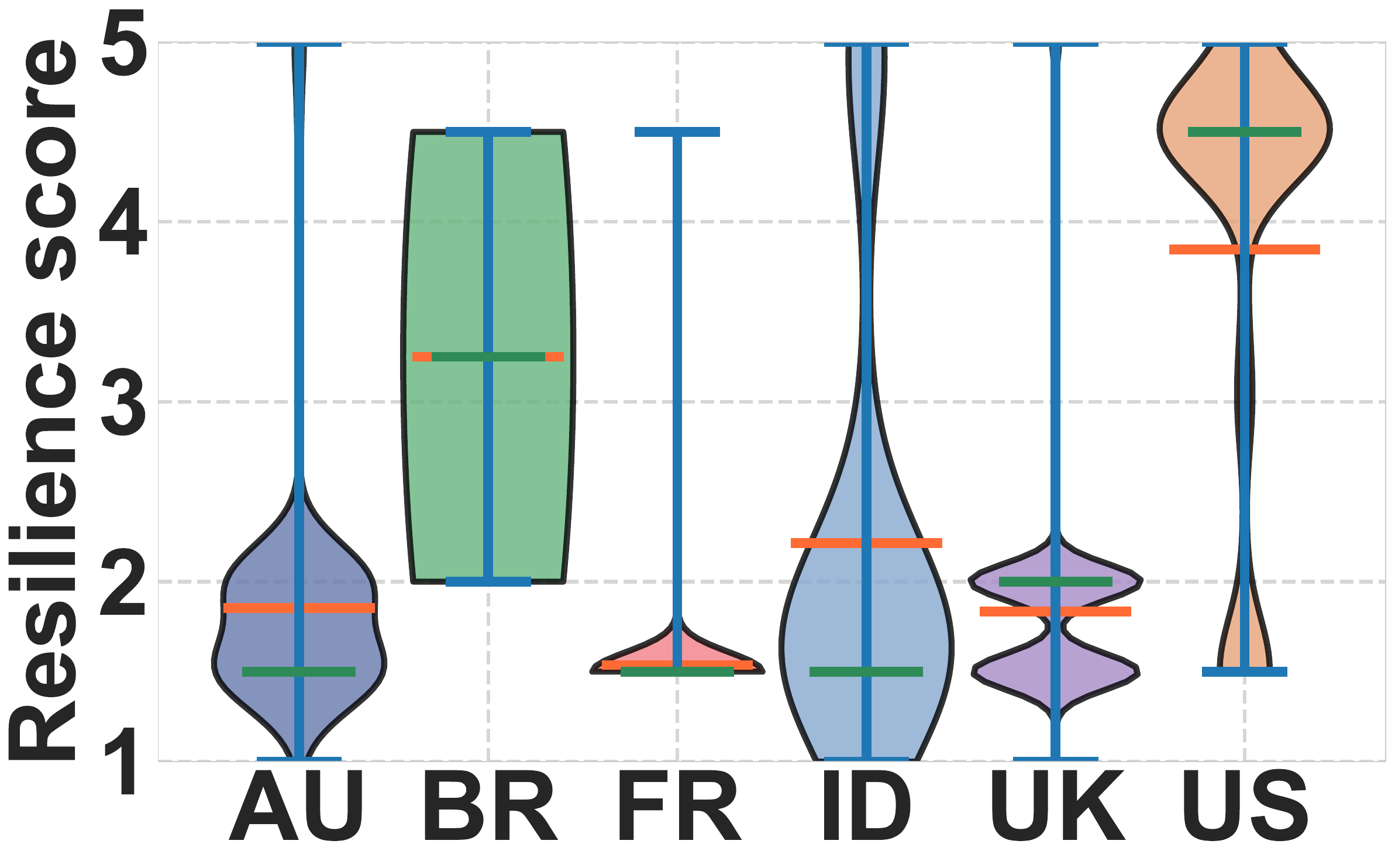}
			\caption{Authori. prioritized}
			\label{fig:dispatch_authoritative_prioritised}
		\end{subfigure}
		\begin{subfigure}{0.24\linewidth}
			\includegraphics[width=\linewidth]{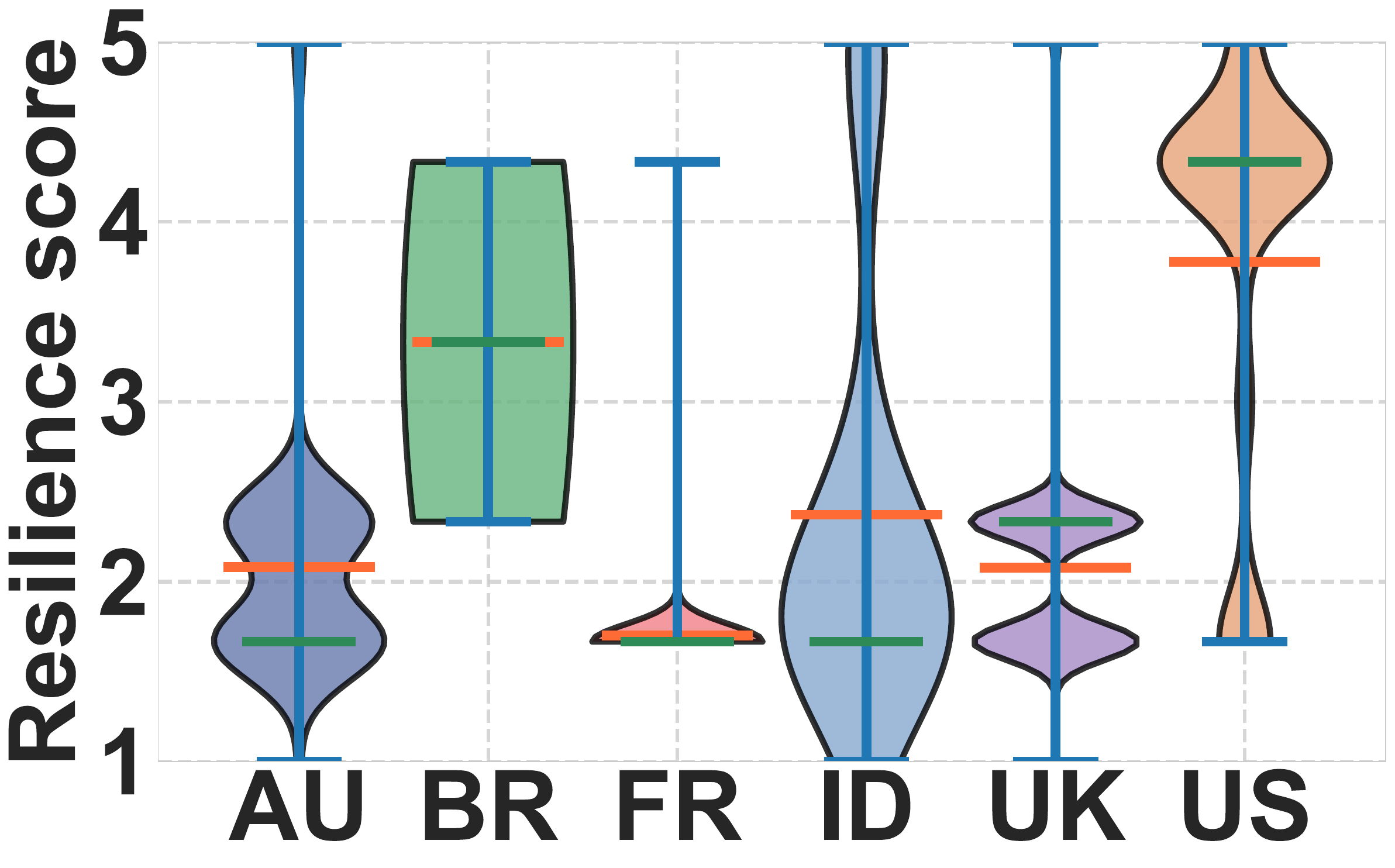}
			\caption{Authori. focused}
			\label{fig:dispatch_authoritative_focused}
		\end{subfigure}
		\caption{Authoritative DNS resilience scores in ``dispatch'' for government domain names in the six studied countries with the weights of ``prim\_disp'' and ``auth\_disp'' scores set to (a) high and low, (b) medium and low (c) low and high, and (d) low and medium.}
		\label{fig:violin_plot_dispatch_variation}
	\end{figure}
	
	In \S\ref{sec:insights}, we present our assessment results for each country by aggregating attribute-level resilience scores to each operational phase and domain levels using equal weights. In this section, we examine how these results vary under alternative weighting schemes that emphasize different operational phases.
	
	Fig.~\ref{fig:violin_plot_overall_variation} shows the overall domain-level scores across the six countries under four weighting strategies: prioritizing placement (Fig.~\ref{fig:placement_prioritised}), configuration (Fig.~\ref{fig:config_prioritised}), and dispatch (Fig.~\ref{fig:dispatch_prioritised}) by assigning them a higher weight ``3'' relative to others ``1'', as well as a scheme that moderately emphasizes placement (weight ``2'').
	Compared to the balanced results in Fig.~\ref{fig:violin_mean}, the U.S. consistently shows the highest resilience across all schemes, while Brazil performs better when dispatch is prioritized. When configuration is emphasized, resilience scores decrease across all countries by approximately one point on the five-point scale. 
	
	We further examine the impact of weighting differences between primary and authoritative functions within each operational phase in Fig.~\ref{fig:violin_plot_placement_variation}, Fig.~\ref{fig:violin_plot_configuration_variation} and Fig.~\ref{fig:violin_plot_dispatch_variation}. The observed variations highlight the flexibility  of our assessment framework, which can be adapted to different evaluation priorities and scenarios.

\end{document}